\documentclass[fleqn,useAMS,usenatbib]{mn2e}

\bibliography{biblio}

\usepackage{mathptmx}

\usepackage{color,graphicx,amsmath,url,lscape,longtable,amssymb}

\title[Debris around FGK stars]{Analysis of the \emph{Herschel} DEBRIS Sun-like star sample}

\author[B.~Sibthorpe et al.]
{\parbox{\textwidth}{B.~Sibthorpe,$^{\! 1,2}$\thanks{E-mail: \texttt{bsibthorpe@gmail.com}}
G.\,M.~Kennedy,$^{\!3}$
M.\,C.~Wyatt,$^{\!4}$
J.-F.~Lestrade,$^{\!5}$
J.\,S.~Greaves,$^{\!6}$
B.\,C.~Matthews,$^{\!7}$
G. Duch\^ene$^{8,9}$
  }
\vspace{0.4cm}\\
\parbox{\textwidth}{
  $^{1}$SRON Netherlands Institute for Space Research, P.O. Box 800, 9700 AV, Groningen, The Netherlands\\
  $^{2}$Airbus Defence and Space, Gunnels Wood Road, Stevenage, Hertfordshire SG1 2AS, UK\\
  $^{3}$Department of Physics, University of Warwick, Gibbet Hill Road, Coventry CV4 7AL, UK\\
  $^{4}$Institute of Astronomy, University of Cambridge, Madingley Road, Cambridge CB3 0HA, UK\\
  $^{5}$Observatoire de Paris - LERMA, CNRS, 61 Av. de l'Observatoire, 75014, Paris, France\\
  $^{6}$School of Physics and Astronomy, Cardiff University, Queens Buildings, The Parade, Cardiff, CF24 3AA, UK\\
  $^{7}$National Research Council of Canada, 5071 West Saanich Rd, Victoria, BC V9E 2E7, Canada\\
  $^{8}$Astronomy Department, University of California Berkeley, Berkeley CA 94720-3411 USA\\
  $^{9}$Universit\'e Grenoble Alpes, CNRS, Institut d'Astrophysique de Grenoble, F-38000 Grenoble, France
}}

\begin{document}

\date{Accepted YYYY MMMM DD. Received YYYY MMMM DD; in original form YYYY MMMM DD}

\pagerange{\pageref{firstpage}--\pageref{lastpage}} \pubyear{2017}

\maketitle

\label{firstpage}

\begin{abstract}
This paper presents a study of circumstellar debris around Sun-like stars using data from the \emph{Herschel} DEBRIS Key Programme. DEBRIS is an unbiased survey comprising the nearest $\sim$90 stars of each spectral type A-M. Analysis of the 275 F-K stars shows that excess emission from a debris disc was detected around 47 stars, giving a detection rate of 17.1$^{+2.6}_{-2.3}$ per cent, with lower rates for later spectral types. For each target a blackbody spectrum was fitted to the dust emission to determine its fractional luminosity and temperature. The derived underlying distribution of fractional luminosity versus blackbody radius in the population showed that most detected discs are concentrated at $f\sim10^{-5}$ and at temperatures corresponding to blackbody radii 7-40 AU, which scales to $\sim40$\,AU for realistic dust properties (similar to the current Kuiper belt). Two outlying populations are also evident; five stars have exceptionally bright emission ($f>5\times10^{-5}$), and one has unusually hot dust $<4$\,AU. The excess emission distributions at all wavelengths were fitted with a steady-state evolution model, showing these are compatible with all stars being born with a narrow belt that then undergoes collisional grinding. However, the model cannot explain the hot dust systems - likely originating in transient events - and bright emission systems - arising potentially from atypically massive discs or recent stirring. The emission from the present-day Kuiper belt is predicted to be close to the median of the population, suggesting that half of stars have either depleted their Kuiper belts (similar to the Solar System), or had a lower planetesimal formation efficiency.
\end{abstract}


%-------------------------------------------------------------------------------
\section{Introduction}
%-------------------------------------------------------------------------------
Debris discs are belts of dusty circumstellar material created during the on-going collision of orbiting planetesimals throughout a star's lifetime \citep{Wyatt2008}. The requirement for planetesimals, themselves a step in the planet formation process, makes debris discs a fundamental component of planetary systems.  Therefore, by studying the incidence, properties and evolution of debris discs it is possible to obtain a greater understanding of planetary systems \citep{Matthews2014a,Moro2015}.

Whilst the term incidence is often used in the literature, it in fact describes the disc detection fraction, and not the true incidence of debris. The detection fraction is the fraction of stars with clear (typically $>3\sigma$) emission in excess of the stellar photosphere at the location of the star, attributed to a debris disc. The sensitivity to such an excess varies significantly as a function of many stellar and disc parameters, as well as the depth, wavelength, angular resolution and calibration accuracy of the survey data. Consequently, it is difficult to directly compare `incidence' rates from various surveys of different stellar samples.

For consistency with previous works the term incidence is used in this paper to describe the detection fractions. However, it must be borne in mind that these are just a rough comparison since surveys utilize observations at different wavelengths, depths and resolutions, and in turn have very different sensitivities, even to the same disc or to the disc parameters around different stars. Knowledge of the biases and completeness thresholds of the data presented are used in an attempt to correct for these factors, and provide correct incidence rates, but still only within the range of the probed parameter space.

In recent years infrared data from the MIPS camera \citep{Rieke2004} on-board the \emph{Spitzer} Space Telescope \citep{Werner2004} has been used to investigate the properties of discs around `Sun-like' stars \citep[e.g.][]{Bryden2006, Trilling2008, Carpenter2009, Kains2011}.  Disc incidence rates of between 10 and 15 per cent have been reported for F, G and K spectral type stars, substantially lower than the $\sim$32 per cent found around A stars \citep{Su2006}.  More recently, however, the Dust Around Nearby Stars (DUNES) survey team \citep{Eiroa2013} reported a higher rate of 20.2$\pm$2 per cent.  This result was derived from new infrared data obtained with the PACS camera \citep{Poglitsch2010} on-board the \emph{Herschel}\footnote{Herschel is an ESA space observatory with science instruments provided by European-led Principal Investigator consortia and with important participation from NASA.} Space Observatory \citep{Pilbratt2010}.  

The variation in disc incidence as a function of observed wavelength makes it difficult to perform direct comparisons between these data sets.  Moreover, various biases, including variation in mean distance to different spectral types, makes it difficult to draw fundamental conclusions about these sources using such statistics, as a function of spectral type.  However, a trend of disc incidence with stellar age has been observed \citep{Su2006,Wyatt2007a,Carpenter2009}. In practice, disc incidence is a combination of the disc fractional luminosity ($f=L_{\rm{disc}}/L_{\rm{star}}$, a wavelength independent property), the detection limits of the data, and the distance of the source.  Therefore, the biases described can be characterised and accounted for using a large unbiased dataset, down to the lowest detection limit for the sample.

This paper presents an analysis of debris disc properties around such an unbiased sample of 275 stars of spectral types F, G and K.  A description of the target sample, their observation and data reduction is given in Section~\ref{sec2}.  This section includes information on source measurement and determination of the disc model parameters when a significant disc excess is detected. Section~\ref{sec3} presents the derived disc model parameters and discusses the incidence rates of discs within the fractional luminosity vs disc radius parameter space. The steady-state evolution of these sources is then presented in Section~\ref{sec4}, followed by a discussion of the results and summary of this work in Sections~\ref{sec5} and \ref{sec6} respectively.
%-------------------------------------------------------------------------------

%-------------------------------------------------------------------------------
\section{Sample and Observations}\label{sec2}
%-------------------------------------------------------------------------------
\subsection{The DEBRIS survey}
The work presented in this paper is based on data from the Disc Emission via a Bias Free Reconnaissance in the Infrared and Submillimetre \cite[DEBRIS;][]{Matthews2010} \emph{Herschel} Key Programme (KPOT\_bmatthew\_1).  DEBRIS is a survey of 446 nearby stars of spectral types A-M.  All targets were observed at 100 and 160\,$\mu$m using the PACS photometer, with additional 70\,$\mu$m follow-up of interesting sources. Particularly bright sources were also followed-up with SPIRE photometry \citep{Griffin2010} at 250, 350 and 500\,$\mu$m. The detection of infrared discs excesses in these data is limited by instrument noise and confusion with background objects. In no cases is excess detection purely limited by uncertainties in the instrument calibration or stellar photospheric flux.

DEBRIS was executed in partnership with the DUNES \emph{Herschel} Key Programme, with sources common to both teams being observed only once by either team, and the resulting data shared. As a result, 98 DEBRIS targets were observed by the DUNES team (project code KPOT\_ceiroa\_1). By design DEBRIS was a flux-limited survey. However, due to the data sharing arrangement with the DUNES programme, and their need to often go beyond the nominal DEBRIS flux density limit, many of the DEBRIS observations are deeper than this limit. This is also the case for DEBRIS-only targets when follow-up observations of interesting sources was performed \citep[e.g.][]{Lestrade2012}. The analysis presented here is designed to make the maximum use of the available data, consequently where additional data are available they are included and used. As a result some targets have flux limits lower than the nominal DEBRIS level. Since targets are at a range of distances and around varied stellar spectral types, disc sensitivity inherently varies between individual targets, even within a flux-limited data set. Therefore including deeper observations where possible does not adversely impact on the unbiased nature of this sample. 

The DEBRIS sample is drawn from the Unbiased Nearby Stars catalogue \citep[UNS;][]{Phillips2010}, with omissions being made only when the predicted 1$\sigma$ cirrus confusion noise level was considered too high to provide a useful debris disc detection ($\ge1.2$\,mJy at 100\,$\mu$m). Two further source in the DEBRIS catalogue, $\epsilon$ Eridani (K001) and $\tau$ Ceti (G002), observed as part of the \emph{Herschel} guaranteed time programme (KPGT\_golofs01\_1) have been included in this sample. Whilst these targets were not observed by DEBRIS, they were in the original DEBRIS target list, and therefore do not bias the sample by their inclusion. Had DEBRIS not been prevented from observing these sources due to duplication with the guaranteed time programme they would have been including in the submitted target list.

The sample used is volume limited, being made up of the nearest $\sim$90 targets of each spectral type that passed the cirrus confusion limit cut. As a result, this sample is free of bias towards any particular stellar parameters, or any prior knowledge of the disc or planetary system. The varied frequency of different spectral types, however, means that different volume limits are used for each spectral type star, with the more common M-types having the smallest limit, and the rarer A-types having the largest limit. Since the disc detection limit varies as a function of target distance, and stellar luminosity, disc detection biases do exist across the range in spectral types. Even so, the variable distance limits do not necessarily imply that this sample is biased against finding large numbers of discs around early type star sub-sample, and vice versa for late types. The data from this survey provide a robust statistical data set from which to study debris discs.

\subsection{The FGK sample}
\begin{figure}
  \includegraphics[width=8.5cm]{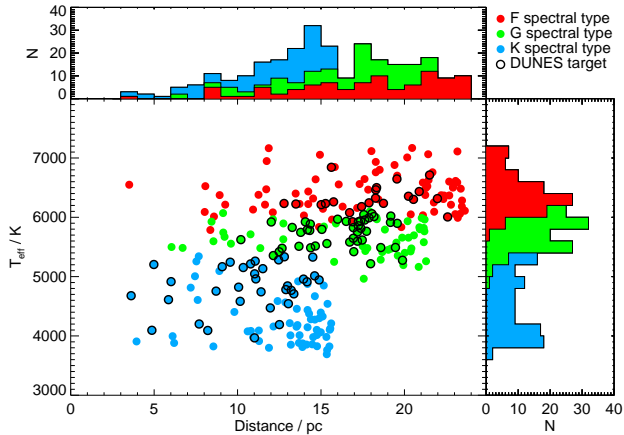}
  \caption{Distribution of stellar distance and effective temperature for the DEBRIS FGK star sample. Targets within the DEBRIS sample, observed by DUNES, are also shown for comparison.}
\label{fig1}
\end{figure}

The DEBRIS sample contains 94, 88 and 91 F, G and K-type stars (hereafter FGKs) respectively, and includes stars observed by both DEBRIS and DUNES. The sample is volume limited, with the largest distance to an F, G and K star being 24, 21 and 16\, pc respectively (Figure~\ref{fig1}). In cases where multiple star systems are present \citep{Rodriguez2015}, only the primary star is included in this work.  Since these are all field stars, and not generally members of clusters or associations, the stellar ages are uncorrelated.  

\begin{figure}
  \includegraphics[trim=6mm 0 0 0,width=8.5cm]{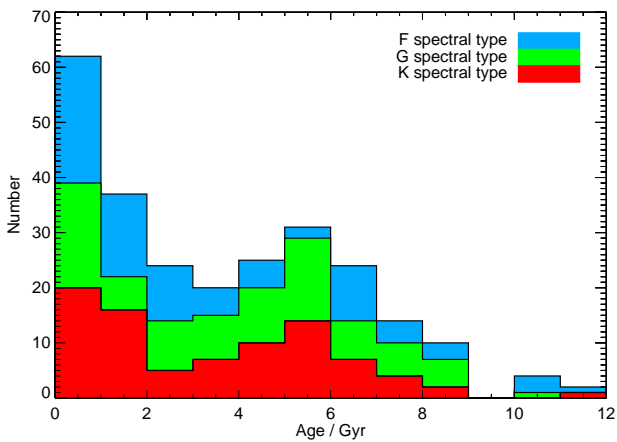}
  \caption{Histogram of stellar ages within the DEBRIS FGK star sample.}
\label{fig2}
\end{figure}

As the methods used to determine the ages of stars typically have different systematic errors associated with them, it was decided to use a single age determination method for all sources. This means that any systematic error in the stellar ages is common to all sources, and therefore could be discounted when assessing trends in the data. The ages are determined using chromospheric activity as an indicator and all ages were taken from the work of \cite{Vican2012}. Whilst the ages in \cite{Vican2012} have been disputed, the benefits of the uniform approach to stellar ages provided by this work, and applicability to the DEBRIS sample, makes them a good choice for this analysis. Age trends are used only in the modelling work presented here, and the age uncertainties do not have a significant impact on the conclusions drawn. The range of stellar ages in this sample is 1\,Myr--11\,Gyr with a median sample age of 3.3\,Gyr; a histogram of stellar ages is given in Figure~\ref{fig2}.


\subsection{Observations and data reduction}
The `mini scan-map' observing mode was used for all PACS observations.  Two scans of each target were performed with a relative scanning angle of 40 deg to mitigate striping artifacts associated with low frequency noise. Scan-maps used a scanning rate of 20~arcsec per second and were constructed of 3~arcmin long scan legs with a separation of 4~arcsec between legs.  The nominal DEBRIS observations used 8 scan-legs per map and performed 2 map repeats per scanning direction.  DUNES led observations typically used 10 scan-legs per map, and performed 2 or more map repeats, depending on the specific source.  The only practical impact of these different observing parameters is a change in the noise level in the images, and hence the disc detection threshold.

The data were reduced using Version 10.0 of the Herschel Interactive Pipeline Environment \citep[HIPE;][]{Ott2010}.  The standard pipeline processing steps were used and maps were made using the photProject task.  The time-ordered data were high-pass filtered, passing scales smaller than 66~arcsec at 70 and 100\,$\mu$m and 102 arcsec at 160\,$\mu$m (equivalent to a filter radius of 16 and 25 frames respectively), to remove low frequency noise in the scan direction. Sources $\ge$2$\sigma$ were then identified in this first stage `dirty' map to create a filter mask. The original data are then filtered a second time, using the derived mask to exclude bright sources which would otherwise result in ringing artifacts, and a final map produced. To maximise the signal to noise ratio of the output maps data from the telescope turn-around phase, at the end of each scan leg, were included in these reductions. Data were included for telescope scan rates down to 5 arcsec per second.

\subsection{Source extraction and photometry}
Flux density measurements of the targets were made using a combination of point spread function (PSF) fitting and aperture photometry. \emph{Herschel} calibration observations were used for PSFs, and because the maps were created in sky coordinates, the PSFs were rotated to the angle appropriate for a given observation. PSF fitting was used by default, with apertures used where the PSF fitting residuals revealed resolved sources. For clean unresolved sources it was found that PSF fitting and aperture photometry yielded very similar results; the preference for PSF fitting is largely to allow mitigation of the effects of confusion from additional point sources.

As PACS observes either 70 or 100\,$\mu$m simultaneously with 160 $\mu$m, both wavelengths were fitted for an observation at the same time, with the source location the same in each image. That is, for a single PACS observation there are four parameters to fit; the x/y position and the flux density at 70/100 and 160 $\mu$m. The same approach was used for SPIRE, but the three wavelengths, 250/350/500 $\mu$m, were fitted simultaneously. In the case of aperture measurements, the location was determined from the PSF fitting, and the appropriate aperture size chosen by hand.

As the locations of the target stars were well known, the PSF fitting routine was initialised at the expected star location (or locations in the case of multiple systems), and then the MPFIT least-squares minimisation routine was used to find the best fitting point source model for each observation. In cases where additional sources were visible (e.g. background galaxies) additional point sources were added to avoid biasing the fluxes. In complex cases where both a resolved disc and confusion were present, a more individually tailored approach was taken \citep[e.g.][]{Wyatt2012,Lestrade2012}.

As noted in the Herschel documentation (PICC-ME-TN-037), the calibration depends somewhat on the data reduction, in particular the filter scale, masking, drizzling and frame selection. Therefore, in order to calibrate photometry for the specific data reduction pipeline used, aperture corrections were independently derived. These use all observations with 70/110 deg cross-linked scans used in the aforementioned calibration document. Aperture photometry with 4, 5, and 8 arcsec radius apertures (for maximal S/N) was made at the position found by fitting a 2 dimensional Gaussian. These fluxes were colour corrected by dividing by colour corrections of 1.016, 1.033, and 1.074. The conversion for these aperture sizes was then derived by dividing each flux by the expected photospheric fluxes for each target at each wavelength. The corrections obtained are 0.4859, 0.5275, and 0.5321, with standard deviations of 1.6 to 1.9 per cent. These compare favourably with the supplied values of 0.476, 0.513, and 0.521.

The PSF fitting results are then compared to aperture photometry using these corrections. The PSF fitted values are systematically low by $\sim$20 per cent due to flux lost in the wings of the PSF by filtering the images. The typical aperture/PSF-fit flux ratio for a large number of targets is derived to be 1.19 at 100\,$\mu$m, and 1.21 at 160\,$\mu$m, with uncertainties of about 0.05 \citep{Kennedy2012a}. At 70\,$\mu$m deriving this factor is more difficult since there are fewer high S/N observations of point sources, since DEBRIS and DUNES preferentially targeted resolved discs for observations in this band. PSF fitting to very high S/N calibration sources was found to be only marginally useful, since the PSF fits are generally poor due to variation in the PSF shape at 70\,$\mu$m. A tentative value of 1.16 was found by \cite{Kennedy2012a}, and there is as yet no evidence that this value is incorrect, so we retain this value.

Estimates of the uncertainty in the measured fluxes were made by measuring the flux density dispersion from apertures placed at several hundred locations in high coverage regions of the maps. Uncertainties are also returned directly from the PSF fitting, but because the chosen pixel size and the drizzling method result in correlated noise, these uncertainties are underestimated by a factor of about 3.6 \citep{Fruchter2002,Kennedy2012a}. It was found that these two methods yielded comparable results, with the main difference being that the PSF fitting results are not necessarily sensitive to larger scale variations in the local background. Thus, the final uncertainties used were the larger of these two methods.

\subsection{Stellar photosphere and disc modelling}\label{sec2.5}
Debris discs can be discovered by infrared (IR) excesses or resolved images. In nearly all cases where a disc is resolved in thermal emission an IR excess is also seen, though this is not always the case for mid-IR imaging \citep{Moerchen2010}. Thus, to detect debris discs around DEBRIS stars a model for the stellar photospheric emission is needed. Optical photometry and near/mid-IR from a variety of sources yields a stellar model, and the extrapolation of this model to longer wavelengths allows a comparison with the \emph{Herschel} photometry and a test of whether an IR excess is present.

The spectral energy distribution (SED) modelling method implemented has been used successfully for DEBRIS and other surveys \citep{Kennedy2012b,Kennedy2012c}. Synthetic photometry of stellar atmosphere models yields flux densities that are compared with observations, and the best fitting model found by a combination of brute-force grid searches and least squares minimisation. The primary goal is to make the best predictions of the stellar fluxes in the IR, so the approach has not been fine tuned in an attempt to provide precise effective temperatures, surface gravities or metallicities. The former typically agree with other results to within a few hundred K, sufficient for our purposes here. Distances are known to all of our target stars, so the main stellar parameters the SED fitting yields are the effective temperature, radius, and luminosity.

Where available, data from the \emph{Spitzer} MIPS at 24 and 70\,$\mu$m and \emph{Spitzer} InfraRed Spectrograph \citep[IRS;][]{Houck2004} spectra from \cite{Lebouteiller2011} were also used in the SED modelling for the DEBRIS FGK sample. As described in Wyatt et al. (2012), synthetic photometry of these spectra in seven artificial bands between 5 and 33 $\mu$m was derived, and then used in the SED fitting in the same way as all other photometry.

Once the best fitting stellar model is found, the ratio of the observed flux, $F_{\lambda}$, to the stellar photospheric flux, $P_{\lambda}$, can be determined ($R_{\lambda}=F_{\lambda}/P_{\lambda}$), which is used in the modelling below. The value of $R_{\lambda}$ does not indicate the significance of any excess however. The significance metric for an IR excess in some band $B$ with observed flux $F_{\rm B}$ and photospheric flux $P_{\rm B}$ is
\begin{equation}
	\chi_{\rm B} = \frac{F_{\rm \lambda} - P_{\rm B}}
	{\sqrt{\sigma_{F_{\rm \lambda}}^2 + \sigma_{P_{\rm B}}^2}},
\end{equation}
where $\sigma_{\rm F_{\rm B}}$ and $\sigma_{\rm P_{\rm B}}$ are the uncertainty on the photometry and photosphere model respectively. If the photometry is very precise and $\sigma_{F_{\rm \lambda}}$ has reached a minimum possible value, set at some fraction of the total flux due to the accuracy of the calibration of the instrument, then the measurement can become ``calibration limited''. This is generally the case with MIPS 24\,$\mu$m for example. If the photometry is less precise then it is said to be ``sensitivity limited''. In this work a detection significance threshold of 3 is used, meaning that a measurement must be 3 standard deviations above the photosphere to be considered a real excess.

Because an individual star typically has several mid- to far-IR measurements with which the presence of an IR excess could be detected, it is possible potentially to look for faint excesses where the significance in no individual band exceeds 3, but collectively an excess appears significant (e.g. two $\chi_{\rm B}=2.9$ excesses). There are some cases like this among DEBRIS stars, however, such an approach was found to be problematic and the excesses discovered this way to be implausible in many cases. Thus, for a star to be deemed to possess an IR excess it is required that at least one band have $\chi_{\rm B}>3$.

If an excess is present, then a model is fit to the star-subtracted infrared photometry to derive some basic properties of the excess emission. The disc model used is simply a blackbody, with a modification to allow for inefficient emission from small grains at (sub)mm wavelengths. The disc model is therefore a Planck function at temperature $T_{\rm disc}$ with some solid angle $\Omega_{\rm disc}$, multiplied by $(\lambda_0/\lambda)^\beta$ beyond the ``turnover'' wavelength $\lambda_0$. The disc temperature, $T_{\rm disc}$ can be converted into a blackbody radius for a given stellar luminosity, $L_\star$, with $r_{\rm bb,disc}=(278.3/T_{\rm disc})^2 \sqrt{L_\star}$. It should be noted that whilst a blackbody model provides a useful representative radius, it can underestimate the true disc radius by up to a factor of 2.5 \citep{Booth2013,Pawellek2014}. Nonetheless, it provides a simple and useful reference disc radius of this general analysis. In some cases the excess emission is poorly modelled by a single blackbody, and for these a second blackbody component is added. The $\lambda_0$ and $\beta$ cannot be constrained for each component individually, so these are the same for both. For a detailed study of the identification and interpretation of these so-called ``two-temperature'' discs, including those identified in the DEBRIS sample, see \cite{Kennedy2014}, \cite{Chen2009}, \cite{Morales2011} and \cite{Ballering2013}. Here the cooler of the two components is used if two are found to be present. A two temperature fit was required for only three of the discs detected, so this approach has no significant impact on the results presented here.

Uncertainties on the fitted models are estimated in two ways. The first is simply the result of the least squares fitting. However, in many cases the parameters are poorly constrained and degeneracies mean that the least squares uncertainties are both underestimated and not representative. Uncertainties obtained by computing the $\Delta_{\chi^2}$ (relative to the best fitting $\chi^2$) from brute force grids are therefore also derived. For the stellar parameters these grids simply show that the least squares results provide useful uncertainties, hence the latter are used.

\begin{table*}
\centering
\caption{Modified blackbody disc model fit parameters, and detection significance, for all detected discs within the DEBRIS F, G and K spectral type sample. For completeness, extended discs are denoted by an $*$, however, the parameters reported in this table are for the equivalent modified blackbody, as used in this work. \emph{Herschel} and \emph{Spitzer} photometric data for the entire sample used in this paper, including non-detections, is provided in Appendix~\ref{tableA1}.}\label{table1}
\begin{tabular}{l r@{}c@{}l r@{}c@{}l r@{}c@{}l c c c}
\hline
\multicolumn{1}{l}{Target}                   & \multicolumn{3}{c}{$L_{\rm{bb,disc}}/L_{*}$} & \multicolumn{3}{c}{$T_{\rm{bb,disc}}$}       & \multicolumn{3}{c}{$r_{\rm{bb,disc}}$}       & \multicolumn{1}{c}{$\beta$}                  & \multicolumn{1}{c}{$\lambda_{\rm{0}}$}       & \multicolumn{1}{c}{$\chi_{\rm{tot}}$}        \\\multicolumn{1}{l}{}                         & \multicolumn{3}{c}{/ $\times10^{-6}$}        & \multicolumn{3}{c}{/ K}                      & \multicolumn{3}{c}{/ AU}                     & \multicolumn{1}{c}{}                         & \multicolumn{1}{c}{/ $\mu$m}                 & \multicolumn{1}{c}{}                         \\\hline
HD 166$^*$&66&&$^{+3.3}_{-2.9}$&86&&$^{+1.6}_{-2.3}$&8.3&&$^{+0.47}_{-0.30}$&0.70&76&36\\
HD 1581&0.58&&$^{+0.18}_{-0.17}$&23&&$^{+7.3}_{-10}$&160&&$^{+360}_{-68}$&1.1&160&4.7\\
HD 5133&8.7&&$^{+1.7}_{-2.8}$&32&&$^{+3.3}_{-4.8}$&42&&$^{+16}_{-7.5}$&0.0&30&9.7\\
HD 7570&9.0&&$^{+2.6}_{-2.6}$&74&&$^{+22}_{-21}$&20&&$^{+19}_{-8.2}$&2.3&45&7.1\\
HD 10647$^*$&290&&$^{+2.9}_{-4.5}$&49&&$^{+0.49}_{-0.89}$&41&&$^{+1.5}_{-0.80}$&0.62&70&170\\
HD 10700$^*$&6.1&&$^{+0.52}_{-0.39}$&63&&$^{+3.4}_{-6.5}$&14&&$^{+3.4}_{-1.4}$&0.10&59&22\\
HD 11171&4.3&&$^{+0.71}_{-0.63}$&58&&$^{+11}_{-8.9}$&56&&$^{+22}_{-16}$&3.0&170&13\\
HD 16673&7.9&&$^{+3.0}_{-2.5}$&98&&$^{+12}_{-27}$&11&&$^{+9.9}_{-2.4}$&2.3&120&5.3\\
HD 17925$^*$&29&&$^{+3.0}_{-3.4}$&73&&$^{+0.99}_{-8.9}$&9.3&&$^{+2.8}_{-0.25}$&3.0&350&20\\
HD 20794&1.6&&$^{+0.70}_{-0.42}$&65&&$^{+11}_{-26}$&15&&$^{+27}_{-3.9}$&0.0&30&5.9\\
HD 22049$^*$&54&&$^{+2.1}_{--0.62}$&40&&$^{+2.3}_{-1.9}$&29&&$^{+2.9}_{-3.0}$&0.66&70&57\\
HD 22484$^*$&11&&$^{+1.1}_{-1.1}$&98&&$^{+6.3}_{-4.8}$&14&&$^{+1.5}_{-1.7}$&3.0&140&17\\
HD 23356$^*$&11&&$^{+2.6}_{-2.5}$&43&&$^{+9.2}_{-20}$&23&&$^{+58}_{-7.4}$&2.2&120&8.8\\
HD 27290$^*$&19&&$^{+1.2}_{-0.86}$&63&&$^{+1.1}_{-3.1}$&50&&$^{+5.3}_{-1.7}$&0.21&71&42\\
HD 30495$^*$&35&&$^{+3.0}_{-2.1}$&68&&$^{+3.3}_{-4.4}$&17&&$^{+2.4}_{-1.5}$&0.52&59&37\\
HD 33262&12&&$^{+2.0}_{-1.7}$&110&&$^{+9.5}_{-9.3}$&7.2&&$^{+1.3}_{-1.1}$&3.0&90&9.3\\
HD 39091&1.6&&$^{+0.73}_{-0.42}$&46&&$^{+12}_{-22}$&46&&$^{+120}_{-17}$&3.0&100&5.0\\
HD 48682$^*$&65&&$^{+4.2}_{--1.5}$&53&&$^{+1.5}_{-3.5}$&37&&$^{+5.5}_{-2.0}$&0.44&70&83\\
HD 55892&6.8&&$^{+2.1}_{-1.9}$&210&&$^{+41}_{-60}$&4.3&&$^{+4.2}_{-1.3}$&3.0&70&4.5\\
HD 56986&9.2&&$^{+3.2}_{-2.5}$&90&&$^{+48}_{-8.2}$&32&&$^{+6.7}_{-18}$&3.0&30&4.4\\
HD 69830&190&&$^{+14}_{-13}$&310&&$^{+9.0}_{-15}$&0.63&&$^{+0.066}_{-0.035}$&3.0&30&23\\
HD 72905&8.1&&$^{+3.2}_{-1.3}$&90&&$^{+8.8}_{-31}$&9.4&&$^{+13}_{-1.6}$&1.1&100&7.8\\
HD 76151&17&&$^{+3.6}_{-3.9}$&83&&$^{+14}_{-18}$&11&&$^{+7.3}_{-3.0}$&1.6&58&10\\
HD 90089$^*$&9.3&&$^{+1.3}_{-0.66}$&31&&$^{+1.2}_{-1.3}$&140&&$^{+13}_{-11}$&2.2&370&20\\
HD 102870&0.81&&$^{+0.15}_{-0.22}$&43&&$^{+13}_{-7.6}$&78&&$^{+37}_{-31}$&3.0&340&6.3\\
HD 109085$^*$&17&&$^{+1.1}_{-0.88}$&40&&$^{+1.9}_{-1.9}$&110&&$^{+11}_{-9.4}$&0.40&320&25\\
HD 110897$^*$&23&&$^{+2.0}_{-3.2}$&56&&$^{+3.5}_{-6.7}$&26&&$^{+7.7}_{-3.0}$&0.093&110&25\\
HD 111631&13&&$^{+4.1}_{-3.1}$&18&&$^{+3.6}_{-4.6}$&74&&$^{+59}_{-23}$&2.5&150&6.5\\
HD 115617$^*$&29&&$^{+1.8}_{-1.4}$&67&&$^{+2.4}_{-3.7}$&16&&$^{+1.9}_{-1.1}$&0.0&30&28\\
HD 128165&5.0&&$^{+4.3}_{-1.7}$&52&&$^{+10}_{-28}$&14&&$^{+52}_{-4.4}$&1.7&99&5.6\\
HD 128167&14&&$^{+9.8}_{-2.5}$&130&&$^{+7.3}_{-65}$&8.0&&$^{+22}_{-0.81}$&0.0&30&8.4\\
HD 131511$^*$&3.7&&$^{+1.3}_{-1.4}$&49&&$^{+17}_{-27}$&24&&$^{+94}_{-11}$&0.0&30&5.0\\
HD 158633$^*$&29&&$^{+4.6}_{-2.9}$&64&&$^{+4.6}_{-13}$&12&&$^{+7.2}_{-1.6}$&0.74&62&28\\
HD 160032&4.8&&$^{+1.0}_{-0.97}$&76&&$^{+11}_{-9.0}$&30&&$^{+8.6}_{-6.8}$&1.9&500&9.0\\
HD 166348&17&&$^{+16}_{-5.8}$&41&&$^{+17}_{-16}$&17&&$^{+28}_{-8.3}$&1.3&70&4.4\\
HD 191849&10&&$^{+6.1}_{-4.4}$&37&&$^{+27}_{-6.7}$&14&&$^{+7.1}_{-9.5}$&3.0&54&6.1\\
HD 199260&16&&$^{+2.9}_{-1.3}$&79&&$^{+3.8}_{-18}$&18&&$^{+12}_{-1.6}$&0.81&70&23\\
HD 206860&9.4&&$^{+2.0}_{-1.6}$&86&&$^{+9.1}_{-8.3}$&11&&$^{+2.5}_{-2.0}$&2.9&98&9.8\\
HD 207129$^*$&97&&$^{+5.3}_{-8.5}$&51&&$^{+1.5}_{-2.6}$&33&&$^{+3.6}_{-1.9}$&0.85&120&28\\
HD 218511&20&&$^{+7.2}_{-5.5}$&31&&$^{+4.4}_{-13}$&31&&$^{+60}_{-7.0}$&1.4&150&8.1\\
HD 219482$^*$&34&&$^{+1.9}_{-1.1}$&90&&$^{+1.8}_{-2.9}$&13&&$^{+0.90}_{-0.51}$&0.82&72&39\\
HD 222368&1.1&&$^{+0.80}_{-0.32}$&60&&$^{+16}_{-31}$&41&&$^{+140}_{-16}$&3.0&110&4.6\\
HIP 1368$^*$&98&&$^{+16}_{-9.7}$&28&&$^{+3.2}_{-4.6}$&33&&$^{+14}_{-6.5}$&0.34&100&13\\
HIP 14954$^*$&3.8&&$^{+0.72}_{-0.54}$&30&&$^{+15}_{-7.6}$&170&&$^{+140}_{-97}$&1.3&71&12\\
HIP 73695&7.7&&$^{+5.9}_{-2.1}$&110&&$^{+15}_{-34}$&9.3&&$^{+11}_{-2.2}$&3.0&70&5.2\\
HIP 88745$^*$&14&&$^{+0.82}_{-1.7}$&50&&$^{+3.2}_{-2.8}$&46&&$^{+5.5}_{-5.4}$&3.0&330&18\\
HIP 105312&1.6&&$^{+1.7}_{-0.67}$&16&&$^{+14}_{-2.5}$&280&&$^{+110}_{-200}$&3.0&100&3.4\\
\hline
\end{tabular}
\end{table*}

\noindent For the disc parameters we compute grids over the four disc parameters ($T_{\rm disc}$, $\Omega_{\rm disc}$, $\lambda_0$, and $\beta$), or over only the first two if there are insufficient IR photometry to constrain the latter two. Of particular interest here are the constraints on the disc fractional luminosity ($f=L_{\rm disc}/L_\star$) and disc radius and/or temperature, so a grid with these parameters is also computed. An overall disc significance metric is also used, which is simply $\chi_{\rm{tot}}=\sqrt{\Delta_{\chi^2}}$, where $\Delta_{\chi^2}$ here is the difference between $\chi^2$ for the best fitting $\Omega_{\rm disc}$ and that for $\Omega_{\rm disc}=0$ (i.e. no disc). The output distributions have been checked and the $\chi_{\rm{tot}}$ histograms are consistent with a Gaussian with unity dispersion, plus a positive population attributable to the discs.

Two independent analyses were performed using these data, the first uses the disc detections and physical parameters derived from the SED fitting described in this section to investigate debris disc incidence rates and the distribution of discs within the fractional luminosity vs disc radius parameter space. The second uses the raw flux density measurements obtained from the \emph{Herschel} maps, as well as data from MIPS at 24 and 70\,$\mu$m, to constrain a disc evolution model, and thereby understand the disc population in a general way.

Both analyses focus on a general characterisation of debris discs around stars of F, G and K spectral types. However, there is great diversity in the range of parameters of each debris system; some sources are known to harbour multiple discs \citep[e.g. ][]{Wyatt2012,Gaspar2014,Matthews2014}, whilst in other cases there are insufficient data to uniquely constrain the system architecture \citep[e.g.][]{Churcher2011}. In order to perform a general analysis of these sources this work makes the assumption that all systems are composed of a single temperature disc. For those SEDs for which two temperatures were required, it is only the cooler component which is considered in this analysis.

%-------------------------------------------------------------------------------
\section{Disc incidence around F, G and K type stars}\label{sec3}
%-------------------------------------------------------------------------------
A total of 47 debris discs were identified in the DEBRIS FGK sample of 275 stars. The stars hosting discs are listed in Table~\ref{table1}, along with the derived parameters for the fitted modified blackbody disc model. A break-down of detected discs by host star spectral type and associated incidence rates are given in Table~\ref{table2}, whilst a complete list of results including measured photometry for all targets is given in Table~\ref{tableA1}. The detection significance, $\chi_{\rm tot}$, is given in Table~\ref{table1}.

Of the 31 disc hosting stars identified by the DUNES team 25 are included in the DEBRIS sample. The remaining 6 were excluded as they either lay beyond the distance limits, or the cirrus confusion was predicted to be above the cut-off, for the DEBRIS survey. This analysis, however, finds discs around only 19 of these 25 sources. No disc is detected around HD 224930 (HIP 171), HD 20807 (HIP 15371), HD 40307 (HIP 27887), HD 43834 (HIP 29271), HD 88230 (HIP 49908) and HD 90839 (HIP 51459); data for these sources are given in Appendix~\ref{ApB}. Five of these sources are identified by \cite{Eiroa2013} as new discs discovered by \emph{Herschel}. An explanation of how this analysis came to a different conclusion to that of the DUNES team for each of these sources, using the same dataset, is given below:
\begin{description}
\item[\textbf{HD 224930 (HIP 171):}] Following subtraction of photospheric flux no significant emission remained at the location of the star. It should be noted, however, that there is a second confusing 3$\sigma$ source nearby at 160\,$\mu$m which could account for the DUNES detection (which is only significant at 160\,$\mu$m).
\item[\textbf{HD 20807 (HIP 15371, zet02 Ret):}] Data at 70 and 100\,$\mu$m show signs of three distinct sources, one at the position of the star and two nearby. The nearby sources were regarded as confusion and fitted and subtracted separately. This is likely the source of the difference with the DUNES detection.
\item[\textbf{HD 40307 (HIP 27887):}] Following subtraction of photospheric flux no significant emission remained at the location of the star. A potentially confusing cirrus can be seen at 160\,$\mu$m, however.
\item[\textbf{HD 43834 (HIP 29271):}] Following subtraction of photospheric flux no significant emission remained at the location of the star. It should be noted that the DUNES detection is based on an excess at 160\,$\mu$m.
\item[\textbf{HD 88230 (HIP 49908):}] Here two sources were fit, the star and a second confusing point source nearby. Following subtraction of photospheric flux no significant emission remained in either PACS band at the location of the star. However, it should be noted that there is significant cirrus emission within the field, although not close to the star position, at 160\,$\mu$m.
\item[\textbf{HD 90839 (HIP 51459):}] Following subtraction of photospheric flux no significant emission remained at the location of the star. In addition, the uncertainty calculated in this work is almost twice that found by \cite{Eiroa2013}.
\end{description}

\cite{Montesinos2016} subsequently presented a further analysis on behalf of the DUNES team which included an additional 54 sources originally observed by DEBRIS. All of these targets are included in this work, and excess detections are in agreement with one exception, HD 216803. \cite{Montesinos2016} identify an excess for this source based on data at 160\,$\mu$m; no excess is detected in this analysis due to a larger uncertainty found for this source at this wavelength in this work, taking it below the detection threshold.

The disc incidence for both the full DEBRIS FGK sample, and individual spectral types, is given in Table~\ref{table2}; uncertainties are calculated in a way suitable for small number statistics using the tables in \cite{Gehrels1986}.

For the combined FGK star sample the incidence is 17.1$^{+2.6}_{-2.3}$ per cent, consistent with the excess rate found by \cite{Trilling2008} using 70\,$\mu$m \emph{Spitzer} data ($16.3^{+2.9}_{-2.8}$). The trend for smaller incidence rates for later spectral types seen by \citeauthor{Trilling2008} is also reproduced, with incidence rates within 3, 3 and 7 per cent for the F, G and K spectral types respectively. This trend is really only significant, however, between F and G/K populations. It should be noted that this comparison is based on the 70\,$\mu$m excess incidence reported by \citeauthor{Trilling2008}. The DEBRIS FGK incidence rates are obtained from model fits to data at all available wavelengths, including 70\,$\mu$m. The differences seen here are attributed primarily to the greater depth of the PACS data and constraints provided by the multi-wavelength analysis.

\begin{table}
  \centering
  \caption{Summary of debris disc detections and associated disc incidence as a function of the host star spectral type. The incidence adjusted for incompleteness in this sample is also given. Uncertainties are calculated in a way suitable for small number statistics using the tables in \protect\cite{Gehrels1986}.}\label{table2}
  \begin{tabular}{ccccc}
    \hline
    Spectral          &                &           &              & Completeness \\
    type              & No. stars      & No. discs & Incidence    & adjusted\\
                      &                &           &              & incidence\\
    \hline
    F                 & 92             & 22        & 23.9$^{+5.3}_{-4.7}$\% & 37.4$^{+6.1}_{-5.1}$\%\\
    G                 & 91             & 13        & 14.3$^{+4.7}_{-3.8}$\% & 24.6$^{+5.3}_{-4.9}$\%\\
    K                 & 92             & 12        & 13.0$^{+4.5}_{-3.6}$\% & 22.5$^{+5.6}_{-4.2}$\%\\
    Total             & 275            & 47        & 17.1$^{+2.6}_{-2.3}$\% & 27.7$^{+2.9}_{-2.9}$\%\\
    \hline 
  \end{tabular}
\end{table}

A more direct comparison can be made with the results of \cite{Eiroa2013}, who use a 3$\sigma$ excess detection at any PACS wavelength as their disc detection requirement. \cite{Eiroa2013} find incidence rates for their 20\,pc limited sub-sample of 20$^{+13}_{-9.3}$, 22$^{+7.4}_{-6.2}$ and 18.5$^{+6.8}_{-5.5}$ per cent for their F, G and K spectral types respectively, and a combined rate of 20$^{+4.3}_{-3.7}$ per cent. Equivalent results from \cite{Montesinos2016} are 24$^{7.5}_{-6.3}$, 20$^{6.1}_{-5.1}$, 18$^{6.4}_{-5.2}$ and 20$^{9.6}_{-8.3}$ per cent. Whilst these rates are, with the exception of the F-types, higher than those found within the DEBRIS sample, the difference is less than 1$\sigma$ in each case, meaning that they are generally in agreement.

When the six sources with contested excess detections (Section~\ref{sec3}) are removed from the DUNES 20\,pc sub-sample, their F, G, and K star incidences decrease to $15^{+12}_{-8.0}$, $16^{+6.9}_{-5.3}$ and $15^{+6.5}_{-5.0}$ per cent, with a combined incidence of $15^{+3.9}_{-3.3}$ per cent, with consistent agreement for the results of \cite{Montesinos2016}. This change largely accounts for the difference in incidence between DUNES and DEBRIS, bringing them well within the associated uncertainties of the two measurements.

\begin{figure}
  \includegraphics[width=8.5cm]{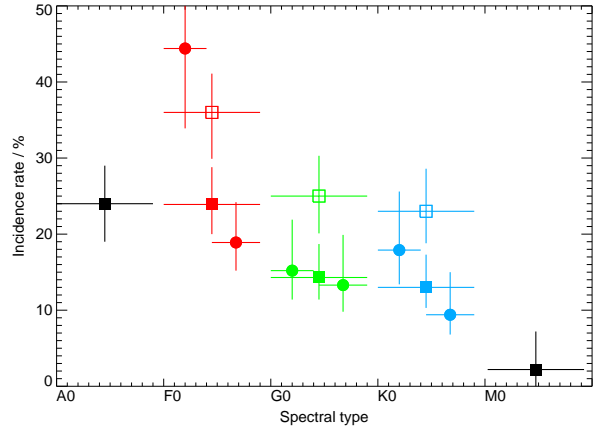}
  \caption{Incidence rate of debris disks within the DEBRIS survey as a function of spectral type. Filled and open squares indicate raw and completeness-corrected incidence rates respectively shown in Table~\ref{table2}. Horizontal errorbars indicate the range of spectral type associated with each estimate. The A- and M-type incidence rates, shown in black, are from Thureau et al. (2014) and Lestrade et al. (in prep.), respectively. For each of the F-, G- and K-type subsamples (red, green and blue symbols, respectively), further incidence rates are computed for earlier and later-type stars, shown by filled circles; i.e., splitting F stars in F0--F4 and F4.5-F9, G stars in G0--G4 and G4.5--G9 and K stars in K0-4 and K4.5--9 sub-samples.}
\label{fig3}
\end{figure}

It is interesting to note that the raw incidence for A-stars in the DEBRIS survey (24 per cent) is similar to that of the F-star sample \citep{Thureau2014}. Similarly, the G and K incidence rates are very close in value. This suggests a possible link between the A and F, and G and K stars, which is distinct for these two subgroups. In addition, the incidence found for M-stars within the DEBRIS survey is $2.2^{+3.4}_{-2.0}$ per cent, just over 2$\sigma$ below that of the K-star sample found here (Lestrade et al. in prep.). Furthermore, within the FGK DEBRIS sample, we also note a difference in raw incidence rates among F0--F4 and F5--F9 subsamples, significant at the 97.8 per cent confidence level (see Figure~\ref{fig3}). No significant difference is observed between early- and late-G stars, while a tentative difference is observed between early- and late-K stars. Overall, despite limited sample sizes, this suggests a gradual decline of incidence rate towards lower stellar mass.

\subsection{Completeness corrected incidences}\label{sec3.1}
The ability to detect excess emission from a debris disc around a star varies for each target, depending on the disc and stellar physical parameters (including distance), as well as the range and depth of available data. Assuming a single component disc system that emits as a blackbody (Section~\ref{sec2.5}), it is possible to determine in which regions of $f$ vs $r_{\rm{bb,disc}}$ parameter space a disc could have been detected for any individual source. The variable detection limits from star-to-star within this parameter space mean that, for any specific combination of $r_{\rm{bb,disc}}$ and $f$, a disc might be detectable for only a fraction of the full DEBRIS FGK sample. This fraction provides a measure of the known completeness, for a given combination of $r_{\rm{bb,disc}}$ and $f$, for this sample. Combining the detection limits provides a function giving a measure of the completeness within a 2-dimensional parameter space.

Figure~\ref{fig4}a shows the $f$ vs $r_{\rm{bb,disc}}$ parameter space, with the location of the confirmed discs plotted therein. The grey contours on Figures~\ref{fig4}b and \ref{fig4}c show the fraction of the sample for which a disc could have been detected if it existed in this region of parameter space, for the entire FGK sample, based on the all the available data. This is taken to be the sample completeness at this point in parameter space. The region at the top of the figures is 100 per cent complete, meaning that a disc with parameters within this space could have been detected around all of the stars in the sample. The completeness contours decrease in steps of 10 per cent down to the shaded region of parameter space, in which no discs could have been detected around any of the sample (cross-hatched region).

Using this completeness function, it is possible to adjust the raw incidence rates given in Table~\ref{table2} in an attempt to account for the known incompleteness of these data. It should be noted that this is only applicable in regions of parameter space wherein at least one source is detectable; no conclusions can be drawn for parameter space with zero completeness. Therefore, any results from this adjustment remain lower limits to the potential true disc incidence.

To calculate the completeness adjusted incidence rates in Table~\ref{table2}, the completeness at the location of each detected source is first calculated. The number of detected sources, adjusted for completeness, is then given by one over the derived completeness at that point in parameter space. For example, if a disc is detected in a region of $f$ vs $r_{\rm{bb,disc}}$ parameter space wherein only 50 per of the sample would have yielded a detection, the completeness fraction is 0.5. Thus, the number of detected discs, adjusted for completeness, would be 2. This is replicated for all detected sources to estimate the number of discs detected, adjusted for sample completeness. This number is then divided by the sample size, to obtain the adjusted incidence rate given in Table~\ref{table2}. The errors are equally scaled by completeness, with scaling only applied when the completeness is greater than 10 per cent, to avoid extremely large adjustments, with equally large uncertainties. The full dataset is broken into three separate samples so as to determine a completeness function for each spectral type separately. These adjustments are illustrative of the effects of incompleteness within this sample, but should not be regarded as fully correcting for completeness.

The trend for smaller incidence rates for later spectral types is maintained, even after attempting to correct for incompleteness. This suggests that the relative incidence between spectral types is reasonably robust, and this trend is real. The similar completeness levels for the three spectral types means that this correction effectively acts as a positive uniform scaling, increasing the average incidence rate to $27.7_{-2.9}^{+2.9}$ per cent. It should be noted that the specific make-up of each sample impacts on the completeness of the sample, and therefore on the completeness correction applied and 10 per cent threshold cut-off. It is this effect that results in a completeness adjusted incidence for the entire sample being lower than what might naively be calculated from the mean of the completeness adjusted incidence rates for each of the three sub-samples from which it is composed. The larger sample size for the combined FGK sample also provides greater statistical robustness to the influence of discs in regions of low completeness regions, which can otherwise bias the reported incidence towards higher values. Such biases are more common in the individual spectral type samples due to their lower levels of completeness at higher fractional luminosity and blackbody disc radius.

%-------------------------------------------------------------------------------
\subsection{Disc fractional luminosity vs radius distribution}\label{sec3.2}
The DEBRIS FGK star sample is a large and unbiased dataset.  These two properties make it possible to study the parameter space of the disc properties, determined by SED model fitting (Section~\ref{sec2.5}), in a more general way than has been possible before.

\begin{figure}
  \includegraphics[trim=0mm 0mm 0mm 30mm,clip,scale=0.5]{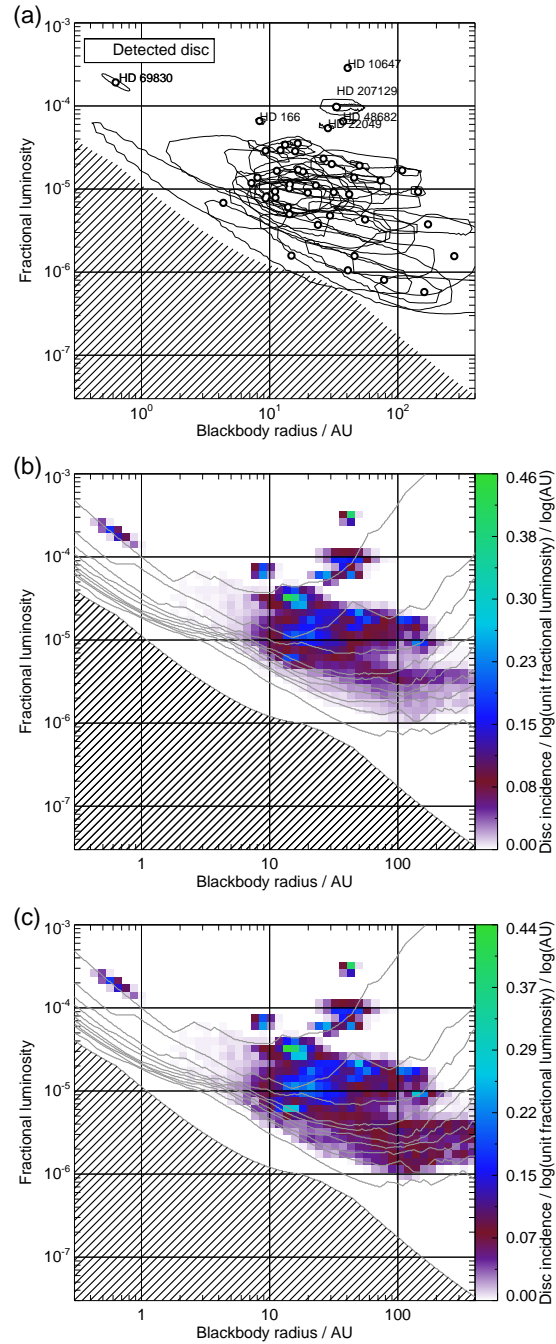}
  \caption{(a) Location of detected debris discs (open black circles) within the fractional luminosity vs blackbody radius parameter space. The line around each detected disc shows the 1$\sigma$ uncertainty for each parameter. The cross-hatched region shows the region of parameter space in which no discs could have been detected with this sample, i.e. zero completeness. (b) The colour scale shows the disc incidence, per log AU per log unit fractional luminosity, as determined from a Monte-Carlo simulation of this sample, with the associated 1$\sigma$ uncertainty contours in fitted disc radius and fractional luminosity used shown in (a). The contour lines show levels of completeness from zero (cross-hatched region) to 100 per cent, in steps of 10 per cent. (c) The colour scale shows the completeness adjusted disc incidence, per log au per log unit fractional luminosity. As with (b), this is calculated from a Monte-Carlo simulation of this sample and the associated 1$\sigma$ uncertainty contours in fitted disc radius and fractional luminosity used shown in (a).}
\label{fig4}
\end{figure}

Figure~\ref{fig4} shows the process by which we estimated the completeness adjusted incidence throughout the range of fractional luminosity vs disc radius parameter space probed in this sample. This process starts with the discs for which significant emission was detected, which are shown on Figure~\ref{fig4}a at the radius and fractional luminosity of the best fit from the SED modelling. However, this modelling also quantified the uncertainties in these parameters, and the same figure also shows the 1$\sigma$ uncertainty contours for the detected discs. These contours are typically asymmetric, with a diagonal `banana' shape running from the top left to bottom right, illustrating the degeneracy inherent in the SED model. The SED model fit information is then used in Figure~\ref{fig4}b to determine the fraction of stars for which a disc is detected in a given region of parameter space. The colour scale gives the disc incidence per log fractional luminosity per log AU, and so is indicative of the number of discs that have been found in different pixels in the image. To make this image, the uncertainties in the parameters for the detected discs were accounted for by spreading each disc across the allowed range of those parameters, weighted according to the probability of the disc having those parameters (which was achieved using 1000 realisations for each disc). This image is then corrected for completeness in Figure~\ref{fig4}c, which is the same as Figure~\ref{fig4}b but divided by the fraction of the sample for which discs could have been detected at this point in parameter space (which is shown by the contours on these figures). The resulting completeness-corrected disc incidence is only shown for regions of parameter space for which completeness is $>10$ per cent, since below this point the uncertainties and associated completeness correction become too large to be useful.  

The completeness adjusted incidence rate, for the parameter space above the 10 per cent completeness level in Figure~\ref{fig4}c, is 28 per cent, the same as that given in Table~\ref{table2}. The only practical difference between this estimate and the one in Table~\ref{table2} is that here the uncertainties of the derived fractional luminosity and blackbody radius for each disc are free to vary within their uncertainties. This introduces a variation in the completeness adjustment applied to each disc. The same incidence obtained by both methods shows that the impact of completeness variability is negligible for this sample.

\subsubsection{Distribution of observed disc properties}\label{sec3.2.1}
The data in Figure~\ref{fig4} show that the disc population can be split into three categories: a smooth `normal' disc population, and two outlying `island' populations, one characterised by small radii (hot) discs and the other by bright discs. The two island populations can be most clearly identified as distinct from the normal disc population in Figures~\ref{fig4}b and \ref{fig4}c. The members of the two island populations are individually labeled in Figure~\ref{fig4}a. It should be noted when studying this plot that it is assumed that discs emit as a blackbody, which can lead to an underestimate of the true physical radius when realistic dust grain emission is considered. 

The small radii population contains only one disc with a radius smaller than 4\,AU, HD 69830. No excess is detected in the \emph{Herschel} data, with the disc only detected at 8--35\,$\mu$m with \emph{Spitzer} and ground-based mid-IR observations. This excess is attributed to very small dust grain emission \citep{Beichman2005}, potentially from a recent single large cometary collision, or interaction with a planetary system \citep{Lovis2006}. This dust is therefore likely to be transient in nature, and therefore have abnormal properties within the context of the wider sample.

The bright disc population consists of five discs with $r_{\rm{bb,disc}}>4$\,AU and $f>5\times10^{-5}$: HD 166, HD 10647 (q1 Eri), HD 207129, HIP 1368 and HD 48682. With the exception of HD 166, with an age of $\sim$200\,Myr, these five bright discs are all fairly old systems, with a mean age of $\sim$2.5\,Gyr. This is contrary to what might be expected when considering steady-state disc evolution \citep{Wyatt2008}. Work by \cite{Lohne2012} finds that the disc around HD 207129 can be explained by steady-state evolution alone, and \cite{Gaspar2013} concluded that this might too be possible for the other four members of this sample. Even so, recent dynamical interaction with a planetary system or other transient events, such as a collision between two particularly large planetesimals, cannot be discounted as an explanation for their late period disc brightness \citep{Wyatt2008}. However, only HD 10647 is known to harbour an exoplanet system \citep{Butler2006}.

The remaining disc detections fall within the `normal' disc population, occupying the $f<5\times10^{-5}$ parameter space. This population shows a generally smooth completeness adjusted incidence rate distribution between $r_{\rm{bb,disc}}=4-300$\,AU, with a clear concentration of debris discs in the $r_{\rm{bb,disc}}=$7-40\,AU range, and a peak rate at $\sim 12$\,AU.

The upper envelope of the disc incidence of the normal disc population resembles an upside-down ‘V’ shape similar to that expected for a population of discs that have been evolving by steady state collisional erosion (Wyatt et al. 2007a). The peak in this V occurs at radii of 10-30\,AU, and the fractional luminosity of this envelope decreases with increasing radius. In the steady state model this results from the long collision timescale at large radii, which means that the fractional luminosity of such large discs is simply a reflection of the amount of mass that they were born with, and how much light that mass can intercept when ground into dust; e.g., if disc masses are independent of their radii then this envelope would decrease $\propto r^{-2}$ \citep{Wyatt2008}. The low disc detection rate in the $r_{\rm{bb,disc}}=$1-10\,AU and $f>10^{-5}$ range suggests that there may be a genuine decrease in disc rates in this region of parameter space. However, the lack of sensitivity in these data to discs at small radii with $f\ll 5 \times10^{-5}$ makes it difficult to assess the population at radii much below 7\,AU. Nevertheless, the short collision timescale for discs with small radii could have resulted in a high decay rate for discs in this region, resulting in their fractional luminosity being reduced to a level at which no discs are detectable within this dataset. Indeed, the steady-state evolution model of \citet[][see Section~\ref{sec4}]{Wyatt2007a} predicts that the upper envelope in the disc incidence should turn over at some radius (causing the aforementioned upside-down V shape). Based on these data, this turn-over appears to occur within the region of peak disc detection rate, i.e. at $r_{\rm{bb,disc}}=$7-40\,AU. Though it is not possible to determine if $<7$\,AU discs have been collisionally depleted, or simply never existed in the first place.

One point to note from Figure~\ref{fig4}c is that, even after adjusting for completeness, the disc incidence
decreases toward lower fractional luminosities; i.e., there are appear to be more discs per log AU per log fractional luminosity at fractional luminosities of $\sim 10^{-5}$ than close to the 10 per cent completeness cut-off limit. This could be a result of truly decreased incidence, which could be indicative of a bimodal disc population, or point to an insufficient quantity of discs to accurately apply this correction method across such a broad parameter space, even given the Monte-Carlo implementation. In any case, the disc incidence in these low completeness regions is an important indicator of the incidence in this region.

The interpretation of the black body radius parameter requires consideration of the fact that this is expected to underestimate the disc's true radius. A study of discs around A-type stars \citep{Booth2013} finds that the blackbody radius underestimates the true radius by a factor of between 1 and 2.5, with tentative evidence for an increase in this factor for later spectral types \citep[see][]{Pawellek2014}. Adopting the upper limit of this range, as is most applicable for this F, G and K star sample, gives a typical radius of approximately 30\,AU, and a range of $\sim$17-100\,AU, based on the data in Figure~\ref{fig4}. This spans the current estimated radius for the Kuiper belt \citep{Vitense2010}, making it typical within this sample. The depth of these data are insufficient to accurately characterise the measured disc population down to the fractional luminosity of the Kuiper belt, however, but do place the Kuiper belt within the range of the `typical' disc radius.

%-------------------------------------------------------------------------------

\section{Steady-state evolution of debris around F, G and K stars}\label{sec4}

While Figure~\ref{fig4} provides a valuable guide to the underlying debris disc population, it is appropriate when fitting a model to this population (which is the purpose of this section) to compare the model more directly with the observations. The excess ratio, i.e. the ratio of the disc to stellar photospheric flux density ($R_{\lambda}=F_{\rm{\lambda,disc}}/F_{\rm{\lambda,star}}$), is a fundamental measurable parameter of debris discs in the infrared. It is a function of disc temperature/radius and fractional luminosity, $f=L_{\rm{disc}}/L_{\rm{star}}$, and is different for each observed wavelength. By studying the distributions of excesses within a sample of stars across multiple wavelengths it is possible to constrain model disc distributions, and thereby better understand the underlying disc population.  

\begin{figure}
  \includegraphics[width=1.0\columnwidth]{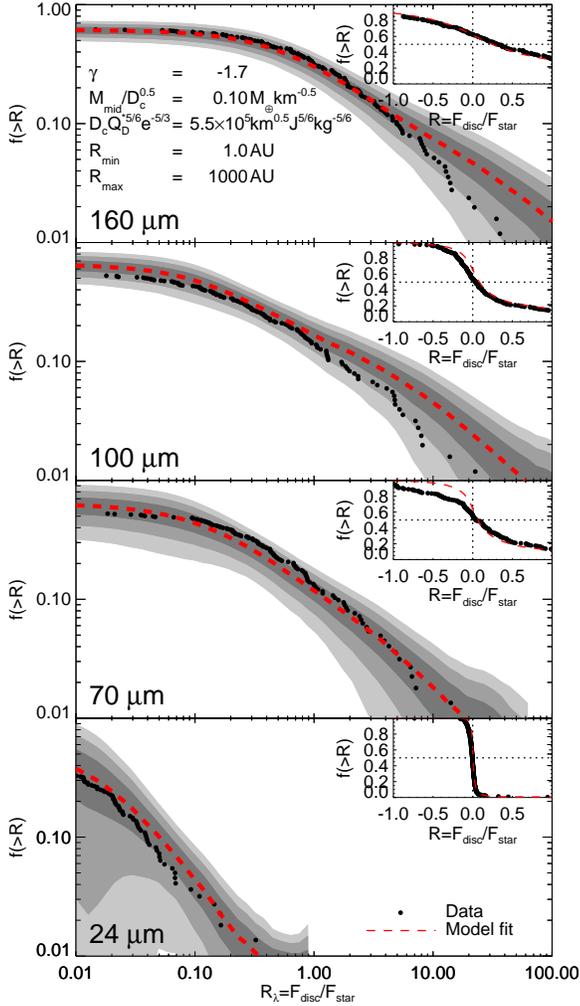}
  \caption{Fraction of FGK star sample with fractional disc excess greater than or equal to $R_{\lambda}$ as a function of $R_{\lambda}$ (black dots) at 24, 70, 100 and 160\,$\mu$m. The red dashed line shows the mean model fit to these data, and the grey shaded contours show the 1, 2 and 3$\sigma$ limits for the model fit. The uncertainty in model fit results from the finite size of the DEBRIS FGK star sampled used in this analysis. The main plots show only the positive region wherein the disc population resides, whilst the inserts show the model fit truncated to $R_{\lambda}=-1$ to $+1$, and plotted with linear axes.}
  \label{fig5}
\end{figure}

Figure~\ref{fig5} shows with black dots the fraction of DEBRIS FGK targets with an excess greater than $R_{\lambda}$ as a function of $R_{\lambda}$. The top two panels show results for the 100 and 160\,$\mu$m DEBRIS data, while the lower two panels show the same plot for the 235 stars (86 per cent of the total sample) for which MIPS 24 and 70\,$\mu$m data are available. The main plot in each panel shows the positive excesses on a log-log scale for clarity, with the sub-plot in each panel showing the distribution with linear axes, truncated to show excesses in the range $R_{\lambda}=$ -1 to 1.

Since the majority of stars do not have detectable discs, these cumulative excess fraction plots all intercept the y-axis at a value close to 0.5, which represents the mean excess of the measured population. The negative excesses are the result of the negative half of the Normal noise distribution, when observing targets hosting faint, or non-existent discs.

\subsection{Disc evolution model}\label{sec4.1}
The work of \cite{Wyatt2007a} provides a simple model for the steady-state evolution of debris discs. The model assumes that all stars are born with a planetesimal belt, and that some of the properties of those belts are common among all stars; that is, all belts have the same maximum planetesimal size, those planetesimals have the same strength, and are stirred to the same level as defined by a mean eccentricity. The planetesimal belts of different stars have different initial masses and radii, and evolve after formation by steady state collisional erosion. Here we use this model to interpret the measured disc excesses in the PACS 100 and 160\,$\mu$m bands, and also in the MIPS 24 and 70\,$\mu$m band for the same targets when available.

A disc is modelled as a single belt of planetesimals at a radius $r$, with width $dr$, in collisional cascade. The size distribution is given by $n(D)\propto D^{2-3q}$, where $q=11/6$ \citep{Dohnanyi1969}, and applies from the largest planetesimal, $D_{\rm{c}}$, down to the blow-out dust grain size, $D_{\rm{bl}}$.  All particles are assumed to be spherical and to act as blackbodies. Given these assumptions, the fractional luminosity is given by $f = \sigma_{\rm{tot}}/4\pi r^2$, where $\sigma_{\rm{tot}}$ is the cross-sectional area of the particles in AU$^2$.  Therefore, with the planetesimal size distribution defined above $f\propto M_{\rm{mid}}D_{\rm c}^{-0.5}$. The blackbody assumption also makes it possible to define the disc temperature, $T=278.3 L_{\rm{star}}^{0.25} r^{-0.5}$ in K, and flux density, $F_{\rm{\nu,disc}}=2.35\times10^{-11}B_\nu(T)\sigma_{\rm{tot}}d^{-2}$ in Jy, where $d$ is the distance to the star in pc and $B_\nu$ is the Planck function in Jy\,sr$^{-1}$.

The long-term evolution of a disc in a steady-state collisional cascade depends only on the collisional lifetime, $t_{\rm{c}}$ of the largest planetesimals, given by,
\begin{equation}
  t_{\rm{c}} = \frac{3.8\rho r^{3.5}(dr/r)D_{\rm{c}}}{M_{\rm{star}}^{0.5}M_{\rm{tot}}}\frac{8}{9G(X_{\rm{c}})},
  \label{tc}
\end{equation}
where $t_{\rm{c}}$ is in Myr, $\rho$ is the particle density in kg\,m$^{-3}$, $D_{\rm{c}}$ is in km, $M_{\rm{star}}$ is the stellar mass in units of $M_{\rm{\odot}}$, $M_{\rm{tot}}$ is the solid disc mass (i.e. excluding gas) in units of $M_{\rm{\oplus}}$, $G(X_{\rm{c}})$ is a factor defined in Equation 9 of \cite{Wyatt2007a}, and $X_{\rm{c}}=D_{\rm{cc}}/D_{\rm{c}}$, where $D_{\rm{cc}}$ is the diameter of the smallest planetesimal that has sufficient energy to destroy a planetesimal of size $D_{\rm{c}}$.  This value can be calculated from the dispersal threshold, $Q_{\rm{D}}^{*}$, defined as the specific incident energy required to catastrophically destroy a particle \citep{Wyatt2002}, given by
\begin{equation}
  X_{\rm{c}} = 1.3\times10^{-3}\left(\frac{Q_{\rm{D}}^{*}rM_{\rm{star}}^{-1}}{2.25e^2}\right)^{1/3},
\end{equation}
where $Q_{\rm{D}}^{*}$ has units of J\,kg$^{-1}$, and $e$ is the particle eccentricity.

This is a simplified formalism of the equation used in \cite{Wyatt2007a}, in which it is assumed that particle eccentricities and inclinations, $I$, are equal.  This assumption is used throughout the modeling presented in this work.

The time dependence of the disc mass can then be calculated by solving the differential equation $dM_{\rm{tot}}/dt = -M_{\rm{tot}}/t_{\rm{c}}$, which gives $M_{\rm{tot}}(t) = M_{\rm{tot}}(0)/(1+t/t_{\rm{c}}(0))$.  This result accounts for the mass evolution resulting from collisional processes, which through the assumed size distribution also sets the evolution of the discs' fractional luminosities and fractional excesses.

For full details of this model see \cite{Wyatt2007a}, and also \cite{Wyatt2007b}, \cite{Kains2011} and \cite{Morey2014} for additional useful examples of its implementation.

\subsection{Model implementation and fitting}\label{sec4.2}
The model described in Section~\ref{sec4.1} was implemented for all stars in the DEBRIS FGK sample. The initial disc parameters are defined by $M_{\rm{tot}}(0)$, $r$, $dr$, and $\rho$, and the disc evolution by $Q_{\rm{D}}^*$, $e$, $I$ and $D_{\rm{c}}$. To simplify the modeling the following parameters were fixed: $\rho=2700$\,kg\,m$^{-3}$, $e/I=1$, $q=11/6$ and $dr=r/2$ , following \cite{Wyatt2007b} and \cite{Kains2011}. All stars were assumed to harbour a disc, and a log-normal distribution was used to define the initial disc masses of the model population. This follows \cite{Andrews2005}, who found such a distribution in a sub-millimeter study of young protoplanetary discs in the Taurus-Auriga star forming region. This distribution was parameterised by the distribution centre, $M_{\rm{mid}}$, and the distribution width. The width was set to 1.14 dex, the value found by \cite{Andrews2005}. The model disc radii are defined by a power law distribution with exponent $\gamma$, between minimum ($R_{\rm{min}}$) and maximum ($R_{\rm{max}}$) radii. Radii of 1 and 1000\,AU were adopted for $R_{\rm{min}}$ and $R_{\rm{max}}$ respectively, based on the data in presented in Section~\ref{sec3.2.1}.

Consequently, there are five remaining parameters in this model: $M_{\rm{mid}}$, $\gamma$, $Q_{\rm{D}}^*$, $e$ and $D_{\rm{c}}$. However, as explained in \citet{Wyatt2007b}, the parameters only affect the observable properties of the disks in certain combinations. Thus without loss of generality we can reduce the number of free parameters to three: $A=D_{\rm{c}}^{1/2}{Q_{\rm{D}}^*}^{5/6}e^{-5/3}$, $B=M_{\rm{mid}}D_{\rm{c}}^{-1/2}$ and $\gamma$. Fixing the combination of parameters given by $A$ ensures that a disc’s collisional evolution timescale is constant, which also sets its fractional luminosity at late times. Fixing the combination of parameters given by $B$ ensures that the disc population is born with the same distribution of fractional luminosity. 

The aim of this modelling was to generate simulated datasets, at all wavelengths simultaneously, that could be compared with the observed FGK star data shown in Figure~\ref{fig5}. To ensure that the model dataset matched the DEBRIS FGK star sample as well as possible the estimated stellar parameters for this sample were used as an input to the model. This differs from previous implementations of the model, wherein the stellar parameters were drawn from a given distribution. The use of the known stellar parameters for this sample, including stellar distance, luminosity, mass, age and effective temperature means that this model dataset reproduces the unavoidable observational biases and sample size limitations of the final dataset. To create a fully representative simulated dataset, the appropriate source measurement and stellar flux density uncertainties were then applied to the output model data, along with a calibration uncertainty for each waveband. This provides a dataset which can be analysed in a self-consistent way and directly compared to the observed data.

The limited size of this stellar sample leads to potentially significant statistical variation in the model output for the same input parameters. The model was therefore run 1000 times for each set of input free parameters, and the cumulative fractional excess plots constructed in the same way as was done for the real measured data. The mean of the runs was then taken as the representative output for the given input parameters, and compared to the observed data.

The degenerate nature of the model makes constraining the free parameters difficult. To fully investigate the parameter space a three dimensional parameter grid was created and the model run for all combinations of input parameters within this grid. The grid was filled with the output $\chi^2$ measurement for each combination of parameters:
\begin{equation}
  \chi^{2}=\sum_{i,k}{\Bigg(\frac{f(>R_{\lambda_i})_{\rm obs}-f(>R_{\lambda_i})_{\rm mod}}{\sigma_{\rm{s}_{i,k}}}\Bigg)^2}.
\end{equation}
Here $f(>R_{\lambda_i})_{\rm obs}$ is the observed fraction of stars with fractional excess at wavelength $\lambda_i$ above a given level, which is measured at the $k$ values of $R_{\lambda_i}$ given by those of the discs observed in the sample. The equivalent distribution for the model is given by $f(>R_{\lambda_i})_{\rm mod}$, which is calculated as the mean of many runs performed for each set of input parameters, while $\sigma_{\rm{s}_{i,k}}$ is the standard deviation of the model distribution (on the basis that this is indicative of the level of uncertainty in the observed distribution due to the small sample size).

This model fitting approach exploits the full dataset, including sources with large negative excesses, rather than implementing an arbitrary $\sigma$ threshold cut. As a result, it is possible to better constrain the model, and determine more representative values for the free parameters, and thus the underlying disc population. Using the data in this way requires a good understanding of the uncertainties, including knowledge that the uncertainties follow a Normal distribution. The smooth curves in the plots in Figure~\ref{fig5} highlight this, following the curve expected from a Normal distribution, with deviations occurring due only to the disc population. This is with the exception of the \emph{Spitzer} 70\,$\mu$m data, which shows a minor deviation from the expected smooth curve below $R_{70\mu\rm{m}}\sim-0.25$. This deviation was regarded as sufficiently small not to significantly adversely influence the derived best fit model, and therefore was not excluded.

\subsection{Best fit parameters}\label{sec4.3}
\begin{figure}
  \includegraphics[scale=0.45]{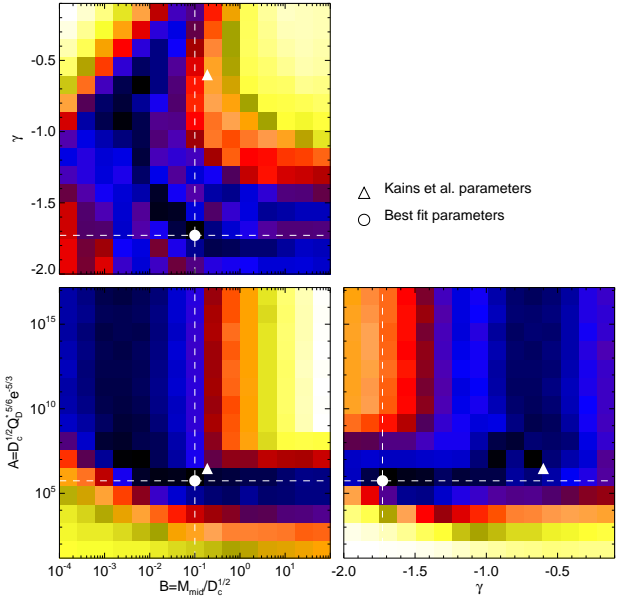}
  \caption{$\chi^{2}$ model fits for all combinations of the fitted model parameters. The panels include the minimum $\chi^2$ found in the output model grid. The colour images show the $\chi^2$ output in a log scale, and the plotted white circle shows the location of the minimum within the given parameter space. The best fit parameters obtained by \citeauthor{Kains2011} are also shown as a white solid triangle on each panel for comparison. The absolute levels in this figure are unimportant, therefore a colour scale is regarded as unnecessary.}
  \label{fig6}
\end{figure}

The dashed line in Figure~\ref{fig5} shows the mean best fit model output compared to the measured data (black dots), and the grey shaded contours show the 1, 2 and 3$\sigma$ variations in the simulated output from each of the model implementations. Note that these uncertainties are applicable to the output of any single model run using these data due to the small sample size, and are not representative of the uncertainties in the observed disc data (although they should be indicative of the uncertainty expected due to the small sample size if the model is an accurate representation of the underlying disc population). The parameters for this best fit model are $A=D_{\rm{c}}^{1/2}{Q_{\rm{D}}^*}^{5/6}e^{-5/3}=5.5\times10^5$ $\rm{km}^{1/2}\rm{J}^{5/6}\rm{kg}^{-5/6}$, $B=M_{\rm{mid}}D_{\rm c}^{-1/2}=0.1$\,$\rm{M}_{\oplus}\rm{km}^{-1/2}$ and $\gamma=-1.7$, with $R_{\rm{min}}$ and $R_{\rm{max}}$ set to 1 and 1000\,AU respectively. The $\Delta \chi^2$ for the parameter space investigated is shown in Figure~\ref{fig6} for each parameter. The minimum $\chi^2$ found is shown by the white circle and the intersection of the dashed lines which span the panels. For reference, the output model parameters found by \citeauthor{Kains2011} are also indicated in each panel by the white triangle.

\begin{figure}
  \includegraphics[trim= 2mm 5mm 0mm 277mm, clip, scale=0.57]{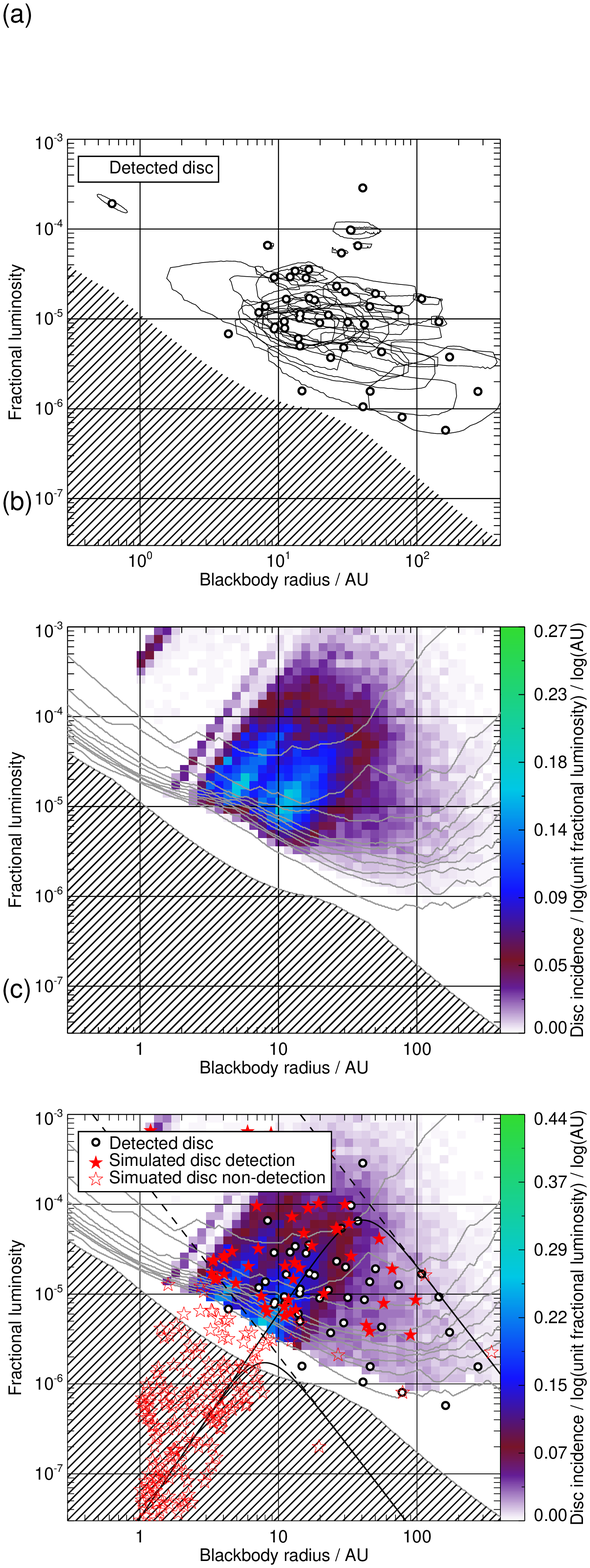}
  \caption{Mean incidence map derived from 1000 simulated model debris disc populations created using the disc evolution model and the derived best fit parameters. As in Figure~\ref{fig4}c the incidence map is truncated at 10 per completeness for the measured FGK sample. Likewise the colour range is matched to that of the real data. The filled red stars show the detected discs from one simulation run, with the incompleteness found in the observed FGK star sample taken into account. The unfilled stars show the non-detections from the same run. The observed discs plotted in Figure~\ref{fig4} are shown for comparison. The two solid lines show the expected fractional luminosity as a function of disc radius, for discs with initial masses 35\,M$_{\oplus}$ (upper curve) and 0.035\,M$_{\oplus}$ (lower curve), after 4.5\,Gyr of evolution. The dashed lines show the initial state of these two cases before any disc evolution has occurred, i.e. at an age of 0\,yr.}
\label{fig7}
\end{figure}

While the model was constrained by a fit to the fractional excess distributions of Figure~\ref{fig5}, to illustrate how the model parameters ($\gamma$ and the combinations of parameters $A$ and $B$) were constrained, it is helpful to refer to the fractional luminosity vs disc radius plot of Figure~\ref{fig7}. This is because any model that fits the fractional excess distributions must result in a fractional luminosity vs radius distribution that is not far from that observed, and changes in the model parameters translate directly into changes in the distribution of model discs shown in the colour scale and red stars of Figure~\ref{fig7}. As already mentioned in Section~\ref{sec3.2.1}, the model population in this figure always looks like an upside-down ‘V’. On the right-hand side of the V, the fractional luminosity decreases with radius $\propto r^{-2}$ at a level that scales with the parameter $B$. This is because $B$ sets the initial fractional luminosity of the disc population, and the long collision timescales at these large radii mean that this part of the population shows little evidence for collisional depletion. On the left-hand side of the V, the fractional luminosity increases with radius $\propto r^{7/3}$ at a level that scales with the parameter $A$. This is because the short collisional lifetime at small radii means that all such discs are collisionally depleted and so tend to a fractional luminosity that depends only on their age and radius \citep{Wyatt2007a}. The parameter $\gamma$ sets the ratio of discs in the left- and right-hand sides ($\gamma>-1$ means a population dominated by large radii discs in terms of number per log radius, and $\gamma<-1$ means a population dominated by small radii discs).

This helps to explain the shape of the $\chi^2$ distribution in Figure~\ref{fig6}, and the ‘L’-shaped degeneracy in the $B$ vs $A$ plot. Starting from the best fit model, this shows that a reasonable fit can be found by increasing $B$ and so the initial masses of the discs. This would result in more bright cold discs, but this can be counteracted by increasing the relative number of small discs by decreasing $\gamma$. Likewise a reasonable fit can be found by decreasing $B$ and increasing $\gamma$. The value of $A$ is reasonably well constrained by the small radii disc population for a value of $B$ close to the best fit model (albeit with the caveats discussed in Section~\ref{sec5.1} about how the small radii disc population might be biased by any warm or hot disc components). However, as $B$ is decreased, eventually $A$ becomes unconstrained and can be arbitrarily large. This is both because the population becomes dominated by large radii discs as $\gamma$ is increased, and also because discs can never have a higher luminosity than their initial value which is set by $B$ irrespective of how large $A$ is.

While Figure~\ref{fig6} makes it appear that these extremes in parameter space provide a reasonable fit to the observations, the fit is clearly improved with the best fit model parameters. The best-fit model readily explains the relatively uniform distribution of discs with large radii, as well as the cluster of discs seen at 7-40\,AU. In the model this cluster arises at the apex of the upside-down ‘V’, which occurs at radii for which the largest planetesimals come into collisional equilibrium on a timescale of the average age of stars in the population.

The best fit model parameters are, with the exception of $\gamma$, close to those found by \citeauthor{Kains2011}. The likely cause of the difference is that \citeauthor{Kains2011} did not have access to longer wavelength data from Herschel which now provides improved constraints on the disc radius distribution. Indeed, using the model parameters of \citeauthor{Kains2011} (which also assumed $r_{\rm max}=160$\,AU) provides a significant overestimate of discs with large excess at 100 and 160\,$\mu$m. This is corrected for in the best fit model, whilst maintaining a good fit to shorter wavelength data, by decreasing the disc radial distribution exponent, $\gamma$, from -0.6 to -1.7. 

\section{Discussion}\label{sec5}

\subsection{Accuracy of Model fit to Observations}\label{sec5.1}
Figure~\ref{fig5} shows that the DEBRIS FGK star observables (i.e., the fractional excess distributions) are in general well fit by the simulated model data created using the disc evolution model described in Section~\ref{sec4}. As noted in Section~\ref{sec4.2}, the observed distribution at 70\,$\mu$m departs from the smooth curve expected for Gaussian noise for negative values of $R_{\lambda}$. However, the fit is good for positive values of $R_{\lambda}$. The model also slightly over-predicts the number of discs at high values of $R_{\lambda}$ in the 100 and 160\,$\mu$m bands. One cause for this could be the assumption that the spectrum resembles a black body at all wavelengths, whereas a faster fall-off in the spectrum is expected due to the lower emission efficiencies of small grains at long wavelengths. Such a fall-off was included in the SED fitting (Section~\ref{sec2.5}), which found several discs with $\lambda_{\rm{0}}<100$\,$\mu$m, which would lead to a lower $\geq 100$\,$\mu$m flux in the disc evolution model had this effect been included in the modelling.

Given the good fit to the observed fractional excess distributions, it is perhaps unsurprising that the model also provides an accurate prediction for the incidence rate for the sample. At the 3$\sigma$ sensitivity limit for these data the incidence rate for the model output is $\sim$19 per cent (i.e., the model predicts that 19 per cent of observed stars should show an excess in at least one waveband), close to the rate of $17.1_{-2.3}^{+2.6}$ per cent found in Section~\ref{sec3}.

The success of the model at fitting the observed fractional excess distributions suggests that the population can be explained by all stars having a single temperature component debris belt. However, while it is not necessary to invoke multiple components to explain the observations, e.g. the warm and cool components that \cite{Gaspar2013} modelled as contributing independently to the 24 and 70/100\,$\mu$m excesses, it is likely that some stars do indeed have two (or more) independent components \citep{Chen2009, Kennedy2014}. Indeed, in three cases a two temperature component fit was required to satisfactorily explain the data, as described above; the warmer component had already subtracted in this analysis. This may bias the distribution of planetesimal belt radii inferred from this model, since the model would require a disc with a relatively small radius to explain a 24\,$\mu$m excess that arises from a warm component belt, even if that belt resides within a system with a cold outer belt. In other words, there is no guarantee that the model will completely reproduce the inferred fractional luminosity vs disc radius distribution seen in Figure~\ref{fig4} which was derived for cold outer belts, i.e., ignoring any warm belt component.

Nevertheless, Figure~\ref{fig7} shows that the fractional luminosity vs disc radius distribution is reasonably well reproduced by the model (which also justifies the use of this figure in Section~\ref{sec4.3} to explain how the model parameters were constrained in the fit). This figure shows the same parameter space as in Figure~\ref{fig4}, but with the incidence shown with the colour scale being derived from a Monte-Carlo run of the disc evolution model, using the obtained best fit parameters as input. The red filled and unfilled stars show the detected and non-detected discs output for a single run of this model respectively. Detection of a disc is determined randomly, weighted by the known completeness level for the DEBRIS FGK star sample within the given region of parameter space. The zero completeness contour used in Figure~\ref{fig4} is included for reference, along with the location of the original measured debris disc detections. The model population maintains the same stellar parameters as the real sample, but with random initial disc properties. The success of the model is evident in that the example model output shown in Figure~\ref{fig7} classed 49 discs as detected, close to the 47 detected in the real sample. It also reproduces the clustering of `detected' sources at approximately 7-40\,AU seen in Figure~\ref{fig4}. 

These successes aside, the regions where the model population provides a poor fit to the observed population are also informative. For example, the model does not accurately reproduce the two outlying island populations discussed in Section~\ref{sec3.2.1}. That the model cannot reproduce the bright hot dust systems like HD 69830 is perhaps unsurprising, given that they are thought to be a transient feature that cannot be fitted by steady state models \citep{Wyatt2007a}. However, the 24\,$\mu$m emission from such transient hot dust systems is still reproduced in the population statistics, and likely contributes (along with the warm component debris belts discussed above) to biasing the model to discs with small radii. This explains, to some extent, why the model predicts a relatively large number of bright discs in the 4-20\,AU size range.

The model also does not provide an accurate estimate of the number of bright ($f>5\times10^{-5}$) outer ($r>4$\,AU) discs in the other outlier island discussed in Section~\ref{sec3.2.1}. In the example shown in Figure~\ref{fig7} the model predicts 9 discs in this region with a median age of $\sim$0.8\,Gyr (with an average of 12 and mean age of 1.8\,Gyr over 100 model runs); this is nearly twice the number observed in DEBRIS FGK sample. However, the model discs in this population are at smaller radii than those observed, perhaps because the model had to compromise to fit the mid-IR and far-IR data simultaneously. This suggests that the observed bright outer discs might be the result of the steady-state evolution of the most massive discs in the underlying population, although their unusually high dust levels could also be the result of recent stochastic events or planet interactions.

It is perhaps interesting to note that the model predicts that there are few discs with a fractional luminosity below the detection limit of these data within the peak 7-40\,AU. This could imply that we have already discovered most of the discs that have radii in this range, and that most of the non-detections correspond to much smaller (collisionally depleted) discs (which is the case for the model in Figure~\ref{fig7}), or much larger (intrinsically faint) discs. However, any conclusions from the modelling on the properties of the discs of stars without detected emission are in part inherited from the assumptions about the underlying distributions of disc radii and masses (which are independent, and a power law and log-normal distribution, respectively). That is, no strong conclusions can be reached on the non-detections without further testing of the model predictions.

\subsection{Underlying distributions}\label{sec5.2}
\begin{figure}
  \includegraphics[width=1.0\columnwidth]{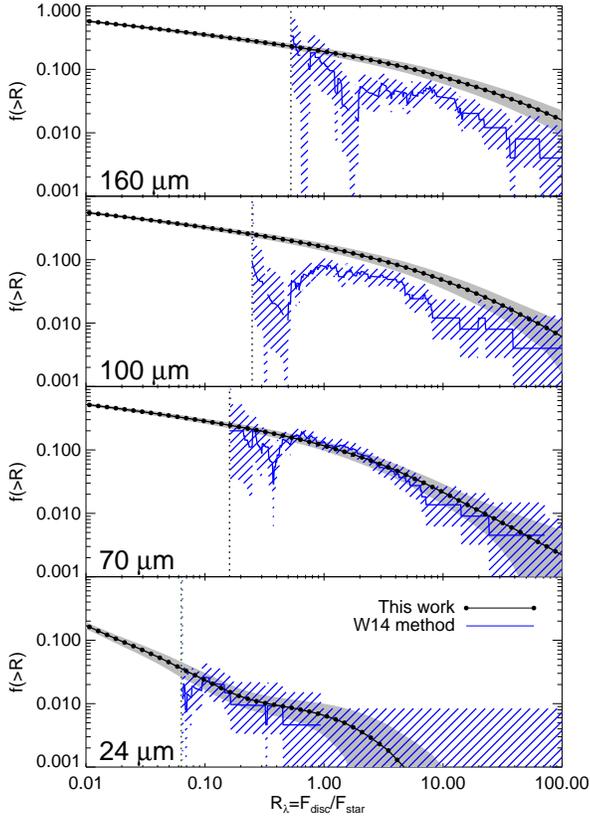}
  \caption{This figure shows the same best fit model output as Figure~\ref{fig5}, but without instrumental, calibration and stellar photosphere uncertainties added (connected dotted line). The grey shaded region shows the 1$\sigma$ uncertainties of this model output, arising from the small size of the DEBRIS FGK star sampled used in this work. The blue line shows the alternative method described in W14 applied to the same data, with the associated 1$\sigma$ uncertainties shown by the diagonal filled blue shaded regions.
  }
  \label{fig8}
\end{figure}

Once the best fit parameters were identified, the model was run again. On this occasion no noise was added to the simulated data before determining the fractional excess. This provides an estimate of the cumulative fractional excess as a function of $R_{\lambda}$ for the underlying disc population (Figure~\ref{fig8}). The 1$\sigma$ uncertainties in the model output arising from the finite sample size are again given in Figure~\ref{fig8} by the grey shaded region. These distributions can be compared with those derived using an alternative method, providing further corroboration of the model. Moreover these distributions quantify some of the model predictions for future observations that probe to lower levels of $R_\lambda$ than the present data.

The alternative measurement of the underlying disc fractional excess distributions applies the method of \citet[hereafter W14]{Wyatt2014} to the input FGK sample data. In this case, to calculate the fraction of stars in a subsample of size $N$ that have an excess level above, say $R_{\lambda}=1$, then a subset of $N^\prime$ stars within that sub-sample for which an excess could have been detected at that level are identified. The fraction of stars with an excess above that level is then the number of detections within that subset $N^\prime_{\rm{det}}$ divided by $N^\prime$. Here a 3$\sigma$ threshold was used to identify whether a detection was possible. The 1$\sigma$ uncertainties associated with this method are also shown by the diagonal line shaded region in Figure~\ref{fig8}. These uncertainties are again calculated for small number statistics using the tables in \cite{Gehrels1986}. This method can be used down to an excess whereupon $N^\prime=0$. This cutoff is typically just above three times the calibration uncertainty, and can be seen in Figure~\ref{fig8}. For more details see \cite{Wyatt2014}.

From the comparison of the distribution of fractional excesses in the underlying best fit model population, with the estimate of these distributions taken directly from the observations (W14), it can be seen that the two distributions agree well at all wavelengths for the range of $R_\lambda$ probed by the observations. The 100 and 160\,$\mu$m model data are at the upper limit of the incidence estimates derived using the W14 method, probably for the same reason that the model slightly over-predicts the fractional excess distributions at these wavelengths in Figure~\ref{fig5} which was discussed at the beginning of Section~\ref{sec5.1}. These two methods for getting the underlying disc fractional excess distributions are complementary: the W14 method provides a measure which requires no assumptions whatsoever, but provides an uneven output with large uncertainties close to the sensitivity threshold and no information below that threshold; the model approach provides a smooth distribution and can also be used to extrapolate to the disc population below the sensitivity threshold, but requires assumptions to be made about the underlying population. The model curves show how the incidence might increase as observations become sensitive to lower $R_{\lambda}$, approaching a detection rate of 100 per cent as $R_{\lambda} \rightarrow 0$ for the model assumption that all stars host a disc.

%-------------------------------------------------------------------------------

%-------------------------------------------------------------------------------
\subsection{The Solar System in Context}\label{sec5.3}
%-------------------------------------------------------------------------------

The analyses and results described in Sections~\ref{sec3} and \ref{sec4} characterise the physical properties and evolution (within the limits of the applied evolutionary model) of the solar-type stars within the DEBRIS sample. These results, however, are generally applicable to the solar-type star population, and can therefore be used to better understand this population as a whole. Also, since the Solar System is near the middle of the age distribution for the DEBRIS FGK star sample, which has a median value of 3.3\,Gyr and an interquartile range of 4.5\,Gyr, with approximately 40 per cent of stars having an age estimate $\ge$4.5\,Gyr \citep{Vican2012}, it is also appropriate to consider the position of our own Kuiper belt within this sample.

For example, the extrapolation of the model population in Section~\ref{sec5.2} can be used to consider how the fractional excess of the present day Kuiper belt compares with those of nearby stars. For reference, the predicted fractional excess from the Kuiper belt is less than $\sim 1$per cent at wavelengths 70-160\,$\mu$m, peaking at $\sim50\,\mu$m (Vitense et al. 2012). Thus, the Kuiper belt is more than an order of magnitude below the threshold of detectability around nearby stars. However, it is fairly average compared with the nearby disc population with this extrapolation, since its thermal emission has fractional excess close to 50 per cent point in the distribution of Figure~\ref{fig8}. This makes it less extreme than the extrapolation of Greaves \& Wyatt (2010), which put the Kuiper belt in the bottom 10 per cent of the distribution.

However, the Kuiper belt is thought to have followed a different evolution to that in the model of Section~\ref{sec4.1}. Rather than the belt mass evolving solely through steady state collisional erosion, the interaction of planetesimals from the belt with the giant planets caused those planets to migrate eventually leading to a system-wide instability that depleted the Kuiper belt \citep{Gomes2005} and resulted in the late heavy bombardment of the inner Solar System \citep{Tera1974}. Between the onset of the LHB and the present day the mass of the Solar-System's planetesimal belt dropped by nearly three orders of magnitude \citep{Gladman2001,Bernstein2004,Levison2008}, with 90 per cent of the disc mass being lost within the first 100\,Myr \citep[Equation 1 of][]{Booth2009}.

To illustrate the consequence of an LHB-like depletion, the two solid lines in Figure~\ref{fig7} show the predicted fractional luminosity, as a function of disc radius, for two different initial disc masses ($35M_\oplus$ and $0.035M_\oplus$) after 4.5\,Gyr of steady state evolution. These use the best fit model parameters from Section~\ref{sec4}. However, to determine the initial luminosity of the disc it was also necessary to assume a maximum planetesimal size. This was assumed to be $D_{\rm{c}}=$5000\,km, meaning that the median disc mass in the population is $7M_\oplus$, and results in an initial fractional luminosity for both cases, before any evolution takes place, that is shown by the dashed lines. Note that the fractional luminosities of the discs that have undergone evolution from their initial values are independent of the assumptions about the maximum planetesimal size, and depend only on the parameter $A$. Figure~\ref{fig7} shows how the pre-LHB Kuiper belt, which is thought to have a mass of $35M_\oplus$ and would have had black body radius of $\sim$10\,AU (assuming scaling factor of 2.5 as before), would have been readily detectable prior to LHB, as also noted in \cite{Booth2009}, but not detectable following LHB depletion.

If such system-wide instabilities and disc clearing are common amongst nearby stars then their belt masses might be expected to exhibit a bimodal distribution. This possibility was not explored in the modelling, which assumed a log-normal distribution of masses for the underlying population moreover with a fixed width. Thus it is not possible to tell if the stars without detected discs are those with close-in belts that have undergone significant collisional erosion (as in the model presented in this paper), or if they are instead those that underwent LHB-like depletions.

%-------------------------------------------------------------------------------

%-------------------------------------------------------------------------------
\section{Summary}\label{sec6}
%-------------------------------------------------------------------------------
\begin{enumerate}

\item
This paper has presented a study of debris discs around a sample of 275 F, G and K spectral type stars. This sample is drawn from the DEBRIS \emph{Herschel} open time key programme and is unbaised towards any stellar property.

\item
The SED of each source was modelled using a modified blackbody function. These fits were made to the DEBRIS data obtained at 100 and 160\,$\mu$m, as well as other ancillary data. All of the data were used in combination to determine the significance of a disc detection. A threshold of 3$\sigma$ was set for a positive detection of a debris disc.

\item
A total of 47 discs were detected. The mean raw disc incidence was $17.1^{+2.6}_{-2.3}$ per cent for fractional luminosities greater than $\sim5\times10^{-6}$, and ranged from 23-13 per cent from spectral types F-K. The measured incidence is in-keeping with previous results from \emph{Spitzer} studies. After adjusting for completeness within the probed disc parameter space the incidence becomes 27.7$^{+2.9}_{-2.9}$ per cent for the whole sample.

\item
The disc incidence as a function of radius and fractional luminosity was mapped out within the spread of disc properties identified in this sample. The incidence map was adjusted for incompleteness, and showed a high concentration of debris discs at
blackbody radii between 7 and 40\,AU, and fractional luminosities in the range $(0.4-4)\times10^{-5}$.

\item
Two outlying populations of discs were also identified: hot discs with a radius smaller than 4\,AU, and bright discs with fractional luminosities larger than $5\times10^{-5}$. The median age of the 5 bright discs is 2.5\,Gyr, suggesting that these
cannot be explained by youth.

\item
A steady-state disc evolution model was fitted simultaneously to the MIPS 24/70, and PACS 100/160\,$\mu$m data for this sample. The steady-state model was found to provide a reasonable fit at all bands. A best fit model was produced with the disc radii defined by a power law ranging from 1-1000\,AU with an exponent of -1.7, and other model parameters constrained to have $M_{\rm mid}D_{\rm c}^{-1/2}=0.1\,M_\oplus$km$^{-1/2}$ (where $M_{\rm mid}$ is the median disc mass in the population and $D_{\rm c}$ is the maximum planetesimal size), and $D_{\rm c}^{1/2}Q_{\rm D}^{\star{5/6}}e^{-5/3}=10^4$\,km$^{1/2}$J$^{5/6}$kg$^{-5/6}$ (where $Q_{\rm D}^\star$ is the dispersal threshold and $e$ the mean eccentricity of the planetesimal orbits).

\item
The success of the steady-state model shows that all stars could be born with a belt of planetesimals that then evolves by collisional erosion. However, it is worth noting that the model still cannot explain the hot dust systems, which likely originate in transient events. Also, the best fit parameters are affected to some extent by the assumption that all stars host just one belt, whereas we know that some stars have mid-IR emission from an additional warmer inner component.

\item
Moreover, the population model is based on the $\sim20$ per cent of stars with detected discs, and thus its predictions for discs below the detection threshold are to a large extent a reflection of the assumptions made about the functional forms for the distributions of radii and masses. Nevertheless, these predictions are valuable, since the model can be readily used to predict the observable properties of the discs of populations of stars with different age, distance and spectral type distributions. Any predictions that future observations show to be incorrect can be used to refine the distributions of disc radii and masses that stars are born with in the population model.

\item
The Kuiper belt was found to be a typical, albeit relatively low mass, example of a debris disc within the sample population. The typical blackbody disc radius in the sample was found to be $\sim$10\,AU, which translates to a true disc radius of $\sim25$\,AU when scaled by a factor of 2.5 to account for realistic grain optical properties, which is only slightly smaller than the nominal present day Kuiper belt radius of $\sim$40\,AU.

\item
The fractional luminosity of the current Kuiper belt is an order of magnitude too faint to have been detected. Its far-IR flux is, however, close to the median of that of the steady state population model. The detected discs have fractional luminosities close to that of the primordial Kuiper belt. This suggests that the majority of stars either had a low planetesimal formation efficiency, or depleted their planetesimal belts in a similar manner to the Solar System (e.g., through dynamical instability in their planetary systems).

\end{enumerate}
%-------------------------------------------------------------------------------

%-------------------------------------------------------------------------------
\section*{Acknowledgments}
This work was supported by the European Union through ERC grant number 279973 (MCW, GMK). GMK was also supported by the Royal Society as a Royal Society University Research Fellow.
%-------------------------------------------------------------------------------

%-------------------------------------------------------------------------------

\appendix
\section{Results table}\label{tableA1}
\onecolumn
\begin{landscape}
\scriptsize
\LTcapwidth=9in
\begin{longtable}{lll r@{}c@{}l r@{}c@{}l r@{}c@{}l r@{}c@{}l r@{}c@{}l r@{}c@{}l r@{}c@{}l r@{}c@{}l r@{}c@{}l r@{}c@{}l r@{}c@{}l r@{}c@{}l c}
\caption{Photometric data and modified blackbody disc model fit parameters for the entire DEBRIS F, G and K spectral type sample used in this paper. The stellar photosphere estimates are given by the S24, S70, S100 and S160 parameters for the 24-160\,$\mu$m bands respectievly. The significance of the excess detection is given by $\chi_{\rm{tot}}$, see Section~\ref{sec2.5}. Whilst the PACS 70\,$\mu$m data for these sources were too incomplete to use individually in the analysis presented in this paper, they are included in this table, since they were used in determining the best fit disc model parameters. For completeness, extended discs are denoted by an $*$, however, the parameters reported in this table are for the equivalent modified blackbody, as used in this work.}\label{tableA1}\\
\hline
\multicolumn{1}{l}{Target} & \multicolumn{1}{l}{UNS ID} & \multicolumn{1}{l}{Spectral Type} & \multicolumn{3}{c}{MIPS24 / mJy}& \multicolumn{3}{c}{S24 / mJy} & \multicolumn{3}{c}{MIPS70 / mJy} & \multicolumn{3}{c}{PACS70 / mJy} & \multicolumn{3}{c}{S70 / mJy} & \multicolumn{3}{c}{PACS100 / mJy} & \multicolumn{3}{c}{S100 / mJy} & \multicolumn{3}{c}{PACS160 / mJy} & \multicolumn{3}{c}{S160 / mJy} & \multicolumn{3}{c}{$L_{\rm{bb,disc}}/L_{*} / \times10^{-6}$} & \multicolumn{3}{c}{$T_{\rm{bb,disc}}$ / K} & \multicolumn{3}{c}{$r_{\rm{bb,disc}}$ / AU} & \multicolumn{1}{c}{$\chi_{\rm{tot}}$}\\
\hline
\endfirsthead
\caption{Photometric data for all sources in the DEBRIS FGK sample (continued).}\\
\hline
\multicolumn{1}{l}{Target} & \multicolumn{1}{l}{UNS ID} & \multicolumn{1}{l}{Spectral Type} & \multicolumn{3}{c}{MIPS24 / mJy} & \multicolumn{3}{c}{S24 / mJy} & \multicolumn{3}{c}{MIPS70 / mJy} & \multicolumn{3}{c}{PACS70 / mJy} & \multicolumn{3}{c}{S70 / mJy} & \multicolumn{3}{c}{PACS100 / mJy} & \multicolumn{3}{c}{S100 / mJy} & \multicolumn{3}{c}{PACS160 / mJy} & \multicolumn{3}{c}{S160 / mJy} & \multicolumn{3}{c}{$L_{\rm{bb,disc}}/L_{*} / \times10^{-6}$} & \multicolumn{3}{c}{$T_{\rm{bb,disc}}$ / K} & \multicolumn{3}{c}{$r_{\rm{bb,disc}}$ / AU} & \multicolumn{1}{c}{$\chi_{\rm{tot}}$}\\
\hline
\endhead
\multicolumn{37}{r}{Continued on next page...} \\
\endfoot
\hline
\endlastfoot
HD 166$^*$&G030A&G8 V&160&$\pm$&1.5&140&$\pm$&2.4&110&$\pm$&3.7&99&$\pm$&7.8&16&$\pm$&0.27&64&$\pm$&4.1&7.5&$\pm$&0.13&26&$\pm$&1.9&2.9&$\pm$&0.050&66&&$^{+3.3}_{-2.9}$&86&&$^{+1.6}_{-2.3}$&8.3&&$^{+0.47}_{-0.30}$&36\\
HD 693&F069A&F8 V Fe-0.8 CH-0.5&270&$\pm$&3.9&260&$\pm$&5.9&37&$\pm$&3.7&&&&30&$\pm$&0.67&18&$\pm$&2.4&14&$\pm$&0.32&6.2&$\pm$&3.9&5.6&$\pm$&0.13&&&&&&&&&&\\
HD 739&F096A&F5 V&160&$\pm$&1.6&160&$\pm$&3.9&20&$\pm$&2.6&&&&19&$\pm$&0.45&8.4&$\pm$&1.9&9.1&$\pm$&0.22&0.51&$\pm$&4.1&3.5&$\pm$&0.084&&&&&&&&&&\\
HD 1237&G070A&G8.5 V (k)&84&$\pm$&0.85&84&$\pm$&1.6&12&$\pm$&2.0&&&&9.6&$\pm$&0.18&4.0&$\pm$&2.7&4.7&$\pm$&0.087&-6.3&$\pm$&3.6&1.8&$\pm$&0.034&&&&&&&&&&\\
HD 1581&F005A&F9.5 V&560&$\pm$&5.6&550&$\pm$&9.5&71&$\pm$&5.7&&&&62&$\pm$&1.1&35&$\pm$&2.6&30&$\pm$&0.52&31&$\pm$&5.0&12&$\pm$&0.20&0.58&&$^{+0.18}_{-0.17}$&23&&$^{+7.3}_{-10}$&160&&$^{+360}_{-68}$&4.7\\
HD 1835&G118A&G5 V CH-0.5&85&$\pm$&0.95&84&$\pm$&1.1&2.3&$\pm$&5.2&&&&9.5&$\pm$&0.13&8.2&$\pm$&2.4&4.6&$\pm$&0.063&16&$\pm$&4.6&1.8&$\pm$&0.024&&&&&&&&&&\\
HD 3443&G044A&G7&200&$\pm$&2.0&210&$\pm$&3.8&19&$\pm$&5.0&&&&24&$\pm$&0.43&9.1&$\pm$&1.6&12&$\pm$&0.21&4.8&$\pm$&2.4&4.5&$\pm$&0.081&&&&&&&&&&\\
HD 3651&K045A&K0 V&200&$\pm$&2.0&200&$\pm$&3.7&12&$\pm$&4.6&22&$\pm$&2.2&22&$\pm$&0.42&6.1&$\pm$&1.6&11&$\pm$&0.21&2.0&$\pm$&2.7&4.2&$\pm$&0.079&&&&&&&&&&\\
HD 4391&G041A&G5 V Fe-0.8&140&$\pm$&1.6&140&$\pm$&2.0&23&$\pm$&2.5&&&&17&$\pm$&0.23&10&$\pm$&1.8&8.1&$\pm$&0.11&-2.6&$\pm$&4.9&3.1&$\pm$&0.043&&&&&&&&&&\\
HD 4628&K016A&K2.5 V&280&$\pm$&2.9&290&$\pm$&4.7&30&$\pm$&14&&&&33&$\pm$&0.54&15&$\pm$&2.4&16&$\pm$&0.27&1.3&$\pm$&3.5&6.4&$\pm$&0.10&&&&&&&&&&\\
HD 4676&F124A&F8 V&210&$\pm$&2.1&210&$\pm$&5.0&30&$\pm$&5.9&&&&24&$\pm$&0.57&16&$\pm$&2.4&12&$\pm$&0.28&13&$\pm$&5.8&4.5&$\pm$&0.11&&&&&&&&&&\\
HD 4747&G089A&G9 V&55&$\pm$&0.56&54&$\pm$&1.2&2.1&$\pm$&3.4&&&&6.2&$\pm$&0.14&8.5&$\pm$&1.9&3.0&$\pm$&0.068&2.8&$\pm$&4.5&1.2&$\pm$&0.026&&&&&&&&&&\\
HD 4813&F038A&F7 V&200&$\pm$&2.0&200&$\pm$&4.6&25&$\pm$&2.8&&&&22&$\pm$&0.52&14&$\pm$&1.6&11&$\pm$&0.25&2.6&$\pm$&3.6&4.2&$\pm$&0.098&&&&&&&&&&\\
HD 4967&K127A&K5&39&$\pm$&0.43&43&$\pm$&0.79&7.0&$\pm$&4.9&&&&4.9&$\pm$&0.091&-1.3&$\pm$&2.2&2.4&$\pm$&0.044&-2.2&$\pm$&3.8&0.93&$\pm$&0.017&&&&&&&&&&\\
HD 5133&K089A&K2.5 V (k)&80&$\pm$&0.81&84&$\pm$&1.3&36&$\pm$&5.3&14&$\pm$&1.6&9.6&$\pm$&0.14&14&$\pm$&1.3&4.7&$\pm$&0.070&16&$\pm$&2.4&1.8&$\pm$&0.027&8.7&&$^{+1.7}_{-2.8}$&32&&$^{+3.3}_{-4.8}$&42&&$^{+16}_{-7.5}$&9.7\\
HD 7439&F122A&F5 V&180&$\pm$&1.8&180&$\pm$&4.9&14&$\pm$&7.1&&&&21&$\pm$&0.55&13&$\pm$&2.1&10&$\pm$&0.27&3.6&$\pm$&4.1&3.9&$\pm$&0.10&&&&&&&&&&\\
HD 7570&F032A&F9 V Fe+0.4&260&$\pm$&2.6&250&$\pm$&4.5&47&$\pm$&3.9&&&&28&$\pm$&0.51&22&$\pm$&2.2&14&$\pm$&0.25&5.0&$\pm$&3.2&5.3&$\pm$&0.095&9.0&&$^{+2.6}_{-2.6}$&74&&$^{+22}_{-21}$&20&&$^{+19}_{-8.2}$&7.1\\
HD 9540&G092A&G8.5 V&60&$\pm$&0.61&62&$\pm$&0.96&-2.3&$\pm$&4.9&&&&7.1&$\pm$&0.11&6.2&$\pm$&2.2&3.4&$\pm$&0.053&-2.6&$\pm$&3.1&1.3&$\pm$&0.021&&&&&&&&&&\\
HD 9826&F020A&F8V&540&$\pm$&5.4&530&$\pm$&9.5&56&$\pm$&5.1&&&&61&$\pm$&1.1&34&$\pm$&2.6&29&$\pm$&0.53&22&$\pm$&3.9&11&$\pm$&0.20&&&&&&&&&&\\
HD 10307&G026A&G1&290&$\pm$&2.9&290&$\pm$&6.2&39&$\pm$&5.5&&&&33&$\pm$&0.70&15&$\pm$&2.1&16&$\pm$&0.34&7.8&$\pm$&5.0&6.3&$\pm$&0.13&&&&&&&&&&\\
HD 10361&K020A&K2 V&250&$\pm$&2.8&210&$\pm$&26&1.2&$\pm$&2.6&&&&24&$\pm$&3.0&10&$\pm$&1.8&12&$\pm$&1.5&6.1&$\pm$&5.2&4.6&$\pm$&0.56&&&&&&&&&&\\
HD 10476&K017A&K1 V&360&$\pm$&3.6&350&$\pm$&6.4&51&$\pm$&7.0&&&&40&$\pm$&0.72&19&$\pm$&2.2&20&$\pm$&0.35&11&$\pm$&4.6&7.5&$\pm$&0.14&&&&&&&&&&\\
HD 10647$^*$&F051A&F9 V&200&$\pm$&2.0&150&$\pm$&4.0&1200&$\pm$&6.6&1000&$\pm$&58&17&$\pm$&0.45&940&$\pm$&54&8.2&$\pm$&0.22&650&$\pm$&42&3.2&$\pm$&0.085&290&&$^{+2.9}_{-4.5}$&49&&$^{+0.49}_{-0.89}$&41&&$^{+1.5}_{-0.80}$&170\\
HD 10700$^*$&G002A&G8.5 V&&&&1600&$\pm$&28&&&&300&$\pm$&6.0&180&&3.2&&&&89&&1.6&110&&7.8&34&&0.61&6.1&&$^{+0.52}_{-0.39}$&63&&$^{+3.4}_{-6.5}$&14&&$^{+3.4}_{-1.4}$&22\\
HD 11171&F119A&F2 III-IV&210&$\pm$&2.1&210&$\pm$&6.2&65&$\pm$&6.6&51&$\pm$&3.1&23&$\pm$&0.70&38&$\pm$&3.0&11&$\pm$&0.34&29&$\pm$&4.3&4.4&$\pm$&0.13&4.3&&$^{+0.71}_{-0.63}$&58&&$^{+11}_{-8.9}$&56&&$^{+22}_{-16}$&13\\
HD 11507&K043A&K7&72&$\pm$&0.72&74&$\pm$&3.2&14&$\pm$&2.8&4.7&$\pm$&1.8&8.5&$\pm$&0.37&3.6&$\pm$&1.1&4.1&$\pm$&0.18&3.4&$\pm$&1.6&1.6&$\pm$&0.071&&&&&&&&&&\\
HD 13445&K038A&K1 V&170&$\pm$&1.7&170&$\pm$&3.8&3.4&$\pm$&5.6&14&$\pm$&2.0&19&$\pm$&0.43&8.6&$\pm$&1.6&9.2&$\pm$&0.21&2.2&$\pm$&2.2&3.5&$\pm$&0.081&&&&&&&&&&\\
HD 13974&G016A&G0V&370&$\pm$&3.7&370&$\pm$&7.0&42&$\pm$&4.6&&&&43&$\pm$&0.79&22&$\pm$&2.2&21&$\pm$&0.39&4.5&$\pm$&3.8&8.0&$\pm$&0.15&&&&&&&&&&\\
HD 14412&G024A&G8 V&110&$\pm$&1.1&110&$\pm$&1.7&11&$\pm$&2.3&&&&13&$\pm$&0.19&6.7&$\pm$&0.95&6.1&$\pm$&0.091&-0.82&$\pm$&1.5&2.4&$\pm$&0.035&&&&&&&&&&\\
HD 16160&K014A&K3 V&320&$\pm$&3.2&330&$\pm$&5.9&32&$\pm$&6.3&&&&38&$\pm$&0.68&13&$\pm$&2.0&18&$\pm$&0.33&4.3&$\pm$&3.3&7.1&$\pm$&0.13&&&&&&&&&&\\
HD 16673&F102A&F8 V Fe-0.4&110&$\pm$&1.1&110&$\pm$&2.6&17&$\pm$&5.9&&&&13&$\pm$&0.30&14&$\pm$&1.8&6.2&$\pm$&0.15&4.2&$\pm$&4.2&2.4&$\pm$&0.056&7.9&&$^{+3.0}_{-2.5}$&98&&$^{+12}_{-27}$&11&&$^{+9.9}_{-2.4}$&5.3\\
HD 16765&F112A&F7 V&110&$\pm$&1.1&110&$\pm$&3.0&9.8&$\pm$&6.7&&&&13&$\pm$&0.34&6.7&$\pm$&2.0&6.2&$\pm$&0.17&-1.9&$\pm$&5.3&2.4&$\pm$&0.064&&&&&&&&&&\\
HD 17051&F046A&F9 V Fe+0.3&170&$\pm$&1.7&170&$\pm$&3.1&22&$\pm$&3.0&&&&19&$\pm$&0.35&5.7&$\pm$&2.0&9.2&$\pm$&0.17&7.5&$\pm$&3.9&3.5&$\pm$&0.066&&&&&&&&&&\\
HD 17206&F024A&F6 V&350&$\pm$&3.5&340&$\pm$&6.1&&&&&&&39&$\pm$&0.69&20&$\pm$&2.6&19&$\pm$&0.33&7.8&$\pm$&3.4&7.2&$\pm$&0.13&&&&&&&&&&\\
HD 17925$^*$&K035A&K1.5 V (k)&190&$\pm$&2.0&180&$\pm$&3.8&69&$\pm$&6.6&72&$\pm$&3.8&21&$\pm$&0.43&57&$\pm$&3.7&10&$\pm$&0.21&38&$\pm$&7.1&3.9&$\pm$&0.080&29&&$^{+3.0}_{-3.4}$&73&&$^{+0.99}_{-8.9}$&9.3&&$^{+2.8}_{-0.25}$&20\\
HD 19305&K107A&K5&&&&46&$\pm$&0.58&-16&$\pm$&5.6&&&&5.3&$\pm$&0.067&4.0&$\pm$&1.7&2.6&$\pm$&0.033&-2.5&$\pm$&5.7&1.0&$\pm$&0.013&&&&&&&&&&\\
HD 20010&F023A&F6 V&&&&630&$\pm$&8.6&&&&&&&71&$\pm$&0.98&37&$\pm$&2.9&35&$\pm$&0.48&16&$\pm$&3.7&13&$\pm$&0.18&&&&&&&&&&\\
HD 20630&G011A&G5 V&370&$\pm$&3.7&360&$\pm$&6.3&36&$\pm$&6.9&&&&41&$\pm$&0.71&23&$\pm$&2.3&20&$\pm$&0.35&9.8&$\pm$&3.2&7.7&$\pm$&0.13&&&&&&&&&&\\
HD 20794&G005A&G8 V&760&$\pm$&7.6&740&$\pm$&14&110&$\pm$&3.9&99&$\pm$&5.0&84&$\pm$&1.6&54&$\pm$&3.2&41&$\pm$&0.77&26&$\pm$&2.7&16&$\pm$&0.30&1.6&&$^{+0.70}_{-0.42}$&65&&$^{+11}_{-26}$&15&&$^{+27}_{-3.9}$&5.9\\
HD 20807&G018A&G0 V&240&$\pm$&2.4&229&$\pm$&3.9&42&$\pm$&3.3&32&$\pm$&2.1&26&$\pm$&0.44&14&$\pm$&1.3&13&$\pm$&0.22&5.2&$\pm$&3.5&4.8&$\pm$&0.083&&&&&&&&&&\\
HD 21197&K122A&K4 V&65&$\pm$&0.66&70&$\pm$&0.90&-3.5&$\pm$&5.0&&&&7.9&$\pm$&0.10&3.4&$\pm$&2.4&3.9&$\pm$&0.050&4.3&$\pm$&3.4&1.5&$\pm$&0.019&&&&&&&&&&\\
HD 21531&K061A&K6 V k&75&$\pm$&0.75&77&$\pm$&1.1&8.3&$\pm$&4.9&&&&8.8&$\pm$&0.13&3.7&$\pm$&0.84&4.3&$\pm$&0.062&3.0&$\pm$&2.4&1.7&$\pm$&0.024&&&&&&&&&&\\
HD 22001&F101A&F3 V&229&$\pm$&2.3&229&$\pm$&4.5&28&$\pm$&2.5&&&&27&$\pm$&0.51&14&$\pm$&2.2&13&$\pm$&0.25&8.1&$\pm$&5.2&5.0&$\pm$&0.096&&&&&&&&&&\\
HD 22049$^*$&K001A&K2 V (k)&2100&$\pm$&46&1700&$\pm$&18&1900&$\pm$&76&1800&$\pm$&52&190&&2.1&&&&92&&1.0&1200&&69&35&&0.39&54&&$^{+2.1}_{--0.62}$&40&&$^{+2.3}_{-1.9}$&29&&$^{+2.9}_{-3.0}$&57\\
HD 22484$^*$&F022A&F8 V&540&$\pm$&5.4&500&$\pm$&8.5&110&$\pm$&5.4&120&$\pm$&8.2&57&$\pm$&0.96&76&$\pm$&6.1&28&$\pm$&0.47&26&$\pm$&2.6&11&$\pm$&0.18&11&&$^{+1.1}_{-1.1}$&98&&$^{+6.3}_{-4.8}$&14&&$^{+1.5}_{-1.7}$&17\\
HD 22496&K079A&K5.0&59&$\pm$&0.60&63&$\pm$&1.4&6.4&$\pm$&4.8&&&&7.2&$\pm$&0.16&6.1&$\pm$&2.1&3.5&$\pm$&0.076&7.7&$\pm$&3.4&1.4&$\pm$&0.030&&&&&&&&&&\\
HD 23356$^*$&K087A&K2.5 V&83&$\pm$&0.84&85&$\pm$&1.3&26&$\pm$&5.7&19&$\pm$&2.2&9.7&$\pm$&0.15&21&$\pm$&2.4&4.7&$\pm$&0.072&9.4&$\pm$&2.7&1.8&$\pm$&0.028&11&&$^{+2.6}_{-2.5}$&43&&$^{+9.2}_{-20}$&23&&$^{+58}_{-7.4}$&8.8\\
HD 23754&F053A&F5 IV-V&390&$\pm$&3.9&390&$\pm$&8.9&49&$\pm$&3.0&&&&44&$\pm$&1.0&28&$\pm$&2.7&21&$\pm$&0.49&4.5&$\pm$&3.3&8.2&$\pm$&0.19&&&&&&&&&&\\
HD 26965&K006A&K0.5 V&790&$\pm$&7.9&800&$\pm$&15&86&$\pm$&2.5&&&&91&$\pm$&1.7&42&$\pm$&2.7&44&$\pm$&0.82&4.3&$\pm$&4.2&17&$\pm$&0.32&&&&&&&&&&\\
HD 27274&K067A&K4.5 V (k)&77&$\pm$&0.78&84&$\pm$&1.2&13&$\pm$&4.5&6.3&$\pm$&2.0&9.6&$\pm$&0.14&3.2&$\pm$&0.91&4.7&$\pm$&0.069&3.1&$\pm$&2.1&1.8&$\pm$&0.027&&&&&&&&&&\\
HD 27290$^*$&F085A&F1 V&320&$\pm$&3.2&300&$\pm$&23&240&$\pm$&5.5&170&$\pm$&8.9&33&$\pm$&2.6&150&$\pm$&7.7&16&$\pm$&1.3&130&$\pm$&14&6.3&$\pm$&0.49&19&&$^{+1.2}_{-0.86}$&63&&$^{+1.1}_{-3.1}$&50&&$^{+5.3}_{-1.7}$&42\\
HD 29875&F081A&F2 V&&&&270&$\pm$&5.8&&&&&&&30&$\pm$&0.66&14&$\pm$&2.3&15&$\pm$&0.32&3.6&$\pm$&4.2&5.7&$\pm$&0.12&&&&&&&&&&\\
HD 30495$^*$&G029A&G1.5 V CH-0.5&190&$\pm$&1.9&190&$\pm$&3.3&130&$\pm$&4.0&130&$\pm$&7.6&21&$\pm$&0.37&85&$\pm$&5.1&10&$\pm$&0.18&40&$\pm$&2.4&4.0&$\pm$&0.070&35&&$^{+3.0}_{-2.1}$&68&&$^{+3.3}_{-4.4}$&17&&$^{+2.4}_{-1.5}$&37\\
HD 30652&F003A&F6V&1100&$\pm$&11&1100&$\pm$&19&130&$\pm$&5.5&&&&120&$\pm$&2.1&61&$\pm$&4.1&59&$\pm$&1.0&20&$\pm$&4.3&23&$\pm$&0.40&&&&&&&&&&\\
HD 32147&K024A&K3+ V&229&$\pm$&2.3&240&$\pm$&3.7&25&$\pm$&2.7&&&&27&$\pm$&0.43&11&$\pm$&1.9&13&$\pm$&0.21&1.6&$\pm$&2.9&5.1&$\pm$&0.081&&&&&&&&&&\\
HD 33262&F012A&F9 V Fe-0.5&330&$\pm$&3.3&310&$\pm$&5.6&68&$\pm$&5.0&&&&35&$\pm$&0.63&32&$\pm$&2.8&17&$\pm$&0.31&7.4&$\pm$&3.1&6.6&$\pm$&0.12&12&&$^{+2.0}_{-1.7}$&110&&$^{+9.5}_{-9.3}$&7.2&&$^{+1.3}_{-1.1}$&9.3\\
HD 33564&F090A&F7 V&190&$\pm$&2.2&190&$\pm$&4.0&0.23&$\pm$&5.1&&&&22&$\pm$&0.46&9.2&$\pm$&1.7&11&$\pm$&0.22&2.2&$\pm$&3.7&4.1&$\pm$&0.085&&&&&&&&&&\\
HD 36435&G095A&G9 V&61&$\pm$&0.62&60&$\pm$&0.82&8.8&$\pm$&4.4&&&&6.8&$\pm$&0.093&3.2&$\pm$&2.3&3.3&$\pm$&0.045&5.7&$\pm$&4.6&1.3&$\pm$&0.017&&&&&&&&&&\\
HD 36705&K117A&K2 V k&&&&110&$\pm$&1.1&-3.1&$\pm$&6.1&&&&12&$\pm$&0.13&5.1&$\pm$&1.7&6.0&$\pm$&0.065&9.8&$\pm$&3.9&2.3&$\pm$&0.025&&&&&&&&&&\\
HD 38393&F006A&F6.5 V&210&$\pm$&2.1&800&$\pm$&15&87&$\pm$&2.9&&&&90&$\pm$&1.7&43&$\pm$&4.5&44&$\pm$&0.81&15&$\pm$&4.7&17&$\pm$&0.31&&&&&&&&&&\\
HD 39091&G085A&G0 V&140&$\pm$&1.4&140&$\pm$&2.6&24&$\pm$&3.1&21&$\pm$&1.4&16&$\pm$&0.30&16&$\pm$&2.4&7.9&$\pm$&0.15&4.3&$\pm$&2.6&3.1&$\pm$&0.056&1.6&&$^{+0.73}_{-0.42}$&46&&$^{+12}_{-22}$&46&&$^{+120}_{-17}$&5.0\\
HD 40307&K065A&K2.5 V&88&$\pm$&0.89&91&$\pm$&1.1&15&$\pm$&4.7&15&$\pm$&2.2&10&$\pm$&0.13&8.3&$\pm$&1.1&5.1&$\pm$&0.064&7.8&$\pm$&3.9&2.0&$\pm$&0.025&&&&&&&&&&\\
HD 43162&G056A&G6.5 V&97&$\pm$&0.98&99&$\pm$&2.5&15&$\pm$&2.4&&&&11&$\pm$&0.28&6.4&$\pm$&1.2&5.5&$\pm$&0.14&6.1&$\pm$&2.2&2.1&$\pm$&0.053&&&&&&&&&&\\
HD 43834&G015A&G7 V&320&$\pm$&3.2&310&$\pm$&5.2&39&$\pm$&5.4&&&&36&$\pm$&0.59&16&$\pm$&2.0&17&$\pm$&0.29&12&$\pm$&4.0&6.7&$\pm$&0.11&&&&&&&&&&\\
HD 46588&F056A&F8 V&150&$\pm$&1.5&160&$\pm$&2.8&20&$\pm$&2.2&&&&18&$\pm$&0.32&8.9&$\pm$&1.3&8.6&$\pm$&0.15&2.9&$\pm$&3.9&3.3&$\pm$&0.060&&&&&&&&&&\\
HD 48682$^*$&F044A&F9 V&200&$\pm$&2.0&200&$\pm$&8.2&330&$\pm$&3.7&&&&23&$\pm$&0.93&260&$\pm$&20&11&$\pm$&0.45&160&$\pm$&20&4.3&$\pm$&0.17&65&&$^{+4.2}_{--1.5}$&53&&$^{+1.5}_{-3.5}$&37&&$^{+5.5}_{-2.0}$&83\\
HD 50692&G065A&G0 V&140&$\pm$&1.4&140&$\pm$&2.5&8.9&$\pm$&4.0&20&$\pm$&2.3&16&$\pm$&0.28&6.6&$\pm$&1.4&7.8&$\pm$&0.14&5.3&$\pm$&2.3&3.0&$\pm$&0.053&&&&&&&&&&\\
HD 52711&G093A&G0 V&120&$\pm$&1.2&120&$\pm$&3.4&13&$\pm$&3.7&&&&13&$\pm$&0.38&7.2&$\pm$&1.2&6.6&$\pm$&0.19&2.9&$\pm$&2.5&2.5&$\pm$&0.072&&&&&&&&&&\\
HD 53705&G059A&G0 V&170&$\pm$&1.7&180&$\pm$&4.7&23&$\pm$&2.5&&&&21&$\pm$&0.53&8.4&$\pm$&3.3&10&$\pm$&0.26&4.2&$\pm$&4.4&3.9&$\pm$&0.10&&&&&&&&&&\\
HD 55575&F045A&F9 V&170&$\pm$&1.7&170&$\pm$&3.5&28&$\pm$&4.0&&&&19&$\pm$&0.39&11&$\pm$&1.7&9.4&$\pm$&0.19&4.8&$\pm$&2.3&3.6&$\pm$&0.074&&&&&&&&&&\\
HD 55892&F098A&F3 V Fe-1.0&270&$\pm$&2.7&250&$\pm$&4.7&34&$\pm$&3.1&&&&29&$\pm$&0.54&14&$\pm$&2.3&14&$\pm$&0.26&7.4&$\pm$&5.1&5.4&$\pm$&0.10&6.8&&$^{+2.1}_{-1.9}$&210&&$^{+41}_{-60}$&4.3&&$^{+4.2}_{-1.3}$&4.5\\
HD 56986&F067A&F2&760&$\pm$&8.5&690&$\pm$&14&&&&&&&78&$\pm$&1.6&39&$\pm$&3.0&38&$\pm$&0.80&32&$\pm$&8.9&15&$\pm$&0.31&9.2&&$^{+3.2}_{-2.5}$&90&&$^{+48}_{-8.2}$&32&&$^{+6.7}_{-18}$&4.4\\
HD 58855&F082A&F6V&150&$\pm$&1.5&160&$\pm$&3.1&17&$\pm$&3.5&&&&18&$\pm$&0.35&9.5&$\pm$&1.4&8.9&$\pm$&0.17&2.1&$\pm$&2.2&3.4&$\pm$&0.065&&&&&&&&&&\\
HD 58946&F058A&F1 V&&&&330&$\pm$&47&&&&&&&38&$\pm$&5.3&18&$\pm$&4.9&18&$\pm$&2.6&8.1&$\pm$&8.1&7.1&$\pm$&0.99&&&&&&&&&&\\
HD 61606&K090A&K3- V&80&$\pm$&0.81&86&$\pm$&1.5&-0.77&$\pm$&5.6&&&&9.9&$\pm$&0.17&-3.2&$\pm$&1.9&4.8&$\pm$&0.085&-2.2&$\pm$&3.6&1.9&$\pm$&0.033&&&&&&&&&&\\
HD 62613&G062A&G8V&83&$\pm$&0.85&82&$\pm$&1.4&10&$\pm$&2.0&8.2&$\pm$&2.3&9.4&$\pm$&0.16&3.0&$\pm$&1.1&4.6&$\pm$&0.078&5.1&$\pm$&2.3&1.8&$\pm$&0.030&&&&&&&&&&\\
HD 68146&F109A&F6.5 V&130&$\pm$&1.3&130&$\pm$&2.2&17&$\pm$&2.1&&&&15&$\pm$&0.25&7.0&$\pm$&3.5&7.4&$\pm$&0.12&7.9&$\pm$&4.6&2.8&$\pm$&0.048&&&&&&&&&&\\
HD 69830&G022A&G8+ V&240&$\pm$&2.4&160&$\pm$&4.1&16&$\pm$&2.1&&&&19&$\pm$&0.47&11&$\pm$&1.3&9.0&$\pm$&0.23&3.5&$\pm$&2.5&3.5&$\pm$&0.088&190&&$^{+14}_{-13}$&310&&$^{+9.0}_{-15}$&0.63&&$^{+0.066}_{-0.035}$&23\\
HD 69897&F061A&F6 V&200&$\pm$&2.0&200&$\pm$&9.4&34&$\pm$&5.4&&&&23&$\pm$&1.1&12&$\pm$&2.5&11&$\pm$&0.52&9.4&$\pm$&5.1&4.3&$\pm$&0.20&&&&&&&&&&\\
HD 71243&F077A&F5 V Fe-0.8&459&$\pm$&5.2&440&$\pm$&7.9&58&$\pm$&4.2&&&&50&$\pm$&0.90&28&$\pm$&2.6&24&$\pm$&0.43&5.5&$\pm$&4.5&9.3&$\pm$&0.17&&&&&&&&&&\\
HD 72905&G036A&G1.5Vb&170&$\pm$&1.7&160&$\pm$&3.2&50&$\pm$&3.1&&&&18&$\pm$&0.36&20&$\pm$&1.5&8.8&$\pm$&0.18&6.7&$\pm$&2.2&3.4&$\pm$&0.068&8.1&&$^{+3.2}_{-1.3}$&90&&$^{+8.8}_{-31}$&9.4&&$^{+13}_{-1.6}$&7.8\\
HD 75732&K060A&K0 IV-V&180&$\pm$&1.8&180&$\pm$&4.6&20&$\pm$&3.2&&&&20&$\pm$&0.52&7.7&$\pm$&1.6&9.9&$\pm$&0.25&4.1&$\pm$&2.3&3.8&$\pm$&0.097&&&&&&&&&&\\
HD 76151&G068A&G3 V&120&$\pm$&1.2&120&$\pm$&2.0&33&$\pm$&3.2&&&&13&$\pm$&0.22&15&$\pm$&1.3&6.5&$\pm$&0.11&6.2&$\pm$&1.8&2.5&$\pm$&0.042&17&&$^{+3.6}_{-3.9}$&83&&$^{+14}_{-18}$&11&&$^{+7.3}_{-3.0}$&10\\
HD 76932&G119A&G2 V Fe-1.8 CH-1&130&$\pm$&1.3&130&$\pm$&2.4&15&$\pm$&2.1&&&&15&$\pm$&0.27&11&$\pm$&2.3&7.4&$\pm$&0.13&7.1&$\pm$&3.0&2.9&$\pm$&0.051&&&&&&&&&&\\
HD 76943&F040A&F5&520&$\pm$&5.2&509&$\pm$&11&18&$\pm$&26&&&&57&$\pm$&1.2&31&$\pm$&2.3&28&$\pm$&0.58&15&$\pm$&4.6&11&$\pm$&0.22&&&&&&&&&&\\
HD 78154&F084A&F7IV-V&280&$\pm$&3.1&310&$\pm$&120&40&$\pm$&5.4&&&&36&$\pm$&14&11&$\pm$&2.5&17&$\pm$&6.7&2.9&$\pm$&5.3&6.7&$\pm$&2.6&&&&&&&&&&\\
HD 78366&G094A&G0 IV-V&110&$\pm$&1.1&110&$\pm$&1.8&17&$\pm$&3.3&13&$\pm$&2.2&12&$\pm$&0.20&6.0&$\pm$&1.0&6.1&$\pm$&0.098&-1.8&$\pm$&1.9&2.3&$\pm$&0.038&&&&&&&&&&\\
HD 79028&G100A&G0 IV-V&240&$\pm$&2.4&240&$\pm$&4.3&25&$\pm$&4.9&&&&27&$\pm$&0.49&13&$\pm$&2.0&13&$\pm$&0.24&5.1&$\pm$&3.5&5.1&$\pm$&0.092&&&&&&&&&&\\
HD 79096&G109A&G9&97&$\pm$&0.98&100&$\pm$&1.9&11&$\pm$&12&&&&11&$\pm$&0.21&3.5&$\pm$&2.1&5.5&$\pm$&0.10&-2.3&$\pm$&2.9&2.1&$\pm$&0.039&&&&&&&&&&\\
HD 79210&K011A&K7&200&$\pm$&2.0&200&$\pm$&12&28&$\pm$&2.4&&&&23&$\pm$&1.4&9.8&$\pm$&2.0&11&$\pm$&0.67&4.0&$\pm$&4.0&4.3&$\pm$&0.26&&&&&&&&&&\\
HD 81997&F048A&F5 V&290&$\pm$&2.9&280&$\pm$&15&34&$\pm$&3.2&&&&32&$\pm$&1.7&16&$\pm$&2.6&16&$\pm$&0.84&4.1&$\pm$&4.6&6.0&$\pm$&0.33&&&&&&&&&&\\
HD 82106&K064A&K3 V&90&$\pm$&0.91&94&$\pm$&1.5&-2.2&$\pm$&6.4&12&$\pm$&2.4&11&$\pm$&0.18&7.4&$\pm$&1.2&5.2&$\pm$&0.085&4.0&$\pm$&2.3&2.0&$\pm$&0.033&&&&&&&&&&\\
HD 82328&F019A&F5.5 IV-V&&&&1200&$\pm$&160&&&&&&&130&$\pm$&18&68&$\pm$&4.5&64&$\pm$&8.6&34&$\pm$&4.0&24&$\pm$&3.3&&&&&&&&&&\\
HD 84117&F031A&F8 V&260&$\pm$&2.9&250&$\pm$&4.8&32&$\pm$&7.5&30&$\pm$&2.7&29&$\pm$&0.55&6.7&$\pm$&1.7&14&$\pm$&0.27&9.4&$\pm$&12&5.4&$\pm$&0.10&&&&&&&&&&\\
HD 84737&G086A&G0 IV-V&250&$\pm$&2.6&250&$\pm$&4.5&36&$\pm$&3.9&&&&28&$\pm$&0.51&8.9&$\pm$&1.7&14&$\pm$&0.25&10&$\pm$&4.0&5.3&$\pm$&0.096&&&&&&&&&&\\
HD 86728&G040A&G4 V&210&$\pm$&2.1&210&$\pm$&4.4&5.5&$\pm$&58&&&&24&$\pm$&0.50&8.7&$\pm$&1.7&12&$\pm$&0.24&4.0&$\pm$&3.0&4.5&$\pm$&0.093&&&&&&&&&&\\
HD 88230&K005A&K5&440&$\pm$&4.4&450&$\pm$&11&43&$\pm$&4.2&&&&51&$\pm$&1.2&23&$\pm$&1.5&25&$\pm$&0.61&12&$\pm$&1.9&9.8&$\pm$&0.24&&&&&&&&&&\\
HD 89125&F113A&F6 V&110&$\pm$&1.1&110&$\pm$&2.3&23&$\pm$&13&&&&13&$\pm$&0.26&5.5&$\pm$&2.1&6.2&$\pm$&0.13&7.0&$\pm$&4.3&2.4&$\pm$&0.049&&&&&&&&&&\\
HD 89269&G108A&G4 V&69&$\pm$&0.70&73&$\pm$&1.2&2.5&$\pm$&43&&&&8.3&$\pm$&0.14&2.7&$\pm$&1.6&4.0&$\pm$&0.069&2.0&$\pm$&4.1&1.5&$\pm$&0.027&&&&&&&&&&\\
HD 89449&F097A&F6 IV-V&240&$\pm$&2.4&240&$\pm$&4.8&29&$\pm$&3.4&&&&27&$\pm$&0.55&10&$\pm$&2.1&13&$\pm$&0.27&10&$\pm$&6.2&5.1&$\pm$&0.10&&&&&&&&&&\\
HD 90089$^*$&F100A&F4 V kF2mF2&150&$\pm$&1.5&150&$\pm$&2.9&40&$\pm$&4.1&&&&17&$\pm$&0.33&54&$\pm$&5.1&8.2&$\pm$&0.16&74&$\pm$&5.4&3.2&$\pm$&0.061&9.3&&$^{+1.3}_{-0.66}$&31&&$^{+1.2}_{-1.3}$&140&&$^{+13}_{-11}$&20\\
HD 90839&F018A&F8 V&280&$\pm$&2.8&280&$\pm$&5.0&34&$\pm$&3.8&&&&31&$\pm$&0.57&19&$\pm$&2.7&15&$\pm$&0.28&9.7&$\pm$&3.8&5.9&$\pm$&0.11&&&&&&&&&&\\
HD 95128&G033A&G0V&270&$\pm$&2.7&270&$\pm$&5.1&33&$\pm$&4.2&&&&31&$\pm$&0.58&12&$\pm$&2.0&15&$\pm$&0.28&-0.45&$\pm$&4.9&5.7&$\pm$&0.11&&&&&&&&&&\\
HD 97101&K056A&K7 V&&&&77&$\pm$&3.9&12&$\pm$&4.9&7.7&$\pm$&2.4&8.7&$\pm$&0.44&5.5&$\pm$&0.94&4.2&$\pm$&0.21&2.8&$\pm$&2.3&1.6&$\pm$&0.083&&&&&&&&&&\\
HD 97584&K102A&K5&65&$\pm$&0.73&71&$\pm$&1.7&6.0&$\pm$&4.4&&&&8.2&$\pm$&0.19&2.9&$\pm$&2.0&4.0&$\pm$&0.094&-7.3&$\pm$&4.4&1.5&$\pm$&0.036&&&&&&&&&&\\
HD 98712&K069A&K6 V ke&55&$\pm$&0.61&65&$\pm$&3.6&0.16&$\pm$&3.3&&&&7.5&$\pm$&0.42&-2.9&$\pm$&1.8&3.7&$\pm$&0.20&-3.0&$\pm$&6.2&1.4&$\pm$&0.079&&&&&&&&&&\\
HD 99491&G079A&G6/8 III/IV&92&$\pm$&0.93&95&$\pm$&4.1&5.9&$\pm$&7.4&&&&11&$\pm$&0.47&5.4&$\pm$&1.8&5.2&$\pm$&0.23&1.7&$\pm$&6.2&2.0&$\pm$&0.087&&&&&&&&&&\\
HD 100180&F121A&F9.5 V&78&$\pm$&0.79&80&$\pm$&1.4&5.2&$\pm$&6.1&&&&9.0&$\pm$&0.15&9.0&$\pm$&2.3&4.4&$\pm$&0.075&6.5&$\pm$&4.4&1.7&$\pm$&0.029&&&&&&&&&&\\
HD 100623&K029A&K0- V&190&$\pm$&1.9&190&$\pm$&3.3&25&$\pm$&2.4&&&&21&$\pm$&0.37&14&$\pm$&1.5&10&$\pm$&0.18&5.5&$\pm$&4.6&4.0&$\pm$&0.070&&&&&&&&&&\\
HD 101177&F120A&F9.5 V&68&$\pm$&0.76&70&$\pm$&1.4&2.0&$\pm$&5.6&&&&8.0&$\pm$&0.16&4.5&$\pm$&1.8&3.9&$\pm$&0.079&-4.9&$\pm$&3.5&1.5&$\pm$&0.031&&&&&&&&&&\\
HD 101501&G013A&G8 V&260&$\pm$&2.6&270&$\pm$&5.1&29&$\pm$&4.5&&&&30&$\pm$&0.58&14&$\pm$&2.1&15&$\pm$&0.28&7.2&$\pm$&4.0&5.7&$\pm$&0.11&&&&&&&&&&\\
HD 101581&K059A&K4.5 V (k)&66&$\pm$&0.67&71&$\pm$&1.8&7.0&$\pm$&5.3&&&&8.1&$\pm$&0.21&2.2&$\pm$&2.0&4.0&$\pm$&0.10&5.0&$\pm$&4.9&1.5&$\pm$&0.039&&&&&&&&&&\\
HD 102365&G012A&G2 V&360&$\pm$&3.6&370&$\pm$&6.4&48&$\pm$&4.1&&&&42&$\pm$&0.72&19&$\pm$&2.2&20&$\pm$&0.35&2.4&$\pm$&7.2&7.8&$\pm$&0.14&&&&&&&&&&\\
HD 102438&G069A&G6 V&87&$\pm$&0.88&88&$\pm$&1.5&9.9&$\pm$&2.2&14&$\pm$&2.3&10&$\pm$&0.17&4.7&$\pm$&1.1&4.9&$\pm$&0.081&2.2&$\pm$&2.3&1.9&$\pm$&0.031&&&&&&&&&&\\
HD 102870&F009A&F9 V&900&$\pm$&9.0&890&$\pm$&17&130&$\pm$&7.8&&&&100&$\pm$&2.0&59&$\pm$&4.1&49&$\pm$&0.96&42&$\pm$&4.8&19&$\pm$&0.37&0.81&&$^{+0.15}_{-0.22}$&43&&$^{+13}_{-7.6}$&78&&$^{+37}_{-31}$&6.3\\
HD 103095&K028A&K1 V Fe-1.5&130&$\pm$&1.3&130&$\pm$&2.5&10&$\pm$&1.9&&&&15&$\pm$&0.29&6.4&$\pm$&1.1&7.4&$\pm$&0.14&2.6&$\pm$&3.1&2.9&$\pm$&0.054&&&&&&&&&&\\
HD 103932&K034A&K4+ V&140&$\pm$&1.4&150&$\pm$&2.3&10&$\pm$&4.9&&&&17&$\pm$&0.26&7.6&$\pm$&0.99&8.3&$\pm$&0.13&4.5&$\pm$&2.1&3.2&$\pm$&0.049&&&&&&&&&&\\
HD 105452&F030A&F1 V&400&$\pm$&4.0&390&$\pm$&7.3&48&$\pm$&4.3&&&&45&$\pm$&0.83&27&$\pm$&2.6&22&$\pm$&0.40&12&$\pm$&5.0&8.3&$\pm$&0.16&&&&&&&&&&\\
HD 106516&F108A&F9 V Fe-1.7 CH-0.7&80&$\pm$&0.81&84&$\pm$&2.0&6.4&$\pm$&6.2&&&&9.5&$\pm$&0.23&2.7&$\pm$&1.9&4.6&$\pm$&0.11&2.1&$\pm$&3.6&1.8&$\pm$&0.043&&&&&&&&&&\\
HD 108954&F103A&F9 V&83&$\pm$&0.83&83&$\pm$&1.3&0.44&$\pm$&4.4&&&&9.4&$\pm$&0.15&3.9&$\pm$&2.3&4.6&$\pm$&0.071&6.1&$\pm$&3.8&1.8&$\pm$&0.027&&&&&&&&&&\\
HD 109085$^*$&F063A&F2 V&610&$\pm$&6.1&310&$\pm$&26&280&$\pm$&4.4&229&$\pm$&13&35&$\pm$&2.9&250&$\pm$&16&17&$\pm$&1.4&229&$\pm$&13&6.6&$\pm$&0.54&17&&$^{+1.1}_{-0.88}$&40&&$^{+1.9}_{-1.9}$&110&&$^{+11}_{-9.4}$&25\\
HD 109358&G007A&G0&560&$\pm$&5.6&560&$\pm$&10&60&$\pm$&5.2&&&&64&$\pm$&1.2&30&$\pm$&2.4&31&$\pm$&0.57&12&$\pm$&5.1&12&$\pm$&0.22&&&&&&&&&&\\
HD 110315&K092A&K4.5 V&68&$\pm$&0.69&74&$\pm$&1.3&20&$\pm$&6.5&&&&8.5&$\pm$&0.15&3.3&$\pm$&2.4&4.1&$\pm$&0.075&5.5&$\pm$&3.9&1.6&$\pm$&0.029&&&&&&&&&&\\
HD 110833&K111A&K3&90&$\pm$&0.91&92&$\pm$&1.3&1.3&$\pm$&4.7&9.6&$\pm$&2.6&11&$\pm$&0.15&4.3&$\pm$&0.92&5.2&$\pm$&0.075&0.98&$\pm$&3.2&2.0&$\pm$&0.029&&&&&&&&&&\\
HD 110897$^*$&F050A&F9 V Fe-0.3&120&$\pm$&1.2&120&$\pm$&2.5&64&$\pm$&2.7&76&$\pm$&7.0&14&$\pm$&0.29&59&$\pm$&4.2&6.6&$\pm$&0.14&39&$\pm$&4.0&2.5&$\pm$&0.054&23&&$^{+2.0}_{-3.2}$&56&&$^{+3.5}_{-6.7}$&26&&$^{+7.7}_{-3.0}$&25\\
HD 111395&G057A&G7 V&99&$\pm$&1.0&100&$\pm$&1.8&20&$\pm$&2.3&&&&11&$\pm$&0.21&4.7&$\pm$&0.92&5.6&$\pm$&0.10&-0.64&$\pm$&2.6&2.1&$\pm$&0.039&&&&&&&&&&\\
HD 111631&K036A&K7&&&&87&$\pm$&0.83&8.7&$\pm$&6.2&&&&10&$\pm$&0.097&9.5&$\pm$&2.2&4.9&$\pm$&0.047&29&$\pm$&4.6&1.9&$\pm$&0.018&13&&$^{+4.1}_{-3.1}$&18&&$^{+3.6}_{-4.6}$&74&&$^{+59}_{-23}$&6.5\\
HD 114710&G010A&G0V&520&$\pm$&5.2&509&$\pm$&9.6&50&$\pm$&5.4&&&&58&$\pm$&1.1&22&$\pm$&2.2&28&$\pm$&0.53&10&$\pm$&5.1&11&$\pm$&0.20&&&&&&&&&&\\
HD 115383&G073A&G0Vs&229&$\pm$&2.3&220&$\pm$&4.6&14&$\pm$&5.1&&&&24&$\pm$&0.52&13&$\pm$&1.3&12&$\pm$&0.25&2.0&$\pm$&2.3&4.6&$\pm$&0.097&&&&&&&&&&\\
HD 115404&K046A&K2.5 V (k)&130&$\pm$&1.3&140&$\pm$&9.4&10&$\pm$&5.4&15&$\pm$&2.3&17&$\pm$&1.1&7.9&$\pm$&1.2&8.1&$\pm$&0.53&3.9&$\pm$&3.6&3.2&$\pm$&0.21&&&&&&&&&&\\
HD 115617$^*$&G008A&G7 V&459&$\pm$&4.6&440&$\pm$&8.4&250&$\pm$&11&200&$\pm$&11&50&$\pm$&0.96&150&$\pm$&10&24&$\pm$&0.47&130&$\pm$&17&9.4&$\pm$&0.18&29&&$^{+1.8}_{-1.4}$&67&&$^{+2.4}_{-3.7}$&16&&$^{+1.9}_{-1.1}$&28\\
HD 116442&G050A&G9 V&&&&69&$\pm$&1.1&&&&&&&7.8&$\pm$&0.13&4.0&$\pm$&2.1&3.8&$\pm$&0.063&1.2&$\pm$&3.3&1.5&$\pm$&0.024&&&&&&&&&&\\
HD 117043&G121A&G6V&88&$\pm$&0.98&87&$\pm$&1.3&0.10&$\pm$&2.2&&&&9.9&$\pm$&0.15&9.6&$\pm$&2.5&4.8&$\pm$&0.071&8.7&$\pm$&3.1&1.8&$\pm$&0.027&&&&&&&&&&\\
HD 118926&K109A&K5&&&&39&$\pm$&0.88&&&&&&&4.5&$\pm$&0.10&5.6&$\pm$&2.6&2.2&$\pm$&0.050&4.8&$\pm$&4.2&0.85&$\pm$&0.019&&&&&&&&&&\\
HD 119756&F075A&F2 V&340&$\pm$&3.5&340&$\pm$&6.7&&&&&&&39&$\pm$&0.76&18&$\pm$&2.0&19&$\pm$&0.37&3.5&$\pm$&3.8&7.2&$\pm$&0.14&&&&&&&&&&\\
HD 120036&K103A&K6.5 V (k)&&&&47&$\pm$&0.54&10&$\pm$&6.1&&&&5.4&$\pm$&0.061&3.6&$\pm$&1.8&2.6&$\pm$&0.030&1.1&$\pm$&4.1&1.0&$\pm$&0.012&&&&&&&&&&\\
HD 120136&F036A&F7V&&&&260&$\pm$&9.7&&&&&&&29&$\pm$&1.1&20&$\pm$&3.4&14&$\pm$&0.53&9.2&$\pm$&6.6&5.5&$\pm$&0.21&&&&&&&&&&\\
HD 120467&K095A&K5.5 V (k)&&&&67&$\pm$&0.60&-3.7&$\pm$&7.0&&&&7.6&$\pm$&0.069&4.1&$\pm$&2.4&3.7&$\pm$&0.034&-2.1&$\pm$&3.2&1.5&$\pm$&0.013&&&&&&&&&&\\
HD 120476&K074A&K3.5 V&&&&74&$\pm$&0.92&2.2&$\pm$&4.7&9.9&$\pm$&3.1&8.4&$\pm$&0.11&5.2&$\pm$&1.2&4.1&$\pm$&0.051&0.21&$\pm$&2.9&1.6&$\pm$&0.020&&&&&&&&&&\\
HD 120690&G097A&G5+ V&100&$\pm$&0.70&100&$\pm$&1.7&9.0&$\pm$&2.7&8.0&$\pm$&2.1&11&$\pm$&0.19&5.6&$\pm$&1.2&5.6&$\pm$&0.092&0.92&$\pm$&1.7&2.1&$\pm$&0.035&&&&&&&&&&\\
HD 122064&K032A&K3V&170&$\pm$&1.7&170&$\pm$&3.2&18&$\pm$&6.6&&&&20&$\pm$&0.37&8.8&$\pm$&0.98&9.6&$\pm$&0.18&4.4&$\pm$&2.4&3.7&$\pm$&0.069&&&&&&&&&&\\
HD 122742&G061A&G6 V&120&$\pm$&1.2&120&$\pm$&2.4&4.6&$\pm$&5.5&15&$\pm$&2.2&14&$\pm$&0.27&6.7&$\pm$&0.99&6.7&$\pm$&0.13&2.3&$\pm$&1.9&2.6&$\pm$&0.050&&&&&&&&&&\\
HD 124580&G123A&G0 V&80&$\pm$&0.81&82&$\pm$&2.2&&&&&&&9.3&$\pm$&0.25&8.8&$\pm$&1.8&4.5&$\pm$&0.12&4.6&$\pm$&4.2&1.7&$\pm$&0.047&&&&&&&&&&\\
HD 125276&F059A&F9 V Fe-1.5 CH-0.7&&&&120&$\pm$&19&&&&&&&14&$\pm$&2.2&3.9&$\pm$&1.3&7.0&$\pm$&1.1&4.0&$\pm$&3.5&2.7&$\pm$&0.42&&&&&&&&&&\\
HD 126053&G063A&G1.5 V&98&$\pm$&0.99&99&$\pm$&1.5&5.2&$\pm$&6.0&14&$\pm$&2.2&11&$\pm$&0.17&3.3&$\pm$&0.93&5.5&$\pm$&0.082&-0.24&$\pm$&1.8&2.1&$\pm$&0.031&&&&&&&&&&\\
HD 126660&F026A&F7V&570&$\pm$&5.7&540&$\pm$&10&71&$\pm$&5.2&&&&61&$\pm$&1.2&33&$\pm$&3.0&30&$\pm$&0.57&9.3&$\pm$&3.8&11&$\pm$&0.22&&&&&&&&&&\\
HD 128165&K072A&K3V&86&$\pm$&0.86&91&$\pm$&1.3&35&$\pm$&4.9&15&$\pm$&2.2&10&$\pm$&0.15&11&$\pm$&1.1&5.1&$\pm$&0.074&3.8&$\pm$&1.9&2.0&$\pm$&0.029&5.0&&$^{+4.3}_{-1.7}$&52&&$^{+10}_{-28}$&14&&$^{+52}_{-4.4}$&5.6\\
HD 128167&F039A&F3Vwvar&&&&290&$\pm$&5.8&&&&&&&33&$\pm$&0.66&36&$\pm$&2.8&16&$\pm$&0.32&16&$\pm$&3.9&6.2&$\pm$&0.12&14&&$^{+9.8}_{-2.5}$&130&&$^{+7.3}_{-65}$&8.0&&$^{+22}_{-0.81}$&8.4\\
HD 128642&G103A&G5&79&$\pm$&0.79&78&$\pm$&1.2&6.9&$\pm$&4.5&9.4&$\pm$&2.3&8.9&$\pm$&0.14&5.8&$\pm$&0.94&4.3&$\pm$&0.068&7.8&$\pm$&2.2&1.7&$\pm$&0.026&&&&&&&&&&\\
HD 129502&F062A&F2 V&540&$\pm$&5.4&509&$\pm$&20&56&$\pm$&3.6&&&&58&$\pm$&2.2&36&$\pm$&2.7&28&$\pm$&1.1&5.5&$\pm$&4.2&11&$\pm$&0.42&&&&&&&&&&\\
HD 130948&G084A&G2V&120&$\pm$&1.2&120&$\pm$&1.9&11&$\pm$&2.9&&&&13&$\pm$&0.21&6.7&$\pm$&1.0&6.5&$\pm$&0.10&-0.056&$\pm$&1.3&2.5&$\pm$&0.040&&&&&&&&&&\\
HD 131156&G006A&G7 V&&&&509&$\pm$&33&&&&&&&59&$\pm$&3.8&35&$\pm$&2.9&28&$\pm$&1.8&17&$\pm$&4.7&11&$\pm$&0.71&&&&&&&&&&\\
HD 131511$^*$&K053A&K0 V&200&$\pm$&2.0&200&$\pm$&3.5&37&$\pm$&5.8&31&$\pm$&2.2&23&$\pm$&0.40&21&$\pm$&3.9&11&$\pm$&0.19&13&$\pm$&3.7&4.2&$\pm$&0.075&3.7&&$^{+1.3}_{-1.4}$&49&&$^{+17}_{-27}$&24&&$^{+94}_{-11}$&5.0\\
HD 131977&K008A&K4 V&420&$\pm$&4.2&420&$\pm$&7.8&53&$\pm$&3.2&&&&48&$\pm$&0.89&22&$\pm$&2.1&24&$\pm$&0.43&5.6&$\pm$&2.8&9.1&$\pm$&0.17&&&&&&&&&&\\
HD 134083&F076A&F5 V&210&$\pm$&2.1&200&$\pm$&3.8&37&$\pm$&4.4&&&&23&$\pm$&0.43&7.0&$\pm$&1.6&11&$\pm$&0.21&3.1&$\pm$&2.8&4.3&$\pm$&0.081&&&&&&&&&&\\
HD 135204&G076A&G9&89&$\pm$&0.90&90&$\pm$&1.5&6.1&$\pm$&6.0&&&&10&$\pm$&0.17&6.7&$\pm$&1.8&5.0&$\pm$&0.085&14&$\pm$&5.2&1.9&$\pm$&0.033&&&&&&&&&&\\
HD 136923&G101A&G9 V&54&$\pm$&0.55&57&$\pm$&0.95&12&$\pm$&4.5&&&&6.5&$\pm$&0.11&4.5&$\pm$&2.0&3.1&$\pm$&0.052&5.9&$\pm$&4.6&1.2&$\pm$&0.020&&&&&&&&&&\\
HD 137763&G104A&G9 V&80&$\pm$&0.81&84&$\pm$&1.9&&&&&&&9.5&$\pm$&0.21&4.0&$\pm$&2.3&4.6&$\pm$&0.10&2.5&$\pm$&5.2&1.8&$\pm$&0.040&&&&&&&&&&\\
HD 139763&K126A&K6 V k&43&$\pm$&0.45&47&$\pm$&1.2&&&&&&&5.4&$\pm$&0.14&3.7&$\pm$&2.2&2.7&$\pm$&0.068&-0.57&$\pm$&5.1&1.0&$\pm$&0.026&&&&&&&&&&\\
HD 140538&G038A&G3 V&150&$\pm$&1.6&170&$\pm$&11&13&$\pm$&6.4&&&&19&$\pm$&1.3&9.0&$\pm$&1.4&9.2&$\pm$&0.62&10&$\pm$&4.2&3.5&$\pm$&0.24&&&&&&&&&&\\
HD 141004&G019A&G0 IV-V&450&$\pm$&4.5&459&$\pm$&8.7&52&$\pm$&5.1&&&&52&$\pm$&0.99&28&$\pm$&2.3&26&$\pm$&0.48&13&$\pm$&4.2&9.8&$\pm$&0.19&&&&&&&&&&\\
HD 141272&G125A&G9 V (k)&44&$\pm$&0.47&47&$\pm$&1.3&0.23&$\pm$&5.4&&&&5.4&$\pm$&0.15&2.0&$\pm$&1.9&2.6&$\pm$&0.073&2.5&$\pm$&4.5&1.0&$\pm$&0.028&&&&&&&&&&\\
HD 142267&G067A&G0IV&110&$\pm$&1.1&110&$\pm$&1.8&11&$\pm$&1.7&&&&12&$\pm$&0.21&4.9&$\pm$&2.0&6.0&$\pm$&0.10&3.2&$\pm$&3.7&2.3&$\pm$&0.039&&&&&&&&&&\\
HD 142373&G052A&G0 V Fe-0.8 CH-0.5&440&$\pm$&4.4&420&$\pm$&7.4&40&$\pm$&5.6&&&&48&$\pm$&0.84&19&$\pm$&2.2&23&$\pm$&0.41&11&$\pm$&3.7&8.9&$\pm$&0.16&&&&&&&&&&\\
HD 142860&F011A&F6V&660&$\pm$&6.7&630&$\pm$&12&73&$\pm$&6.5&&&&72&$\pm$&1.3&33&$\pm$&2.7&35&$\pm$&0.64&13&$\pm$&4.8&13&$\pm$&0.24&&&&&&&&&&\\
HD 143761&G064A&G0 V&200&$\pm$&2.0&200&$\pm$&3.5&31&$\pm$&4.1&&&&23&$\pm$&0.40&8.8&$\pm$&1.6&11&$\pm$&0.19&-1.7&$\pm$&4.0&4.3&$\pm$&0.075&&&&&&&&&&\\
HD 144579&K098A&K0 V Fe-1.2&89&$\pm$&0.89&93&$\pm$&1.2&13&$\pm$&2.8&9.7&$\pm$&3.3&11&$\pm$&0.14&3.3&$\pm$&1.1&5.2&$\pm$&0.069&-0.41&$\pm$&2.1&2.0&$\pm$&0.027&&&&&&&&&&\\
HD 146361&G122A&G1 IV-V (k)&&&&190&$\pm$&2.6&&&&&&&21&$\pm$&0.30&10&$\pm$&2.3&10&$\pm$&0.14&14&$\pm$&4.8&4.0&$\pm$&0.056&&&&&&&&&&\\
HD 147379&K039A&K7&&&&84&$\pm$&0.68&17&$\pm$&5.1&&&&9.7&$\pm$&0.078&1.6&$\pm$&3.0&4.7&$\pm$&0.038&-6.0&$\pm$&4.2&1.8&$\pm$&0.015&&&&&&&&&&\\
HD 147584&F016A&F9 V&290&$\pm$&2.9&280&$\pm$&5.2&&&&&&&32&$\pm$&0.59&14&$\pm$&1.8&15&$\pm$&0.29&0.74&$\pm$&3.7&5.9&$\pm$&0.11&&&&&&&&&&\\
HD 151288&K031A&K5&100&$\pm$&1.0&100&$\pm$&19&16&$\pm$&1.6&&&&12&$\pm$&2.1&4.0&$\pm$&1.7&5.7&$\pm$&1.0&0.80&$\pm$&3.1&2.2&$\pm$&0.40&&&&&&&&&&\\
HD 153597&F034A&F6Vvar&270&$\pm$&2.7&270&$\pm$&5.0&47&$\pm$&4.9&&&&30&$\pm$&0.57&15&$\pm$&1.2&15&$\pm$&0.28&-1.6&$\pm$&2.5&5.7&$\pm$&0.11&&&&&&&&&&\\
HD 154345&G088A&G8V&71&$\pm$&0.71&73&$\pm$&1.0&5.3&$\pm$&5.1&6.9&$\pm$&1.3&8.3&$\pm$&0.11&3.2&$\pm$&0.93&4.0&$\pm$&0.055&-1.2&$\pm$&1.6&1.6&$\pm$&0.021&&&&&&&&&&\\
HD 154577&K080A&K2.5 V (k)&66&$\pm$&0.67&68&$\pm$&0.98&&&&&&&7.9&$\pm$&0.11&11&$\pm$&3.3&3.8&$\pm$&0.055&4.2&$\pm$&7.5&1.5&$\pm$&0.021&&&&&&&&&&\\
HD 157214&G035A&G0V&220&$\pm$&2.2&220&$\pm$&3.8&24&$\pm$&4.2&&&&25&$\pm$&0.43&14&$\pm$&1.6&12&$\pm$&0.21&5.6&$\pm$&2.8&4.6&$\pm$&0.080&&&&&&&&&&\\
HD 157881&K021A&K5&170&$\pm$&1.7&180&$\pm$&8.1&18&$\pm$&3.0&&&&20&$\pm$&0.92&9.9&$\pm$&1.9&10&$\pm$&0.45&12&$\pm$&3.5&3.9&$\pm$&0.17&&&&&&&&&&\\
HD 158633$^*$&K062A&K0V&110&$\pm$&1.1&110&$\pm$&1.5&57&$\pm$&1.7&&&&13&$\pm$&0.18&36&$\pm$&2.7&6.3&$\pm$&0.085&14&$\pm$&3.0&2.4&$\pm$&0.033&29&&$^{+4.6}_{-2.9}$&64&&$^{+4.6}_{-13}$&12&&$^{+7.2}_{-1.6}$&28\\
HD 160032&F099A&F4 V&240&$\pm$&2.4&240&$\pm$&5.7&49&$\pm$&4.5&&&&27&$\pm$&0.65&31&$\pm$&2.6&13&$\pm$&0.32&24&$\pm$&4.9&5.1&$\pm$&0.12&4.8&&$^{+1.0}_{-0.97}$&76&&$^{+11}_{-9.0}$&30&&$^{+8.6}_{-6.8}$&9.0\\
HD 160922&F118A&F5V&240&$\pm$&2.4&240&$\pm$&4.6&32&$\pm$&5.2&&&&27&$\pm$&0.52&16&$\pm$&1.8&13&$\pm$&0.25&10&$\pm$&3.1&5.1&$\pm$&0.098&&&&&&&&&&\\
HD 162003&F115A&F5IV-V&300&$\pm$&3.0&300&$\pm$&6.1&47&$\pm$&5.0&&&&34&$\pm$&0.69&17&$\pm$&2.7&17&$\pm$&0.34&3.8&$\pm$&4.5&6.4&$\pm$&0.13&&&&&&&&&&\\
HD 165499&G075A&G0 V&160&$\pm$&8.8&180&$\pm$&3.3&18&$\pm$&2.5&&&&20&$\pm$&0.38&11&$\pm$&1.5&9.8&$\pm$&0.18&-1.1&$\pm$&2.1&3.8&$\pm$&0.070&&&&&&&&&&\\
HD 166348&K068A&K6 V (k)&63&$\pm$&0.64&68&$\pm$&0.99&21&$\pm$&5.3&&&&7.8&$\pm$&0.11&13&$\pm$&2.6&3.8&$\pm$&0.055&8.2&$\pm$&5.5&1.5&$\pm$&0.021&17&&$^{+16}_{-5.8}$&41&&$^{+17}_{-16}$&17&&$^{+28}_{-8.3}$&4.4\\
HD 166620&K044A&K2 V&150&$\pm$&1.5&150&$\pm$&2.2&9.3&$\pm$&4.6&18&$\pm$&2.3&17&$\pm$&0.26&7.6&$\pm$&1.2&8.4&$\pm$&0.13&5.6&$\pm$&5.4&3.2&$\pm$&0.048&&&&&&&&&&\\
HD 167425&F116A&F9.5 V&83&$\pm$&0.93&88&$\pm$&56&18&$\pm$&5.7&&&&10&$\pm$&6.4&4.9&$\pm$&1.8&4.9&$\pm$&3.1&3.4&$\pm$&4.1&1.9&$\pm$&1.2&&&&&&&&&&\\
HD 168151&F114A&F5V&210&$\pm$&2.1&210&$\pm$&3.9&24&$\pm$&3.2&&&&24&$\pm$&0.44&10&$\pm$&1.7&11&$\pm$&0.21&4.0&$\pm$&3.5&4.4&$\pm$&0.083&&&&&&&&&&\\
HD 170153&F002A&F7Vvar&1000&$\pm$&10&1000&$\pm$&19&130&$\pm$&5.0&&&&120&$\pm$&2.1&53&$\pm$&3.7&56&$\pm$&1.0&28&$\pm$&5.7&22&$\pm$&0.41&&&&&&&&&&\\
HD 179930&K125A&K7&52&$\pm$&0.58&58&$\pm$&2.3&0.26&$\pm$&5.1&&&&6.6&$\pm$&0.26&7.3&$\pm$&2.3&3.2&$\pm$&0.13&1.2&$\pm$&3.6&1.3&$\pm$&0.050&&&&&&&&&&\\
HD 180161&G106A&G8V&60&$\pm$&0.61&61&$\pm$&0.99&8.6&$\pm$&4.4&&&&7.0&$\pm$&0.11&6.2&$\pm$&1.9&3.4&$\pm$&0.055&-2.9&$\pm$&5.4&1.3&$\pm$&0.021&&&&&&&&&&\\
HD 181321&G091A&G1 V&81&$\pm$&0.83&78&$\pm$&1.1&0.51&$\pm$&2.5&&&&8.9&$\pm$&0.12&4.5&$\pm$&2.6&4.3&$\pm$&0.059&5.6&$\pm$&4.0&1.7&$\pm$&0.023&&&&&&&&&&\\
HD 185395&F065A&F3+ V&&&&340&$\pm$&409&&&&&&&38&$\pm$&46&17&$\pm$&2.2&19&$\pm$&22&11&$\pm$&6.0&7.2&$\pm$&8.7&&&&&&&&&&\\
HD 186408&G120A&G1.5 V&&&&120&$\pm$&1.9&&&&&&&14&$\pm$&0.22&6.2&$\pm$&2.4&6.8&$\pm$&0.11&-3.2&$\pm$&3.5&2.6&$\pm$&0.041&&&&&&&&&&\\
HD 189245&F095A&F8.5 V Fe-0.6 CH-0.5&120&$\pm$&1.3&120&$\pm$&2.3&11&$\pm$&2.3&&&&14&$\pm$&0.26&5.2&$\pm$&2.2&6.6&$\pm$&0.13&1.8&$\pm$&4.7&2.5&$\pm$&0.049&&&&&&&&&&\\
HD 189567&G077A&G2 V&110&$\pm$&1.2&120&$\pm$&2.1&23&$\pm$&2.6&11&$\pm$&2.3&13&$\pm$&0.24&5.0&$\pm$&0.95&6.4&$\pm$&0.12&4.0&$\pm$&1.9&2.5&$\pm$&0.045&&&&&&&&&&\\
HD 190007&K063A&K4 V (k)&90&$\pm$&0.92&96&$\pm$&1.3&11&$\pm$&3.0&&&&11&$\pm$&0.15&5.4&$\pm$&1.7&5.4&$\pm$&0.075&6.1&$\pm$&7.4&2.1&$\pm$&0.029&&&&&&&&&&\\
HD 190422&F123A&F9 V CH-0.4&77&$\pm$&0.78&76&$\pm$&1.6&-0.38&$\pm$&6.4&&&&8.6&$\pm$&0.19&7.7&$\pm$&1.9&4.2&$\pm$&0.090&0.59&$\pm$&4.1&1.6&$\pm$&0.035&&&&&&&&&&\\
HD 191849&K012A&K7.0&190&$\pm$&1.9&190&$\pm$&7.7&35&$\pm$&2.5&&&&22&$\pm$&0.88&19&$\pm$&2.5&11&$\pm$&0.43&18&$\pm$&4.1&4.1&$\pm$&0.17&10&&$^{+6.1}_{-4.4}$&37&&$^{+27}_{-6.7}$&14&&$^{+7.1}_{-9.5}$&6.1\\
HD 192310&K027A&K2+ V&260&$\pm$&2.6&260&$\pm$&9.9&24&$\pm$&3.7&&&&30&$\pm$&1.1&11&$\pm$&1.8&14&$\pm$&0.55&5.2&$\pm$&5.6&5.6&$\pm$&0.21&&&&&&&&&&\\
HD 194640&G098A&G8 V&75&$\pm$&0.76&79&$\pm$&1.1&7.2&$\pm$&6.3&&&&9.0&$\pm$&0.12&5.0&$\pm$&2.1&4.4&$\pm$&0.061&4.7&$\pm$&3.8&1.7&$\pm$&0.023&&&&&&&&&&\\
HD 196761&G037A&G8 V&100&$\pm$&1.0&110&$\pm$&1.9&3.7&$\pm$&2.8&12&$\pm$&2.2&12&$\pm$&0.21&4.7&$\pm$&1.1&5.9&$\pm$&0.10&1.1&$\pm$&2.3&2.3&$\pm$&0.040&&&&&&&&&&\\
HD 196877&K057A&K5.0&55&$\pm$&0.57&54&$\pm$&2.3&3.6&$\pm$&1.9&&&&6.2&$\pm$&0.26&-1.4&$\pm$&1.6&3.0&$\pm$&0.13&-5.7&$\pm$&3.6&1.2&$\pm$&0.050&&&&&&&&&&\\
HD 197076&G117A&G1 V&75&$\pm$&0.75&79&$\pm$&1.2&-4.0&$\pm$&5.2&&&&9.0&$\pm$&0.14&4.7&$\pm$&2.6&4.4&$\pm$&0.068&2.3&$\pm$&3.8&1.7&$\pm$&0.026&&&&&&&&&&\\
HD 197692&F027A&F5 V&420&$\pm$&4.2&420&$\pm$&7.9&52&$\pm$&7.8&&&&48&$\pm$&0.89&25&$\pm$&2.7&23&$\pm$&0.43&7.6&$\pm$&3.2&8.9&$\pm$&0.17&&&&&&&&&&\\
HD 199260&F105A&F6 V&120&$\pm$&1.2&120&$\pm$&2.3&48&$\pm$&2.2&48&$\pm$&2.9&13&$\pm$&0.26&27&$\pm$&1.8&6.5&$\pm$&0.12&11&$\pm$&4.2&2.5&$\pm$&0.048&16&&$^{+2.9}_{-1.3}$&79&&$^{+3.8}_{-18}$&18&&$^{+12}_{-1.6}$&23\\
HD 200525&F079A&F9.5&170&$\pm$&1.9&160&$\pm$&4.3&2.0&$\pm$&5.1&&&&18&$\pm$&0.49&12&$\pm$&2.2&8.9&$\pm$&0.24&7.5&$\pm$&5.6&3.4&$\pm$&0.092&&&&&&&&&&\\
HD 200779&K116A&K6 V&56&$\pm$&0.63&62&$\pm$&1.1&3.0&$\pm$&5.2&&&&7.1&$\pm$&0.13&3.0&$\pm$&2.3&3.5&$\pm$&0.063&8.3&$\pm$&5.6&1.3&$\pm$&0.024&&&&&&&&&&\\
HD 200968&G072A&G9.5 V (k)&66&$\pm$&0.68&78&$\pm$&6.1&&&&&&&9.0&$\pm$&0.71&4.1&$\pm$&2.3&4.4&$\pm$&0.35&-5.9&$\pm$&6.9&1.7&$\pm$&0.13&&&&&&&&&&\\
HD 202275&F066A&F7&370&$\pm$&3.7&360&$\pm$&7.4&45&$\pm$&4.6&&&&41&$\pm$&0.83&24&$\pm$&2.3&20&$\pm$&0.41&2.1&$\pm$&4.7&7.7&$\pm$&0.16&&&&&&&&&&\\
HD 202560&K004A&K7.0&550&$\pm$&6.1&560&$\pm$&9.7&57&$\pm$&5.5&&&&64&$\pm$&1.1&40&$\pm$&3.2&31&$\pm$&0.54&24&$\pm$&15&12&$\pm$&0.21&&&&&&&&&&\\
HD 203244&G110A&G8 V&61&$\pm$&0.68&61&$\pm$&0.95&0.20&$\pm$&4.7&&&&6.9&$\pm$&0.11&4.6&$\pm$&2.0&3.4&$\pm$&0.052&-2.7&$\pm$&3.2&1.3&$\pm$&0.020&&&&&&&&&&\\
HD 203608&F007A&F9 V Fe-1.4 CH-0.7&509&$\pm$&5.1&500&$\pm$&9.0&52&$\pm$&6.1&&&&57&$\pm$&1.0&31&$\pm$&2.8&28&$\pm$&0.50&9.1&$\pm$&4.8&11&$\pm$&0.19&&&&&&&&&&\\
HD 205390&K101A&K1.5 V&73&$\pm$&0.74&77&$\pm$&1.2&11&$\pm$&5.3&14&$\pm$&2.3&9.0&$\pm$&0.14&4.2&$\pm$&1.0&4.4&$\pm$&0.071&7.1&$\pm$&2.3&1.7&$\pm$&0.028&&&&&&&&&&\\
HD 206860&G080A&G0 V CH-0.5&110&$\pm$&1.2&110&$\pm$&2.1&26&$\pm$&2.7&&&&13&$\pm$&0.24&17&$\pm$&1.3&6.1&$\pm$&0.12&3.5&$\pm$&2.5&2.4&$\pm$&0.045&9.4&&$^{+2.0}_{-1.6}$&86&&$^{+9.1}_{-8.3}$&11&&$^{+2.5}_{-2.0}$&9.8\\
HD 207098&F015A&kA5hF0mF2 III&1000&$\pm$&10&1100&$\pm$&20&&&&&&&120&$\pm$&2.3&62&$\pm$&4.2&58&$\pm$&1.1&30&$\pm$&4.0&22&$\pm$&0.43&&&&&&&&&&\\
HD 207129$^*$&G053A&G0 V Fe+0.4&170&$\pm$&1.7&160&$\pm$&3.4&440&$\pm$&7.8&320&$\pm$&28&18&$\pm$&0.39&350&$\pm$&20&8.8&$\pm$&0.19&220&$\pm$&17&3.4&$\pm$&0.073&97&&$^{+5.3}_{-8.5}$&51&&$^{+1.5}_{-2.6}$&33&&$^{+3.6}_{-1.9}$&28\\
HD 209100&K003A&K4 V (k)&&&&1100&$\pm$&9.1&110&$\pm$&6.1&&&&120&$\pm$&1.0&57&$\pm$&3.4&60&$\pm$&0.51&24&$\pm$&3.0&23&$\pm$&0.20&&&&&&&&&&\\
HD 210027&F013A&F5V&660&$\pm$&6.6&650&$\pm$&12&71&$\pm$&5.3&&&&73&$\pm$&1.4&33&$\pm$&2.5&36&$\pm$&0.69&8.6&$\pm$&3.4&14&$\pm$&0.26&&&&&&&&&&\\
HD 210302&F064A&F6 V&220&$\pm$&2.2&220&$\pm$&6.0&8.6&$\pm$&5.7&&&&25&$\pm$&0.68&12&$\pm$&1.9&12&$\pm$&0.33&-0.29&$\pm$&4.0&4.6&$\pm$&0.13&&&&&&&&&&\\
HD 211970&K076A&K5.0&44&$\pm$&0.45&47&$\pm$&0.84&11&$\pm$&5.8&&&&5.4&$\pm$&0.097&4.3&$\pm$&2.0&2.6&$\pm$&0.048&6.5&$\pm$&3.4&1.0&$\pm$&0.018&&&&&&&&&&\\
HD 212330&G113A&G2 IV-V&250&$\pm$&2.5&240&$\pm$&4.2&29&$\pm$&4.3&&&&27&$\pm$&0.47&11&$\pm$&2.1&13&$\pm$&0.23&5.4&$\pm$&3.9&5.1&$\pm$&0.089&&&&&&&&&&\\
HD 213845&F111A&F5 V&160&$\pm$&1.6&160&$\pm$&3.6&-1.4&$\pm$&3.0&&&&18&$\pm$&0.41&15&$\pm$&2.3&8.7&$\pm$&0.20&6.4&$\pm$&3.4&3.4&$\pm$&0.077&&&&&&&&&&\\
HD 214749&K077A&K4.5 V k&71&$\pm$&0.72&78&$\pm$&1.4&8.2&$\pm$&5.9&&&&8.9&$\pm$&0.16&2.4&$\pm$&1.9&4.3&$\pm$&0.078&1.6&$\pm$&2.7&1.7&$\pm$&0.030&&&&&&&&&&\\
HD 214953&F126A&F9.5 V&95&$\pm$&1.1&99&$\pm$&2.2&6.6&$\pm$&5.2&&&&11&$\pm$&0.25&2.7&$\pm$&2.0&5.4&$\pm$&0.12&5.5&$\pm$&5.1&2.1&$\pm$&0.046&&&&&&&&&&\\
HD 215648&F043A&F7V&480&$\pm$&4.8&509&$\pm$&10&55&$\pm$&3.7&&&&58&$\pm$&1.2&25&$\pm$&2.8&28&$\pm$&0.56&8.3&$\pm$&4.0&11&$\pm$&0.22&&&&&&&&&&\\
HD 216133&K088A&K7&&&&35&$\pm$&1.7&&&&&&&4.0&$\pm$&0.19&2.5&$\pm$&1.9&2.0&$\pm$&0.095&2.6&$\pm$&3.2&0.76&$\pm$&0.037&&&&&&&&&&\\
HD 216803&K019A&K4+ V k&220&$\pm$&2.2&229&$\pm$&3.2&26&$\pm$&3.1&&&&26&$\pm$&0.37&12&$\pm$&2.1&13&$\pm$&0.18&6.4&$\pm$&4.7&4.9&$\pm$&0.070&&&&&&&&&&\\
HD 217107&G102A&G8&110&$\pm$&1.1&110&$\pm$&2.1&6.3&$\pm$&5.6&&&&13&$\pm$&0.24&2.6&$\pm$&1.7&6.1&$\pm$&0.12&-0.66&$\pm$&3.3&2.4&$\pm$&0.045&&&&&&&&&&\\
HD 217357&K022A&K5&140&$\pm$&1.4&140&$\pm$&2.4&22&$\pm$&2.4&&&&16&$\pm$&0.27&9.9&$\pm$&1.3&7.8&$\pm$&0.13&3.5&$\pm$&3.3&3.0&$\pm$&0.052&&&&&&&&&&\\
HD 218511&K114A&K5.5 V (k)&55&$\pm$&0.56&56&$\pm$&3.0&0.32&$\pm$&5.3&12&$\pm$&1.2&6.4&$\pm$&0.35&16&$\pm$&2.5&3.1&$\pm$&0.17&17&$\pm$&3.2&1.2&$\pm$&0.066&20&&$^{+7.2}_{-5.5}$&31&&$^{+4.4}_{-13}$&31&&$^{+60}_{-7.0}$&8.1\\
HD 219482$^*$&F087A&F6 V&140&$\pm$&1.4&120&$\pm$&2.4&79&$\pm$&2.0&72&$\pm$&4.8&14&$\pm$&0.27&39&$\pm$&2.4&6.8&$\pm$&0.13&16&$\pm$&1.7&2.6&$\pm$&0.050&34&&$^{+1.9}_{-1.1}$&90&&$^{+1.8}_{-2.9}$&13&&$^{+0.90}_{-0.51}$&39\\
HD 219571&F117A&F4 V&509&$\pm$&5.1&490&$\pm$&9.1&54&$\pm$&4.9&&&&55&$\pm$&1.0&33&$\pm$&2.7&27&$\pm$&0.50&2.8&$\pm$&4.7&10&$\pm$&0.19&&&&&&&&&&\\
HD 221503&K112A&K5&49&$\pm$&0.55&53&$\pm$&1.0&19&$\pm$&9.1&&&&6.1&$\pm$&0.12&0.20&$\pm$&2.4&3.0&$\pm$&0.059&-1.0&$\pm$&3.8&1.2&$\pm$&0.023&&&&&&&&&&\\
HD 222237&K052A&K3+ V&110&$\pm$&1.1&110&$\pm$&1.7&16&$\pm$&1.8&&&&13&$\pm$&0.19&5.0&$\pm$&1.1&6.2&$\pm$&0.093&1.9&$\pm$&1.6&2.4&$\pm$&0.036&&&&&&&&&&\\
HD 222335&G087A&G9.5 V&57&$\pm$&0.58&58&$\pm$&1.0&0.86&$\pm$&5.5&&&&6.6&$\pm$&0.12&2.6&$\pm$&2.0&3.2&$\pm$&0.057&6.2&$\pm$&3.9&1.2&$\pm$&0.022&&&&&&&&&&\\
HD 222368&F021A&F7V&530&$\pm$&5.3&530&$\pm$&12&71&$\pm$&5.5&&&&60&$\pm$&1.3&43&$\pm$&3.1&29&$\pm$&0.64&13&$\pm$&3.4&11&$\pm$&0.25&1.1&&$^{+0.80}_{-0.32}$&60&&$^{+16}_{-31}$&41&&$^{+140}_{-16}$&4.6\\
HD 224953&K118A&K5.0&&&&32&$\pm$&1.2&0.25&$\pm$&4.9&&&&3.8&$\pm$&0.14&3.2&$\pm$&2.2&1.8&$\pm$&0.070&8.7&$\pm$&5.5&0.71&$\pm$&0.027&&&&&&&&&&\\
HD 234078&K091A&K5&50&$\pm$&0.50&55&$\pm$&2.0&11&$\pm$&5.1&&&&6.2&$\pm$&0.23&0.82&$\pm$&2.0&3.1&$\pm$&0.11&7.7&$\pm$&3.3&1.2&$\pm$&0.044&&&&&&&&&&\\
HIP 171&G020A&G5&210&$\pm$&2.1&220&$\pm$&4.2&27&$\pm$&4.9&&&&25&$\pm$&0.49&12&$\pm$&1.2&12&$\pm$&0.24&9.5&$\pm$&2.8&4.7&$\pm$&0.091&&&&&&&&&&\\
HIP 1368$^*$&K115A&K7&46&$\pm$&0.51&51&$\pm$&1.2&22&$\pm$&8.7&17&$\pm$&3.3&6.0&$\pm$&0.14&42&$\pm$&6.2&2.9&$\pm$&0.070&60&$\pm$&8.3&1.1&$\pm$&0.027&98&&$^{+16}_{-9.7}$&28&&$^{+3.2}_{-4.6}$&33&&$^{+14}_{-6.5}$&13\\
HIP 2762&F092A&F8.5&220&$\pm$&2.2&220&$\pm$&5.5&21&$\pm$&6.0&&&&25&$\pm$&0.62&18&$\pm$&2.4&12&$\pm$&0.30&6.5&$\pm$&3.5&4.6&$\pm$&0.12&&&&&&&&&&\\
HIP 13375&K108A&K5&&&&29&$\pm$&0.61&&&&&&&3.4&$\pm$&0.071&2.4&$\pm$&1.9&1.7&$\pm$&0.035&-4.0&$\pm$&3.7&0.65&$\pm$&0.013&&&&&&&&&&\\
HIP 14954$^*$&F110A&F8.5&240&$\pm$&2.4&250&$\pm$&5.2&49&$\pm$&3.8&&&&28&$\pm$&0.59&44&$\pm$&3.5&14&$\pm$&0.28&30&$\pm$&3.4&5.3&$\pm$&0.11&3.8&&$^{+0.72}_{-0.54}$&30&&$^{+15}_{-7.6}$&170&&$^{+140}_{-97}$&12\\
HIP 18280&K124A&K7&43&$\pm$&0.49&48&$\pm$&1.2&7.4&$\pm$&6.0&&&&5.5&$\pm$&0.14&4.3&$\pm$&1.8&2.7&$\pm$&0.070&-2.5&$\pm$&3.7&1.0&$\pm$&0.027&&&&&&&&&&\\
HIP 23452&K023A&K7&140&$\pm$&1.4&140&$\pm$&4.4&14&$\pm$&3.0&&&&16&$\pm$&0.50&9.4&$\pm$&2.1&8.1&$\pm$&0.24&6.1&$\pm$&2.9&3.1&$\pm$&0.095&&&&&&&&&&\\
HIP 27188&K082A&K7&51&$\pm$&0.57&56&$\pm$&1.3&5.2&$\pm$&5.1&&&&6.4&$\pm$&0.15&5.0&$\pm$&2.3&3.1&$\pm$&0.071&-9.6&$\pm$&4.0&1.2&$\pm$&0.028&&&&&&&&&&\\
HIP 31634&K119A&K7&&&&53&$\pm$&2.5&&&&&&&6.1&$\pm$&0.29&2.2&$\pm$&2.0&3.0&$\pm$&0.14&11&$\pm$&4.6&1.2&$\pm$&0.054&&&&&&&&&&\\
HIP 37279&F001A&F5&&&&14000&$\pm$&370&&&&&&&1600&$\pm$&42&810&$\pm$&46&780&$\pm$&20&350&$\pm$&19&300&$\pm$&7.8&&&&&&&&&&\\
HIP 37288&K099A&K7&&&&38&$\pm$&2.0&&&&&&&4.4&$\pm$&0.23&4.4&$\pm$&2.0&2.1&$\pm$&0.11&3.0&$\pm$&3.6&0.84&$\pm$&0.045&&&&&&&&&&\\
HIP 38382&G055A&G0&250&$\pm$&2.5&240&$\pm$&6.2&17&$\pm$&5.5&&&&28&$\pm$&0.70&14&$\pm$&2.0&13&$\pm$&0.34&7.3&$\pm$&3.1&5.2&$\pm$&0.13&&&&&&&&&&\\
HIP 42220&K083A&K7&&&&58&$\pm$&1.7&&&&&&&6.6&$\pm$&0.20&4.3&$\pm$&2.3&3.2&$\pm$&0.096&6.6&$\pm$&3.9&1.3&$\pm$&0.037&&&&&&&&&&\\
HIP 42748&K121A&K7&&&&47&$\pm$&1.3&&&&&&&5.4&$\pm$&0.15&4.9&$\pm$&1.9&2.6&$\pm$&0.071&9.2&$\pm$&3.2&1.0&$\pm$&0.027&&&&&&&&&&\\
HIP 43820&K049A&K5&&&&130&$\pm$&3.1&&&&&&&15&$\pm$&0.36&6.5&$\pm$&2.1&7.1&$\pm$&0.17&-0.17&$\pm$&3.3&2.8&$\pm$&0.068&&&&&&&&&&\\
HIP 44722&K097A&K7&&&&48&$\pm$&1.4&&&&&&&5.5&$\pm$&0.17&3.0&$\pm$&1.5&2.7&$\pm$&0.081&1.7&$\pm$&4.5&1.0&$\pm$&0.032&&&&&&&&&&\\
HIP 47080&G017A&G8&&&&260&$\pm$&5.3&&&&&&&29&$\pm$&0.61&12&$\pm$&2.0&14&$\pm$&0.29&9.0&$\pm$&2.8&5.5&$\pm$&0.11&&&&&&&&&&\\
HIP 52600&K086A&K7&&&&60&$\pm$&2.0&&&&&&&6.9&$\pm$&0.23&2.0&$\pm$&1.9&3.4&$\pm$&0.11&-0.18&$\pm$&4.7&1.3&$\pm$&0.044&&&&&&&&&&\\
HIP 55203&F004A&G0&990&$\pm$&9.9&980&$\pm$&36&110&$\pm$&8.1&&&&110&$\pm$&4.1&56&$\pm$&3.8&54&$\pm$&2.0&25&$\pm$&4.4&21&$\pm$&0.77&&&&&&&&&&\\
HIP 61094&K081A&K7&&&&40&$\pm$&2.3&&&&&&&4.6&$\pm$&0.27&2.6&$\pm$&1.9&2.3&$\pm$&0.13&-3.3&$\pm$&4.7&0.88&$\pm$&0.051&&&&&&&&&&\\
HIP 61941&F014A&F5.5&1400&$\pm$&14&1300&$\pm$&29&&&&&&&150&$\pm$&3.3&79&$\pm$&4.8&74&$\pm$&1.6&39&$\pm$&5.2&28&$\pm$&0.61&&&&&&&&&&\\
HIP 63366&G116A&G9&&&&45&$\pm$&0.48&-5.4&$\pm$&5.2&&&&5.2&$\pm$&0.055&-0.42&$\pm$&2.1&2.5&$\pm$&0.027&-13&$\pm$&5.2&0.97&$\pm$&0.010&&&&&&&&&&\\
HIP 64241&F055A&F5.5&380&$\pm$&3.8&390&$\pm$&7.2&52&$\pm$&5.1&&&&44&$\pm$&0.82&24&$\pm$&2.6&21&$\pm$&0.40&14&$\pm$&4.1&8.2&$\pm$&0.15&&&&&&&&&&\\
HIP 66459&K041A&K5&&&&60&$\pm$&1.4&&&&&&&6.9&$\pm$&0.17&2.1&$\pm$&1.6&3.4&$\pm$&0.081&3.4&$\pm$&4.2&1.3&$\pm$&0.032&&&&&&&&&&\\
HIP 67090&K070A&K5&&&&34&$\pm$&1.3&&&&&&&3.9&$\pm$&0.15&1.6&$\pm$&1.7&1.9&$\pm$&0.071&8.2&$\pm$&3.8&0.73&$\pm$&0.028&&&&&&&&&&\\
HIP 70218&K096A&K6 V&51&$\pm$&0.57&54&$\pm$&0.77&6.5&$\pm$&4.9&&&&6.1&$\pm$&0.087&2.9&$\pm$&1.7&3.0&$\pm$&0.042&-1.7&$\pm$&4.4&1.1&$\pm$&0.016&&&&&&&&&&\\
HIP 73470&K055A&K7&&&&72&$\pm$&1.7&&&&&&&8.3&$\pm$&0.20&4.7&$\pm$&1.8&4.0&$\pm$&0.097&-1.2&$\pm$&3.7&1.6&$\pm$&0.038&&&&&&&&&&\\
HIP 73695&G025A&G2&450&$\pm$&5.0&430&$\pm$&8.9&76&$\pm$&5.8&&&&49&$\pm$&1.0&29&$\pm$&2.7&24&$\pm$&0.49&18&$\pm$&10&9.3&$\pm$&0.19&7.7&&$^{+5.9}_{-2.1}$&110&&$^{+15}_{-34}$&9.3&&$^{+11}_{-2.2}$&5.2\\
HIP 75312&G081A&G2&250&$\pm$&2.5&250&$\pm$&4.6&42&$\pm$&5.8&&&&29&$\pm$&0.52&12&$\pm$&1.9&14&$\pm$&0.25&6.3&$\pm$&3.9&5.4&$\pm$&0.097&&&&&&&&&&\\
HIP 86036&G034A&G0&250&$\pm$&2.5&250&$\pm$&4.5&30&$\pm$&4.6&&&&28&$\pm$&0.52&12&$\pm$&1.6&14&$\pm$&0.25&4.3&$\pm$&2.9&5.3&$\pm$&0.097&&&&&&&&&&\\
HIP 88745$^*$&F037A&F7&270&$\pm$&2.7&270&$\pm$&5.5&110&$\pm$&5.6&94&$\pm$&10&31&$\pm$&0.63&88&$\pm$&10&15&$\pm$&0.30&80&$\pm$&15&5.9&$\pm$&0.12&14&&$^{+0.82}_{-1.7}$&50&&$^{+3.2}_{-2.8}$&46&&$^{+5.5}_{-5.4}$&18\\
HIP 93017&F029A&G0&250&$\pm$&2.5&250&$\pm$&4.4&27&$\pm$&5.1&&&&28&$\pm$&0.50&17&$\pm$&2.4&14&$\pm$&0.24&2.0&$\pm$&3.9&5.3&$\pm$&0.094&&&&&&&&&&\\
HIP 99461&K010A&K2.5&&&&459&$\pm$&6.0&51&$\pm$&8.5&&&&53&$\pm$&0.69&24&$\pm$&2.3&26&$\pm$&0.34&8.3&$\pm$&4.6&10&$\pm$&0.13&&&&&&&&&&\\
HIP 105312&G083A&G7&110&$\pm$&1.1&110&$\pm$&1.9&11&$\pm$&6.7&12&$\pm$&2.1&13&$\pm$&0.21&9.8&$\pm$&1.1&6.2&$\pm$&0.10&8.9&$\pm$&3.9&2.4&$\pm$&0.040&1.6&&$^{+1.7}_{-0.67}$&16&&$^{+14}_{-2.5}$&280&&$^{+110}_{-200}$&3.4\\
HIP 107310&F106A&F6&340&$\pm$&3.4&350&$\pm$&6.9&39&$\pm$&4.8&&&&39&$\pm$&0.78&22&$\pm$&2.5&19&$\pm$&0.38&14&$\pm$&4.3&7.3&$\pm$&0.15&&&&&&&&&&\\
HIP 110109&G031A&G0&220&$\pm$&2.5&229&$\pm$&4.3&25&$\pm$&5.7&&&&26&$\pm$&0.49&8.9&$\pm$&1.8&12&$\pm$&0.24&-3.3&$\pm$&6.3&4.8&$\pm$&0.091&&&&&&&&&&\\
HIP 110778&G107A&G5&160&$\pm$&1.8&170&$\pm$&3.0&0.27&$\pm$&9.0&&&&19&$\pm$&0.34&6.1&$\pm$&2.1&9.1&$\pm$&0.17&0.76&$\pm$&4.9&3.5&$\pm$&0.064&&&&&&&&&&\\
GJ 55.3A&F093A&F5 V&&&&200&$\pm$&3.0&&&&&&&23&$\pm$&0.35&9.2&$\pm$&2.2&11&$\pm$&0.17&0.75&$\pm$&5.4&4.3&$\pm$&0.065&&&&&&&&&&\\
GJ 4A&K050A&K7&&&&64&$\pm$&0.75&15&$\pm$&6.0&&&&7.3&$\pm$&0.087&1.8&$\pm$&2.1&3.6&$\pm$&0.042&-8.4&$\pm$&4.8&1.4&$\pm$&0.016&&&&&&&&&&\\
\hline
\end{longtable}
\end{landscape}
\twocolumn

\section{Confused sources}\label{ApB}
In Section 3, six sources are identified as having no disc, contradicting the existing publications. This Appendix contains the \emph{Herschel}-PACS data for these sources and the PSF model subtracted residuals for each image. The asterisk shows the expected source location, and the plus sign shows the fitted location. All primary sources are specified by the UNS designation. Any background sources identified are also fit to provide clean photometry, these are identified by the same ID, with a X1, X2 etc. suffix, up to the number of background sources included in the fitting. The black dashed circle shows the beam size for these data. In all cases both pairs of data are shown, i.e. 70 or 100\,$\mu$m and the associated 160\,$\mu$m images.

\begin{figure}
  \includegraphics[width=\columnwidth]{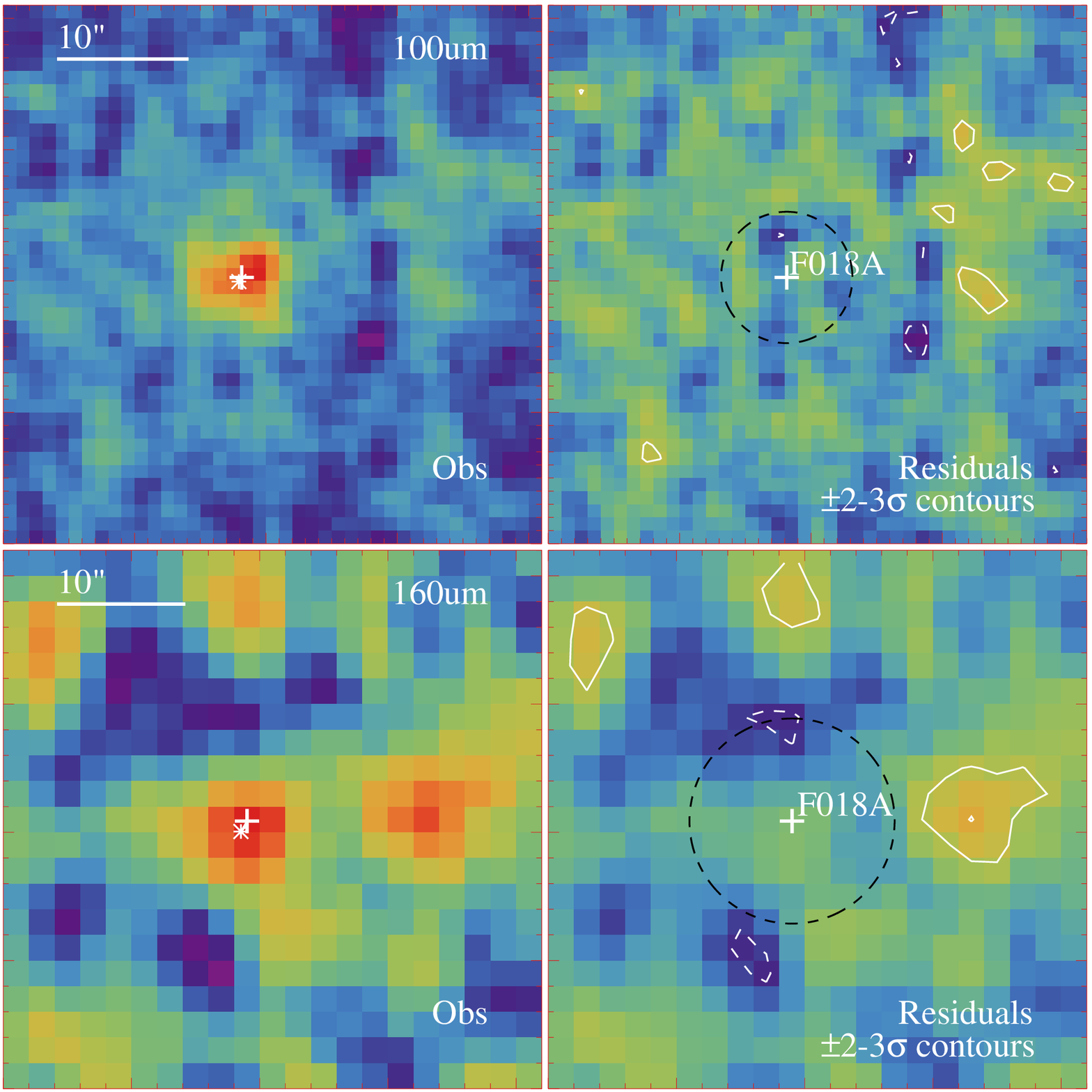}
  \caption{Data and model fit for at 100 and 160\,$\mu$m}
\end{figure}

\begin{figure}
  \includegraphics[width=\columnwidth]{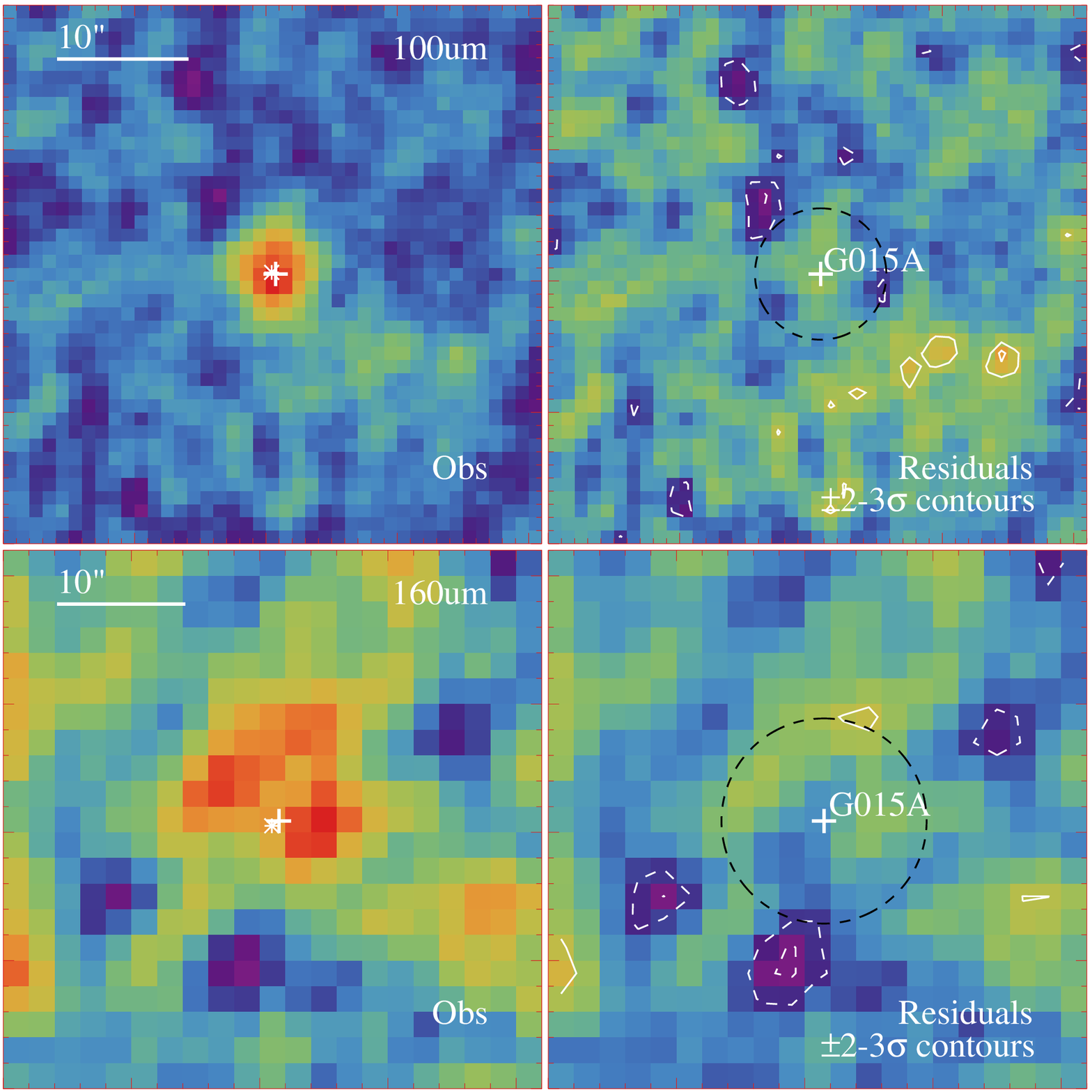}
  \caption{Data and model fit for HD 90839 (HIP 51459, F018) at 100 and 160\,$\mu$m}
\end{figure}

\begin{figure}
  \includegraphics[width=\columnwidth]{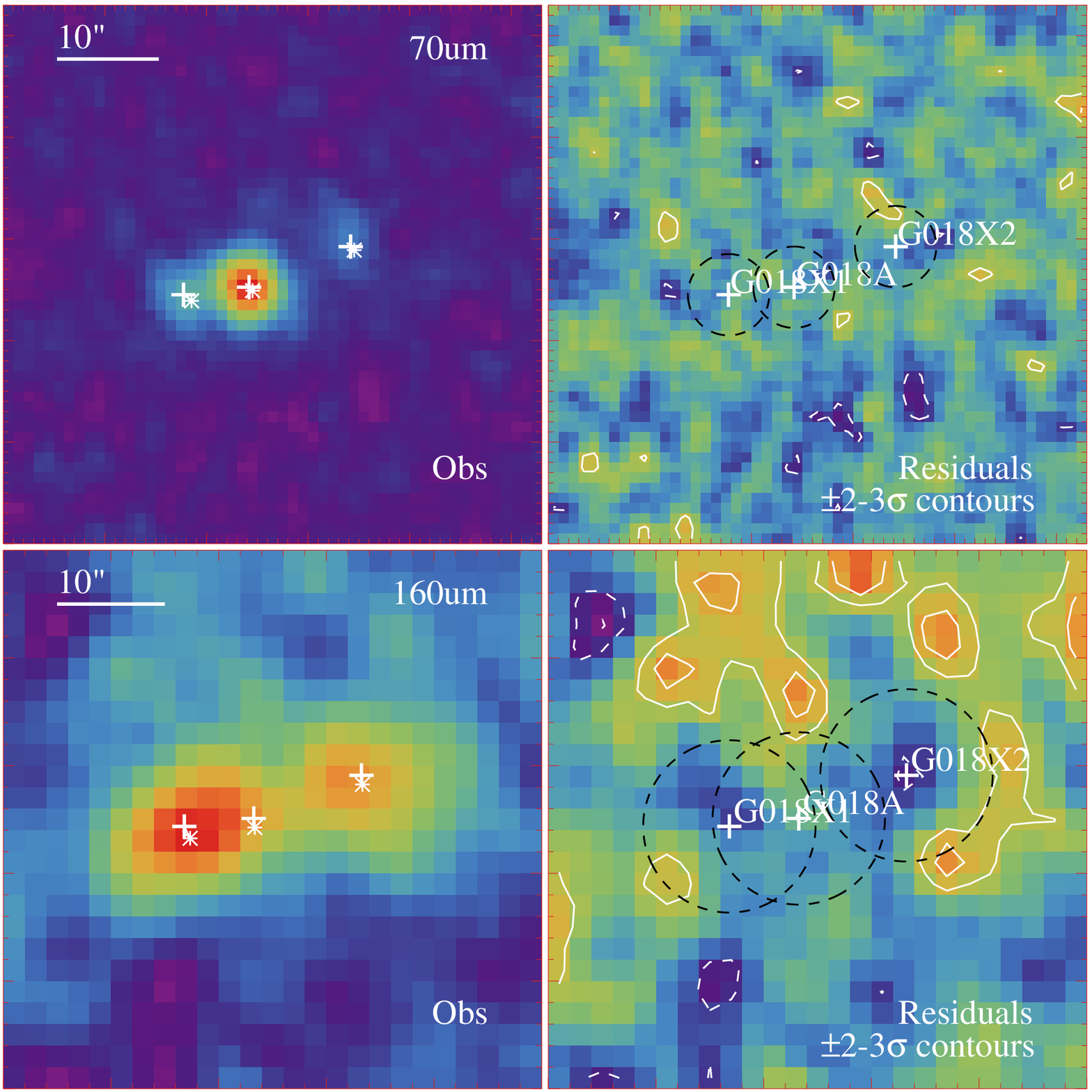}
  \caption{Data and model fit for HD 20907 (HIP 15371, G018) at 70 and 160\,$\mu$m}
\end{figure}

\begin{figure}
  \includegraphics[width=\columnwidth]{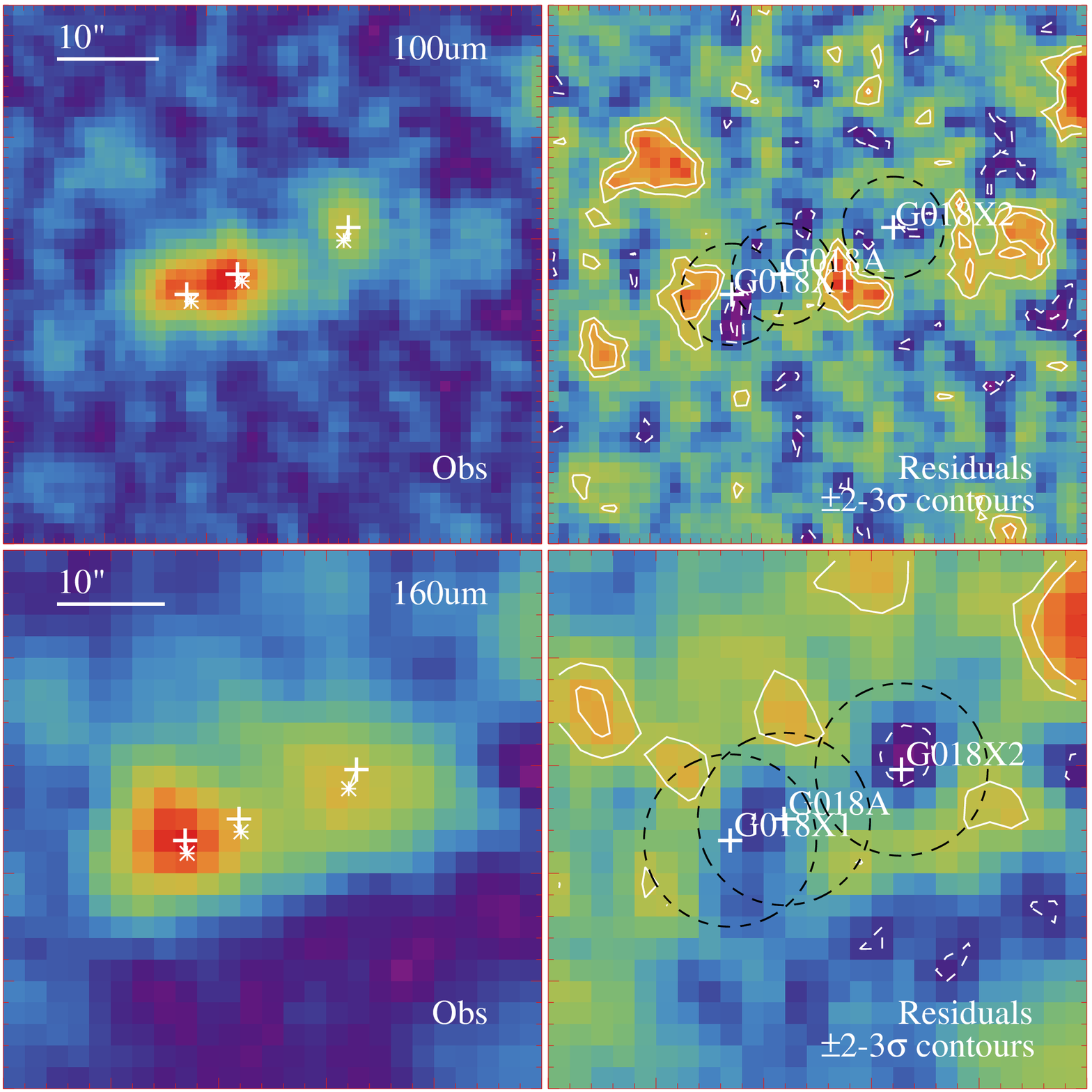}
  \caption{Data and model fit for HD 20907 (HIP 15371, G018) at 100 and 160\,$\mu$m}
\end{figure}

\begin{figure}
  \includegraphics[width=\columnwidth]{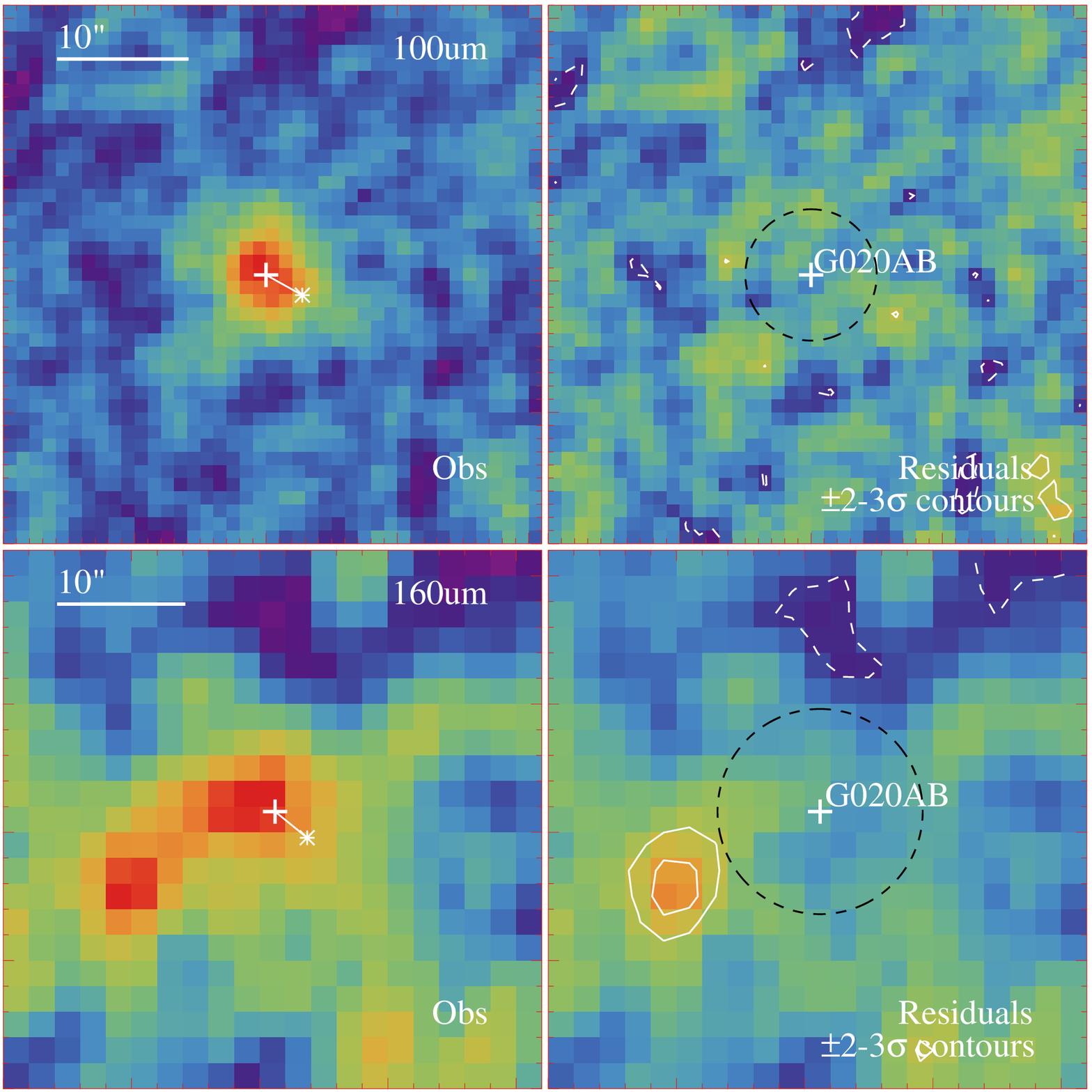}
  \caption{Data and model fit for HD 224930 (HIP 171, G020) at 100 and 160\,$\mu$m}
\end{figure}

\begin{figure}
  \includegraphics[width=\columnwidth]{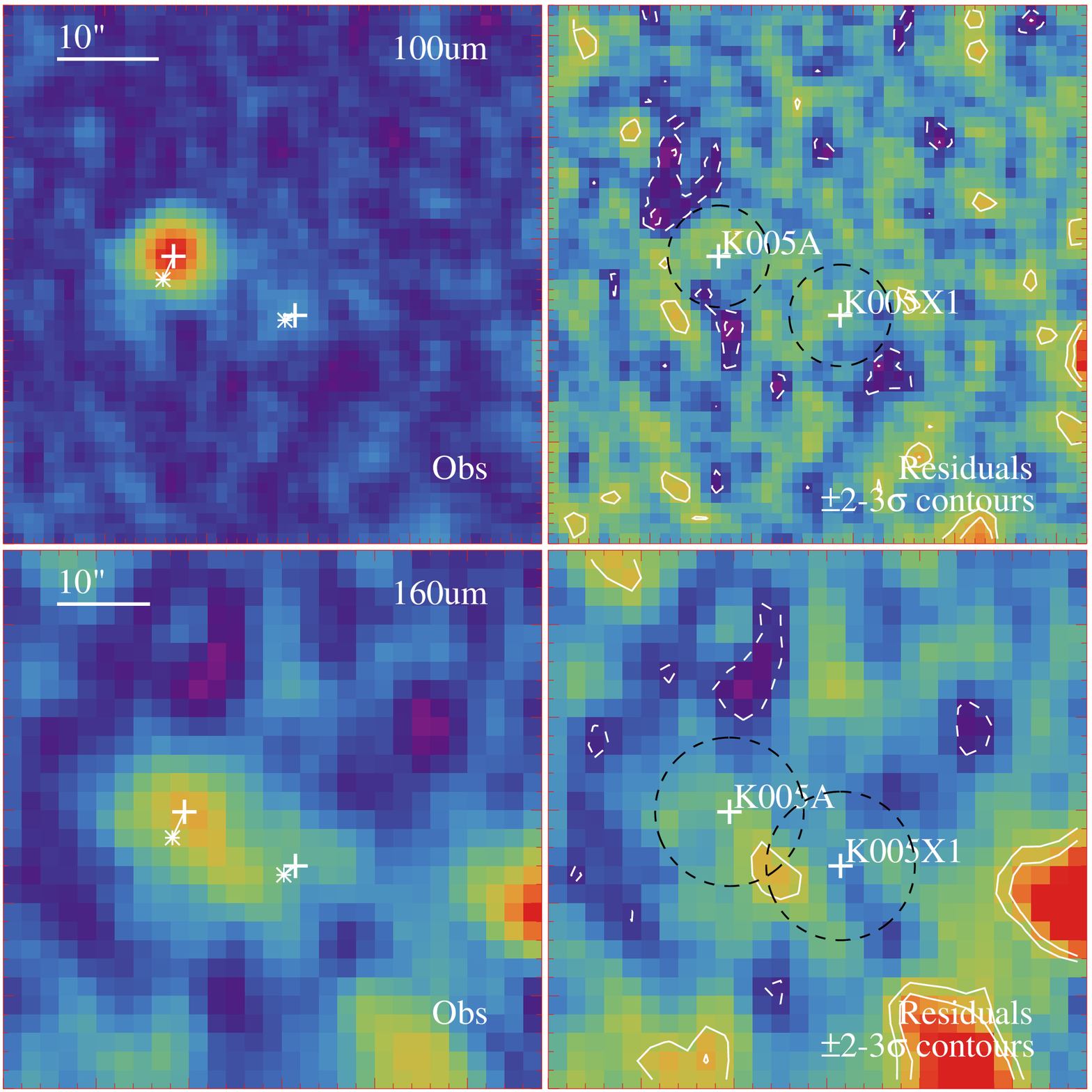}
  \caption{Data and model fit for HD 88230 (HIP 49908, K005) at 100 and 160\,$\mu$m}
\end{figure}

\begin{figure}
  \includegraphics[width=\columnwidth]{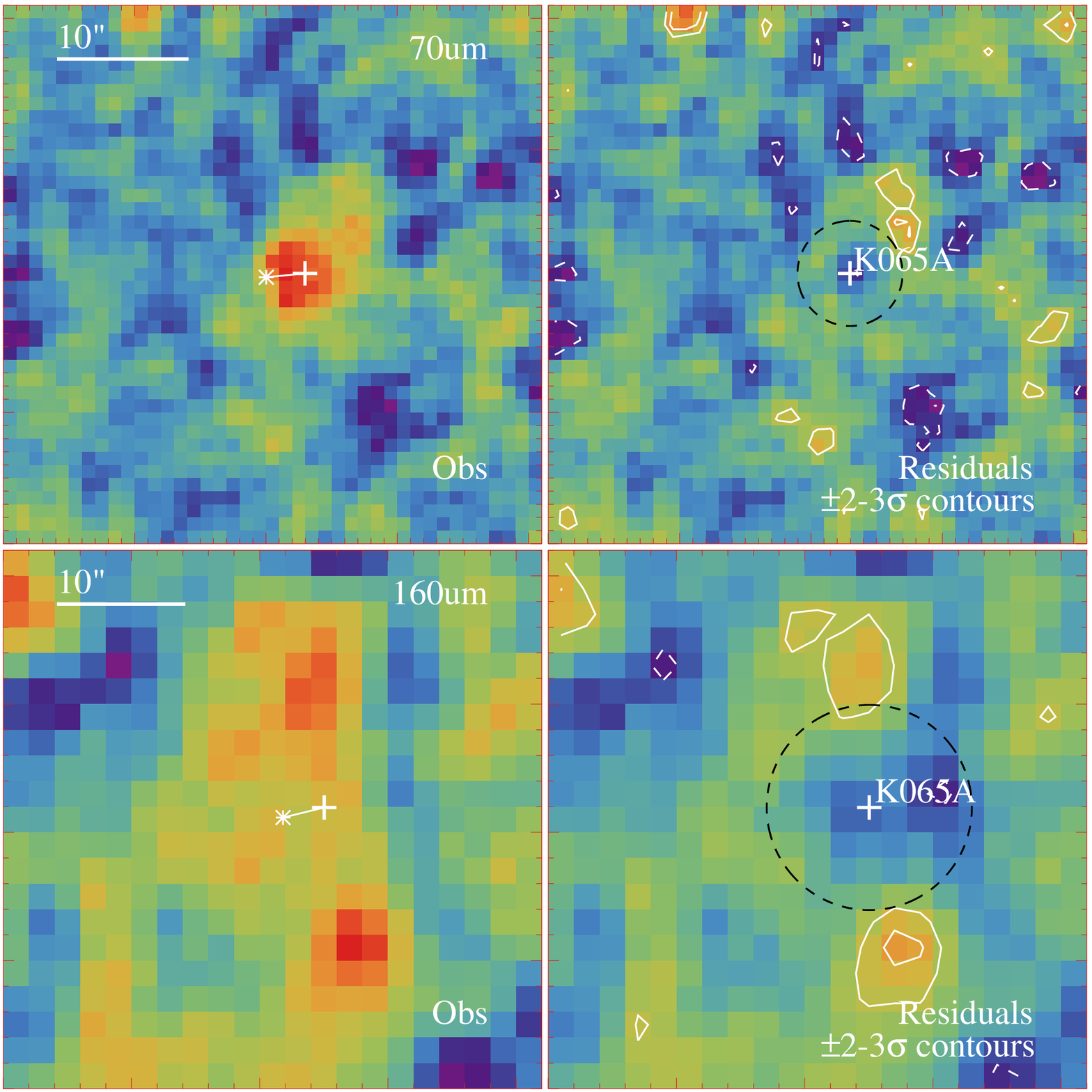}
  \caption{Data and model fit for HD 40307 (HIP 27887, K065) at 70 and 160\,$\mu$m}
\end{figure}

\begin{figure}
  \includegraphics[width=\columnwidth]{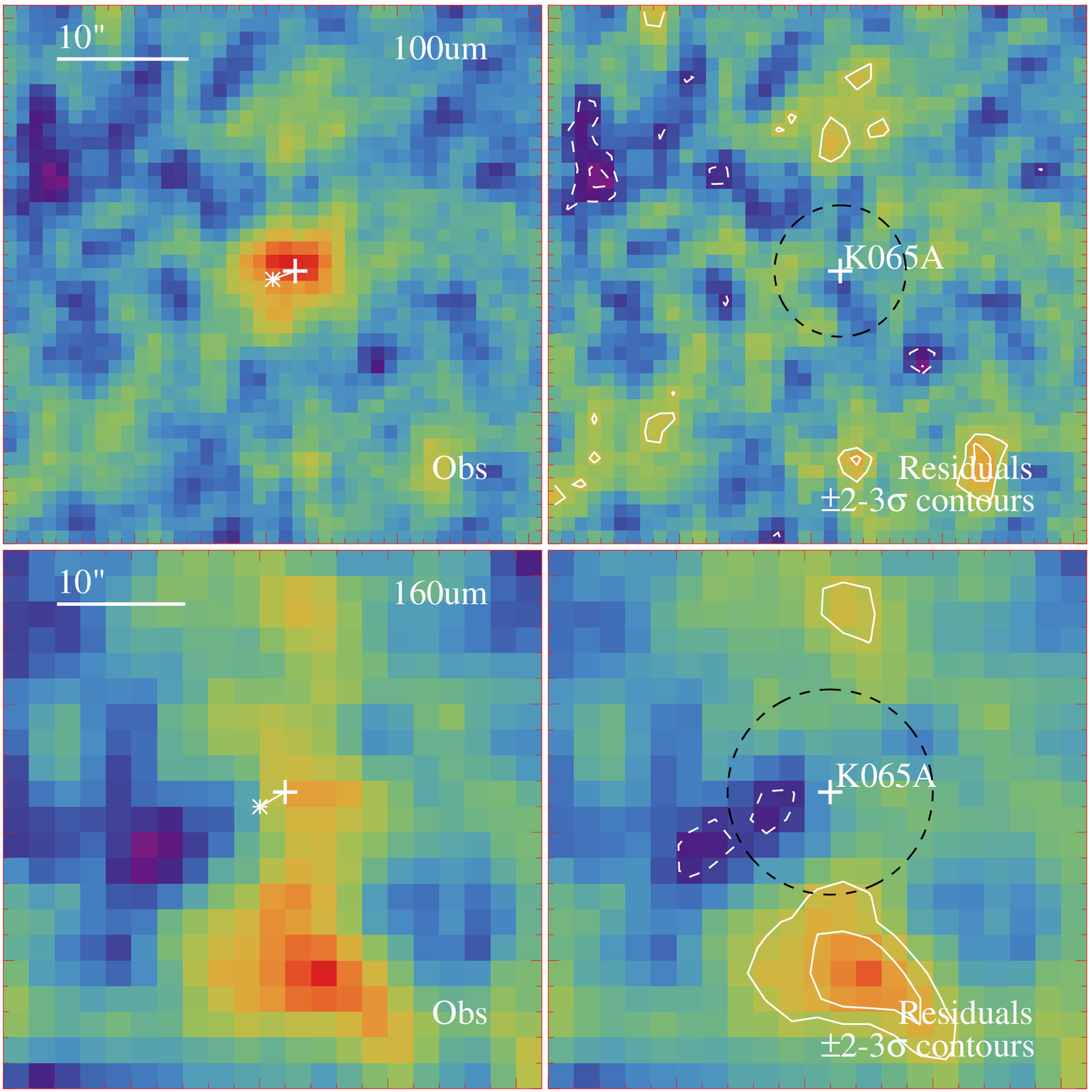}
  \caption{Data and model fit for HD 40307 (HIP 27887, K065) at 100 and 160\,$\mu$m}
\end{figure}

%\bsp
\label{lastpage}
\end{document}